\documentclass[aps,twocolumn,epsf,psfig,dvips]{revtex4}
\usepackage{feynmp}
\usepackage{amssymb}
\usepackage{epsfig}
\usepackage{graphicx}
\usepackage{subfigure}
\usepackage{mathrsfs}
\usepackage{hyperref}
\usepackage{cases}
\usepackage{bbm}
\usepackage{xcolor}
\usepackage{multirow}
\usepackage{makecell}
\usepackage{array}
\usepackage{threeparttable}

\usepackage{lmodern}
\usepackage{textcomp}

\begin{document}



\title{Singular low-energy states of tilted Dirac semimetals induced by the fermion-fermion interactions}

\date{\today}

\author{Jie-Qiong Li}
\affiliation{Department of Physics, Tianjin University, Tianjin 300072, P.R. China}

\author{Dong-Xing Zheng}
\affiliation{Department of Physics, Tianjin University, Tianjin 300072, P.R. China}

\author{Jing Wang}
\altaffiliation{Corresponding author: jing$\textunderscore$wang@tju.edu.cn}
\affiliation{Department of Physics, Tianjin University, Tianjin 300072, P.R. China}

\begin{abstract}
We attentively investigate the effects of short-range fermion-fermion interactions
on the low-energy properties of both two-dimensional type-I and type-II tilted Dirac
semimetals by means of the renormalization group framework. Practicing the standard
renormalization group procedures via taking into account all one-loop corrections
gives rise to the coupled energy-dependent evolutions of all interaction parameters,
which are adopted to carefully examine whether and how the fermion-fermion interactions
influence the low-energy physical behaviors of tilted Dirac fermions. After carrying out
the detailed analysis of coupled flows, we figure out the tilting parameter dictates the
low-energy states of tilted Dirac fermions in conjunction with starting values of
fermion-fermion couplings. With proper variations of these two kinds of parameters,
the tilted Dirac fermions can either flow towards the Gaussian fixed point or undergo
certain instability that is conventionally accompanied by a phase transition in the
low-energy regime. In addition, all potential instabilities can be clustered into five
distinct classes owing to the competitions between the tilting parameter and initial
fermionic interactions. Moreover, the dominant phases accompanied by the instabilities
are determined via computing and comparing the susceptibilities of eight potential phases.
\end{abstract}


\maketitle


\section{Introduction}

Both electronic states and physical properties of Dirac materials including
two-dimensional (2D) graphene~\cite{Novoselov2004Science,Novoselov2005Nature,
Castro2009RMP,Peres2010RMP}, Weyl semimtals (WSMs)~\cite{Burkov2011PRL,Yang2011PRB,Wan2011PRB,Huang2015PRX,Xu2015Science,
Xu2015Nature,Lv2015NP,Weng2015PRX,Roy2018PRX}, and Dirac semimetals (DSMs)~\cite{Vafek2014ARCMP,Wehling2014AP,Wang2012PRB,Young2012PRL,
Steinberg2014PRL,Liu2014NM,Liu2014Science,Xiong2015Science,Roy2009PRB,
Roy2016,Roy-2014-2016,Savary2014PRB,Moon2014PRX,Montambaux} have been poured extensively
attention in the contemporary condensed matter physics. In a sharp contrast to the
conventional Fermi metals featuring a finite Fermi surface~\cite{Altland2006Book},
they generally only possess discrete Dirac nodal points and exhibit a linear quasiparticle
dispersion with gapless low-energy excitations~\cite{Castro2009RMP,Vafek2014ARCMP,Wehling2014AP}.
As a result, the density of states (DOS) vanishes at Dirac
points~\cite{Castro2009RMP,Vafek2014ARCMP} and then induces a multitude of interesting
phenomena including non-trivial topological properties
~\cite{Hasan2010RMP,Qi2011RMP} and non-centrosymmetric WSMs with time-reversal
protected states~\cite{Shekhar2015Nature,Lv2015PRX,Xu2015Science,
Lv2015NP,Yang2015Nature,Xu2015Nature,Xu2016NC}. It is of remarkable significance to
highlight that Dirac/Weyl cones can be stretched and thus tilted by breaking the
fundamental Lorentz symmetry with additional
forces~\cite{Lee2018PRB,Lee2019PRB} or so-called $t$-Lorentz symmetry~\cite{Jafari2019PRB-t}.
In other words, this is equivalent to inducing
anisotropic fermion velocities of energy dispersions for DSMs and/or WSWs.

Recently, tilted Dirac materials have attracted considerable attention in
condensed matter fields. For instance, tilted Dirac cones have been realized in
the two-dimensional (2D) organic compound
$\alpha-(\mathrm{BEDT-TTF})_{2}\mathrm{I}_{3}$ and certain mechanically deformed
graphene~\cite{Katayama2006JPSJ,Kobayashi2007JPSJ,Goerbig2008PRB}. In addition,
the three-dimensional (3D) tilted Weyl cones have been proposed in
$\mathrm{WTe}_{2}$~\cite{Soluyanov2015Nature}, the Fulde-Ferrell ground state
of a spin-orbit coupled fermionic superfluid~\cite{Xu2015PRL}, or a cold-atom optical lattice~\cite{Xu2016PRA}. All these kinds of materials are therefore designated
as tilted DSMs (WSMs), which conventionally can be broken into two distinct types
relying heavily upon the tilted angles. Type-I tilted DSMs (WSMs) still maintain
analogous Dirac (Weyl) cones as long as the tilted angle is insufficient to destroy
the point-like Fermi surface~\cite{Castro2009RMP,Peres2010RMP,Jafari}.
In comparison, the nodal point would be replaced by two straight lines
indicating the open Fermi surface and nonzero DOS, once the tilted angle
is adequately large~\cite{Lee2018PRB}.
We therefore obtain another breed of tilted materials dubbed type-II tilted DSMs (WSMs)~\cite{Soluyanov2015Nature}, which have been recently realized in both $\mathrm{PdTe_2}$~\cite{Noh2017PRL,Fei2017PRB} and $\mathrm{PtTe_2}$~\cite{Yan2017NC}.

Attesting to unique Dirac cones and unconventional low-energy excitations of
tilted DSMs, they become one of the hottest topics in condensed matter
physics~\cite{Shekhar2015NP,Parameswaran2014PRX,Potter2014NC,Baum2015PRX,
Arnold2016NC,Zhang2016NC,Lee2018PRB,Lee2019PRB,Fritz2017PRB,Fritz2019arXiv,
Jafari2018PRB,Alidoust2019arXiv,Yang2018PRB,Trescher2015PRB,Proskurin2015PRB,
Brien2016PRL,Zyuzin2016JETPL,Ferreiros2017PRB}. In particular, the effects of
long-range Coulomb interactions on the low-energy properties of tilted DSMs
were investigated by many groups~\cite{Lee2018PRB,Lee2019PRB,Fritz2017PRB,Fritz2019arXiv}.
Unfortunately, it is well known that the long-range Coulomb interaction would be
easily screened in the realistic systems by adopting some metallic substrate to
enhance the dielectric constant~\cite{Castro2009RMP,Kotov2012RMP,Sarma2011RMP,Katsnelson2006PRB}.
Accordingly, the short-range fermion-fermion interactions, which can induce a plethora of
unusual behaviors in fermionic systems~\cite{Murray2014PRB,Herbut,Herbut-2,Wang2017PRB_QBCP,Wang2018},
tend to play a vital role in pinning down the low-energy physical implications once the
Coulomb interaction is screened in the realistic or excluded by external
forces~\cite{Katsnelson2006PRB,Castro2009RMP,Kotov2012RMP,Sarma2011RMP}.
However, the fermion-fermion interactions are hitherto inadequately taken into account
in previous studies of 2D tilted DSMs. This indicates the physical information that is
closely associated with these locally fermionic interactions
may be partially discarded or cannot be fully captured in the low-energy
regime. In order to improve our understandings for these tilted materials,
it is therefore considerably instructive to carefully investigate whether
and how the fermion-fermion interactions affect the low-energy physical
properties of the 2D tilted DSMs?

Stimulated by these, we within this work endeavor to explore how four distinct
sorts of short-range fermion-fermion interactions affect the low-energy
fates of physical states for 2D type-I and type-II tilted-Dirac materials.
In order to treat all types of these fermion-fermion interactions on the same
footing, it is convenient to adopt the powerful energy-shell renormalization group (RG) approach~\cite{Wilson1975RMP,Polchinski1992arXiv,Shankar1994RMP}. Performing
all one-loop calculations and carrying out the standard RG analysis give rise
to the entangled RG evolutions, which are closely associated with all interaction
parameters and thus carry the full physical information. Several interesting
physical behaviors have been extracted from these coupled RG equations.

To be concrete, we realize that the low-energy states of tilted DSMs are sensitive
to both the tilting parameter $\zeta$ and initial values of fermion-fermion interactions
that are characterized by $|\lambda_{i}(0)|$ with $i=0,1,2,3$ (in order to facilitate
our analysis, we exploit $\lambda_i(0)$ hereafter to denote $\lambda_i(0)$ with $i=0,1,2,3$)
corresponding to different sorts of fermionic interactions. Tuning the values of $\zeta$
and $\lambda_{i}(0)$, the tilted DSMs can be attracted and flow to the Gaussian fixed
point or certain instability associated with some phase transition at a critical energy
scale. For type-I DSMs, both the increase of $\zeta$ and $|\lambda_{i}(0)|$ would be
helpful to activate an instability in the low-energy regime. In comparison, the increase
of $|\lambda_{i}(0)|$ is in favor to trigger certain potential instability for type-II
tilted DSMs but instead tuning up $\zeta$ is harmful to the development of instability.
In addition, these two quantities $\zeta$ and $|\lambda_{i}(0)|$ strongly compete and
exhibit different powers in both type-I and type-II DSMs attesting to qualitatively
different topologies of Fermi surfaces. For the type-I tilted DSMs, $|\lambda_{i}(0)|$
plays a major role in pining down the low-energy states except $\zeta\rightarrow1$ at
which $\zeta$ is responsible for the possible instability. Whereas, the tilting parameter
$\zeta$ dominates the low-energy fates of the type-II tilted DSMs once $\zeta$ is sufficient
large and instead $|\lambda_{i}(0)|$ wins the competition if $\zeta$ is small or
$\zeta\rightarrow1$. Furthermore, we figure out that all underlying instabilities that are induced in
both type-I and type-II tilted DSMs can be divided into five different classes depending
on both the interplays among fermion-fermion interactions and tilting parameter in the
low-energy regime. Specifically, four classes of instabilities can be expected in the
type-I tilted system. Rather, the type-II tilted Dirac fermions
allow two distinct classes of instabilities. The basic results for type-I and type-II
tilted Dirac fermions are provided in Table~\ref{Type-I} and Table~\ref{Type-II}, respectively.
At last, we show that ferromagnet, antiferromagnet, and spin bond density are
more preferable in the vicinity of these instabilities after evaluating and comparing
the susceptibilities of potential phases.

The rest of this paper is structured as follows. In Sec.~\ref{Sec_model},
we introduce our microscopic model and the related effective quantum field theory.
The Sec.~\ref{Sec_eqs} is accompanied to compute all the one-loop corrections to the
interaction parameters, which are utilized to derive the coupled flow equations via
performing the standard RG analysis. We provide detailed analysis of singular low-energy
states caused by the fermion-fermion interactions for type-I and type-II tilted Dirac
fermions in Sec.~\ref{Sec_type-I} and Sec.~\ref{Sec_type-II}, respectively. In
Sec.~\ref{Sec_instability}, we divide all underlying instabilities induced by
fermion-fermion interactions into five distinguished classes depending upon the
specific values of both tilting parameter and starting values of fermionic couplings
and address the dominant phases around all the potential instabilities. Finally,
we provide a brief summary of our primary results in Sec.~\ref{Sec_summary}.

%


\section{Effective theory}\label{Sec_model}

Within this work, we focus on the 2D tilted DSMs at the chemical potential $\mu=0$,
whose non-interacting effective action in the low-energy regime can be written
as~\cite{Lee2018PRB,Goerbig2008PRB}
\begin{eqnarray}
S_{0}
&=&\sum_{\xi,\alpha}\int\frac{dp_0}{2\pi}\int\frac{d^2\mathbf{p}}{(2\pi)^2}
\psi_{\xi\alpha}^{\dag}(p_0,\mathbf{p})(-ip_0+\xi\zeta v_1 p_1\nonumber\\
&&+\xi v_1 p_1\sigma_{1}+p_2\sigma_{2})\psi_{\xi\alpha}(p_0,\mathbf{p}),\label{Eq_S-0}
\end{eqnarray}
with the valley degeneracy $\xi=\pm1$, spin degeneracy $\alpha=\pm1$,
and Pauli matrices $\sigma_{i}$ satisfying the anticommutation algebra
$\{\sigma_i,\sigma_j\}=2\delta_{ij}$  ($i,j=1,2,3$). Here, the fermionic spinors
$\psi_{\xi\alpha}(p_0,\mathbf{p})$ and $\psi^\dagger_{\xi\alpha}(p_0,\mathbf{p})$
are exploited to characterize the excited quasiparticles around the nodal points
at $\mathbf{K}$ ($\mathbf{-K}$) in the first Brillouin zone. The dimensionless
parameter $\zeta$ that is served as a tilting variable is able to tilt the Dirac
cones and reshape the structure of Fermi surface as long as $\zeta$ is nonzero.
As a result, it gives rises to two inequivalent fermion velocities $v_1$ and $v_2$
along the $x$ and $y$ directions, respectively.

Subsequently, one can straightforwardly extract the free fermionic propagator
from the non-interacting action, namely
\begin{eqnarray}
G_0(ip_0,\mathbf{p})&=&\frac{1}{-ip_0+\xi\zeta v_1 p_1+\xi v_1 p_1\sigma_{1}
+p_2\sigma_{2}}.\label{Eq_G0}
\end{eqnarray}
With the help of this free action~(\ref{Eq_S-0}), the energy eigenvalues can
be derived as follows,
\begin{eqnarray}
\epsilon_{\pm}(\mathbf{p})
&=&\xi\zeta (v_1p_{1})\pm\sqrt{(v_1p_{1})^{2}+(v_2p_{2})^{2}},
\label{Eq_eigen-energy}
\end{eqnarray}
which intimately hinge upon the value of $\zeta$ and principally govern the
overall structure of Fermi surface. As a consequence, tilted DSMs can be
manifestly clustered into distinct sorts via tuning the tilting parameter
$\zeta$~\cite{Lee2018PRB}. At $|\zeta|<1$, the DSMs still possess a point-like
Fermi surface only with simply tilted Dirac cones and thus belong to the type-I
Dirac systems. In sharp comparison, $|\zeta|>1$ would completely sabotage the
point-like structure of Fermi surface and yield an open Fermi surface that
consists of two crossed lines, namely $(v_2p_2)=\pm\sqrt{\zeta^2-1}
(v_1p_1)$~\cite{Lee2018PRB,Lee2019PRB}.
Under such circumstance, it goes to the type-II Dirac systems.

To proceed, we would like to bear in mind that the primary concern of this work is
to examine the low-energy behaviors of both two-dimensional type-I and type-II tilted
Dirac semimetals against fermion-fermion interactions. For this purpose, we are going
to bring out the four potential sorts of short-range fermion-fermion interactions~\cite{Nandkishore2013PRB,Potirniche2014PRB,Wang2017PRB},
\begin{widetext}
\begin{eqnarray}
S_{\mathrm{int}}&=&\sum_{i=0}^{3}\sum_{\alpha,\xi,\alpha^{\prime},\xi^{\prime}}
\lambda_{i}\int_{-\infty}^{+\infty}
\frac{dp_{0}^{\prime}dp_{0}^{\prime
\prime}dp_{0}^{\prime\prime\prime}}
{(2\pi)^{3}}\int\frac{d^{2}\mathbf{p}^{\prime}d^{2}\mathbf{p}^{\prime\prime}
d^{2}\mathbf{p}^{\prime\prime\prime}}
{(2\pi)^{6}}\psi_{\xi\alpha}^{\dagger}(p_{0}^{\prime},\mathbf{p}
^{\prime})\sigma_i\psi_{\xi\alpha}
(p_{0}^{\prime\prime},\mathbf{p}^{\prime\prime})\psi_{\xi^{\prime}
\alpha^{\prime}}^{\dag}(p_{0}^{\prime\prime\prime},\mathbf{p}^{\prime\prime\prime})
\sigma_i\nonumber\\
&&\times
\psi_{\xi^{\prime}\alpha^{\prime}}(p_{0}^{\prime}+p_{0}^{\prime\prime}-p_{0}^{\prime\prime\prime},
\mathbf{p}^{\prime}+\mathbf{p}^{\prime\prime}-\mathbf{p}^{\prime\prime\prime}),\label{Eq_S-int}
\end{eqnarray}
\end{widetext}
where the coupling $\lambda_i$ combined with its corresponding Pauli matrix
$\sigma_i$ are exploited together to designate certain breed of fermion-fermion interaction.
For completeness, we run $i$ from $0$ to $3$ and hence collect four distinct types
of fermionic interactions.

\begin{figure*}[htbp]
\centering
\includegraphics[width=2.12in]{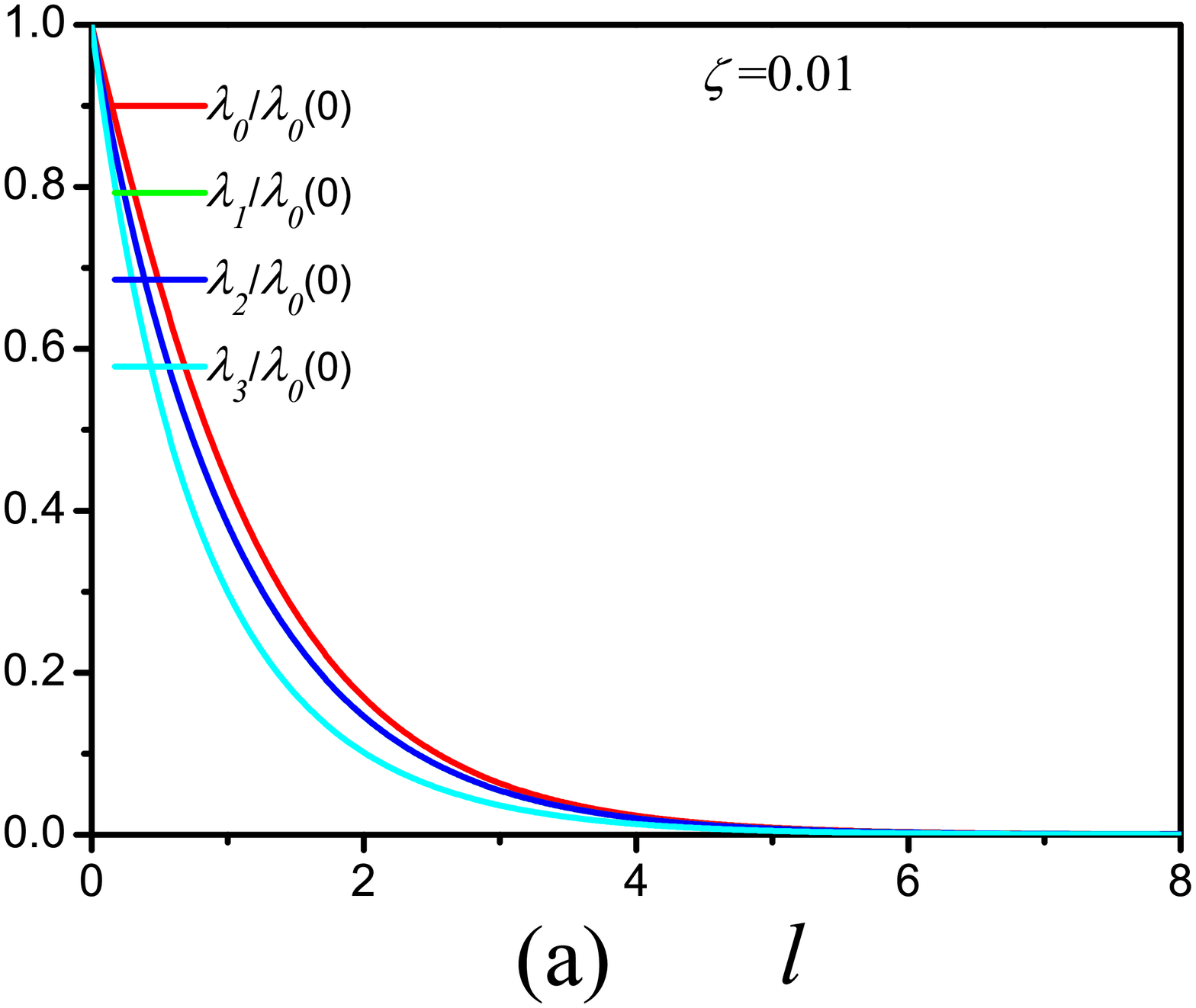}
\hspace{-1.43cm}
\includegraphics[width=2.12in]{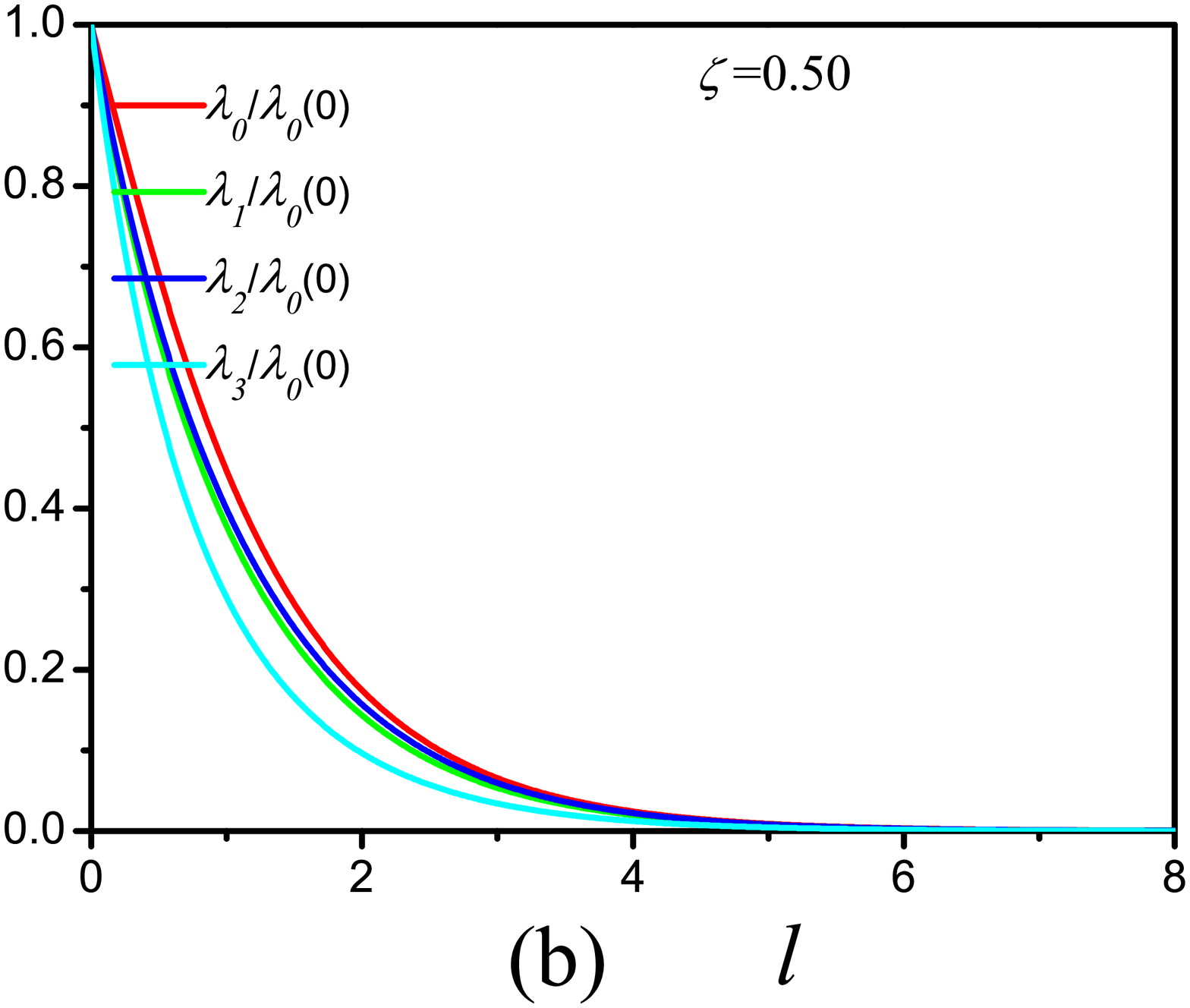}
\hspace{-1.43cm}
\includegraphics[width=2.12in]{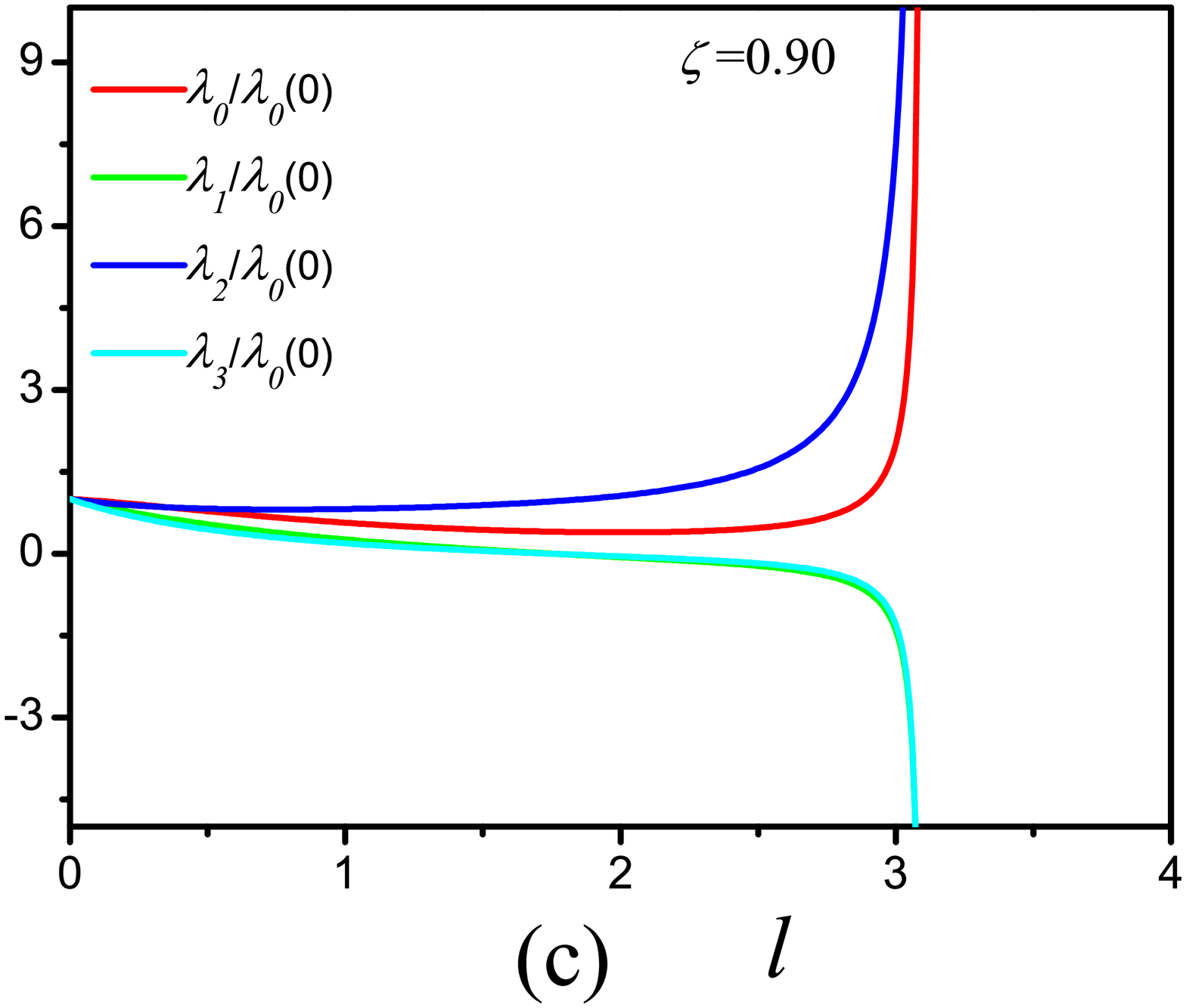}
\hspace{-1.43cm}
\includegraphics[width=2.12in]{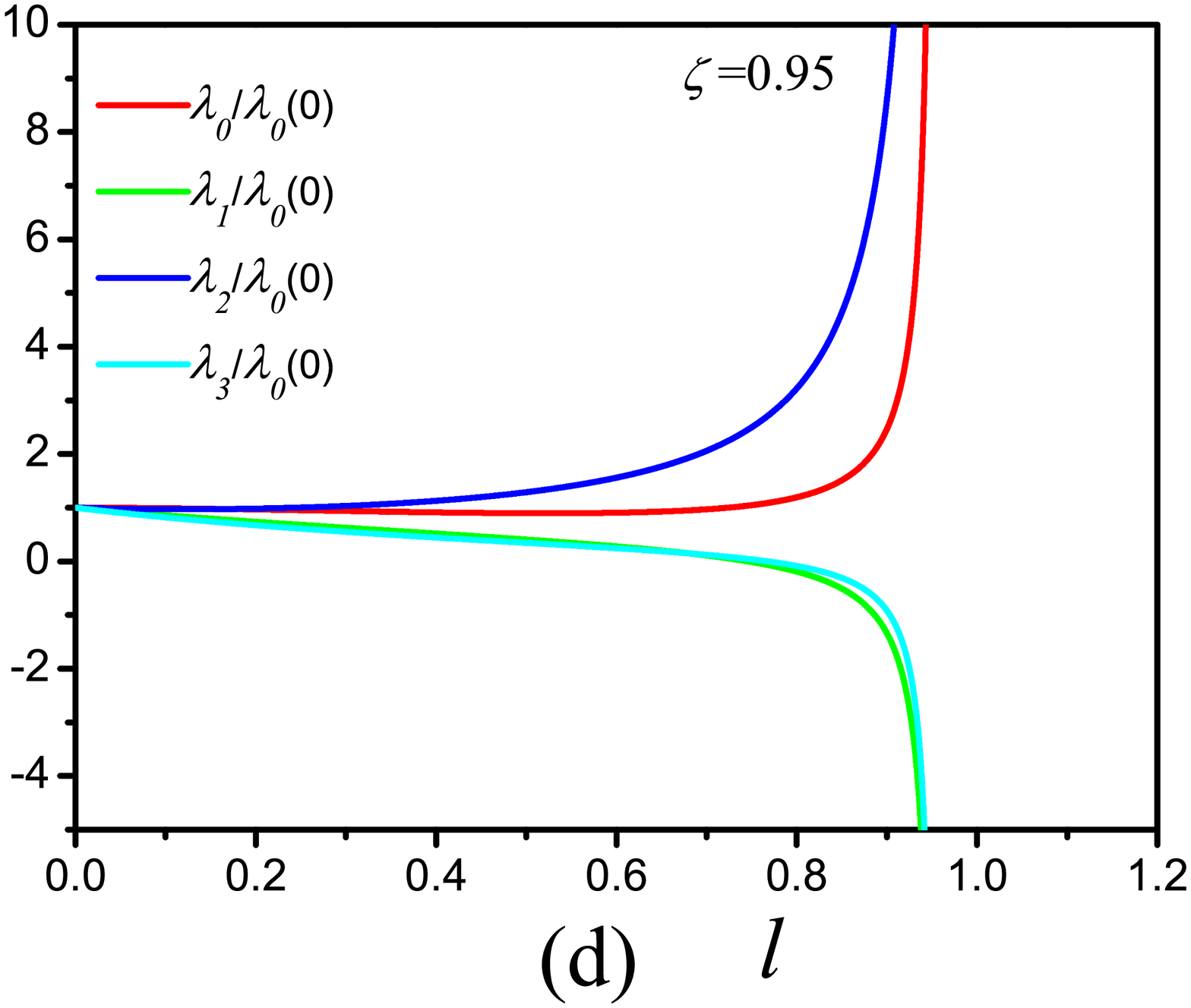}\\
\vspace{-0.05cm}
\caption{(Color online) Energy-dependent evolutions of fermion-fermion strengths
$\lambda_i(l)/\lambda_0(l=0)$ with distinct typical values of tilting parameter
$\zeta$ and $\lambda_i(l=0)=10^{-4}$ as well as $v_1=v_2=10^{-2}$ (the basic results
for $\lambda_i(0)<0$ are similar and hence not shown here).}\label{Fig-1}
\end{figure*}

At this stage, we are left with the effective theory after taking into account the
non-interacting action~(\ref{Eq_S-0}) in tandem with fermionic interactions~(\ref{Eq_S-int}),
\begin{eqnarray}
S_{\mathrm{eff}}=S_0(p_j\rightarrow \widetilde{p}_{j}/v_j)+S_{\mathrm{int}}(p_j\rightarrow \widetilde{p}_{j}/v_j),\label{Eq_S-eff}
\end{eqnarray}
where the index $j=1,2$. In order to simplify our notions and further analysis,
we hereafter introduce two reduced momenta $\widetilde{p}_{1} \equiv v_1p_1$ and
$\widetilde{p}_{2} \equiv v_2p_2$ to our effective theory. Based on this effective
theory, we are going to study the low-energy properties under the presences of
potential short-range fermion-fermion interactions and the competitions among them.


\section{RG evolutions}\label{Sec_eqs}

As aforementioned in Sec.~\ref{Sec_model}, the Fermi surfaces of 2D tilted DSMs
can either be closed or open relying closely upon the concrete value of tilting
parameter. Generally, following the spirt of RG approach~\cite{Shankar1994RMP},
one can practice the RG process by eliminating and reshaping the thin momentum shells
to approach the Fermi surface and construct the RG evolutions of related parameters
for certain physical system, whose Fermi surface is well-defined (finite) and
closed~\cite{Altland2006Book}. Based on the unusual structures of tilted Dirac
fermions, we are now forced to give up momentum-shell RG but instead implement
energy-shell RG. It implies that one needs to integrate out a thin energy shell
one by one during the RG analysis~\cite{Shankar1994RMP,Lee2018PRB,Huh2008PRB,She2010PRB,Wang2011PRB}.

In order to work in the energy-shell framework, we hereby parallel the
strategies advocated in Ref.~\cite{Lee2018PRB} and keep in mind the difference of
energy dispersions between type-I and type-II tilted DSMs, whose equal-energy
curves correspond to ellipses and hyperbolas, respectively. As a result,
one is required to parametrize these two distinct sorts of equal-energy
curves separately. To this end, we exploit the following transformations from $(\widetilde{p}_{1},\widetilde{p}_{2})$ with $\widetilde{p}_i=v_ip_i$ to ($E$, $\theta$)~\cite{Lee2018PRB,Lee2019PRB},
\begin{eqnarray}
\mathrm{Type-I\!:} \widetilde{p}_{1}\!&=&\!
\frac{|E|\cos\theta-\xi\zeta E}{1-\zeta^{2}},\,
\widetilde{p}_{2}=\frac{|E|\sin\theta}{\sqrt{1-\zeta^{2}}},\label{Eq_parametrize-type-I}\\
\mathrm{Type-II\!:} \widetilde{p}_{1}\!&=&\!
\frac{\xi\zeta E\pm|E|\cosh\theta}{\zeta^{2}-1},\,
\widetilde{p}_{2}=\frac{|E|\sinh\theta}{\sqrt{\zeta^{2}-1}},\label{Eq_parametrize-type-II}
\end{eqnarray}
where $E$ and $\theta$ designate the eigenvalues of energy $\epsilon_{\pm}(\mathbf{p})$~(\ref{Eq_eigen-energy}) and ``\emph{effective
angle}" between $\widetilde{p}_{1}$ and $\widetilde{p}_{2}$. Owing to qualitative distinctions
of equal-energy curves between type-I and type-II tilted Dirac fermions, it is therefore
worth pointing out that the variable $\theta$ at certain $E$ is restricted to $0\leq\theta< 2\pi$ and $-\infty<\theta< \infty$ for type-I and type-II situations, respectively. To proceed,
we exploit the following Jacobian transformation $d\widetilde{p}_{1}d\widetilde{p}_{2}=|J|dEd\theta$
at $E>0$ (the results for $E<0$ can be obtained similarly),
\begin{eqnarray}
J_{\mathrm{type-I}}&=&\left|
  \begin{array}{cc}
    \frac{\partial \widetilde{p}_{1}}{\partial E} & \frac{\partial \widetilde{p}_{1}}{\partial\theta}\\\\
     \frac{\partial \widetilde{p}_{2}}{\partial E} & \frac{\partial \widetilde{p}_{2}}{\partial \theta}  \\
  \end{array}
\right|
=\left|
  \begin{array}{cc}
    \frac{\cos\theta-\xi\zeta}{1-\zeta^{2}} & -\frac{E\sin\theta}{1-\zeta^{2}} \\\\
     \frac{\sin\theta}{\sqrt{1-\zeta^{2}}} & \frac{E\cos\theta}{\sqrt{1-\zeta^{2}}},  \\
  \end{array}
\right|,\\
J_{\mathrm{type-II}}&=&\left|
  \begin{array}{cc}
    \frac{\partial \widetilde{p}_{1}}{\partial E} & \frac{\partial \widetilde{p}_{1}}{\partial\theta}\\\\
     \frac{\partial \widetilde{p}_{2}}{\partial E} & \frac{\partial \widetilde{p}_{2}}{\partial \theta}  \\
  \end{array}
\right|
=\left|
  \begin{array}{cc}
    \frac{\cosh\theta\pm\xi\zeta}{\zeta^{2}-1} & \pm\frac{E\sinh\theta}{\zeta^{2}-1} \\\\
     \frac{\sinh\theta}{\sqrt{\zeta^{2}-1}} & \frac{E\cosh\theta}{\sqrt{\zeta^{2}-1}}  \\
  \end{array}
\right|,
\end{eqnarray}

Based on these the momentum integrals in the effective action~(\ref{Eq_S-eff})
are accordingly casted into~\cite{Lee2018PRB,Lee2019PRB}
\begin{eqnarray}
&&\int_\mathrm{Type-I} \!\!\!\!\!\!\!\!\!\!\!\!\!d^2\widetilde{\mathbf{p}}
\!=\!\frac{1}{2}\int^\Lambda_{-\Lambda}\frac{|E|dE}{(1-\zeta^2)^{\frac{3}{2}}}
\int^{2\pi}_0\!\!\!\!\!d\theta (1-\eta_E\xi\zeta\cos\theta),\label{Eq_intergal-I}\\
&&\int_\mathrm{Type-II} \!\!\!\!\!\!\!\!\!\!\!\!\!d^2\widetilde{\mathbf{p}}
\!=\!\frac{1}{2}\int^\Lambda_{-\Lambda}\frac{|E|dE}{(\zeta^2-1)^{\frac{3}{2}}}
\left[\int^{\infty}_{-\infty}\!\!\!\!\!d\theta (|\zeta|\cosh\theta
+\eta_E\eta_{\zeta}\xi)\right.\nonumber\\
&&\left.\hspace{1.35cm}+\int^{\infty}_{-\infty}\!\!\!\!\!d\theta (|\zeta|\cosh\theta-\eta_E\eta_{\zeta}\xi)\right].\label{Eq_intergal-II}
\end{eqnarray}
Here, integrals over $\theta$ for a fixed $E$ in ~(\ref{Eq_intergal-I})
and ~(\ref{Eq_intergal-II}) denote the ellipse and the left (or right) branch of
equal-energy hyperbola for type-I and type-II cases, respectively.
In addition, the $\eta_X$ with $X=E,\zeta$ collects the signs of $E$ and $\zeta$,
namely $\eta_X\equiv\mathrm{sgn(X)}$. Moreover, the parameter $\Lambda$ characterizes
the ultra cutoff that is directly associated with the lattice constant.

Before moving further, one also needs the rescaling transformations that are intimate
bridges connecting two successive RG steps. To proceed, we can choose the $-ip_0$
term in the non-interacting action~(\ref{Eq_S-0}) as the fixed point following the
spirit of standard RG approach~\cite{Shankar1994RMP}, which is invariant under the whole RG process.
As a consequence, one with the help of Eqs.~(\ref{Eq_intergal-I}) and (\ref{Eq_intergal-II}) can straightforwardly derive the RG transformations for fields and other quantities
~\cite{Shankar1994RMP,Wang2011PRB,Lee2018PRB,Lee2019PRB,Huh2008PRB,She2010PRB},
\begin{eqnarray}
p_0&\rightarrow& p'_0=e^{-l}p_0,\label{Eq_scaling-1}\\
E&\rightarrow& E'=e^{-l}E,\\
\theta &\rightarrow& \theta'=\theta,\\
\psi &\rightarrow& \psi'=e^{2l}\psi,\label{Eq_scaling-2}
\end{eqnarray}
where the variable parameter $l$ specifies an energy scale that is closely linked with
the cutoff, namely $E=\Lambda e^{-l}$.

At this stage, we are in an appropriate situation to perform RG analysis.
Concretely, we follow the strict procedures of RG approach~\cite{Shankar1994RMP,Lee2018PRB}
and carry out tedious but straightforward calculations for all one-loop corrections provided in Appendix~\ref{Appendix_1L-corrections}. At first, it is worth pointing out that one-loop diagrams
contributed by fermion-fermion interactions do not give rise to any corrections owning to
conservations of momentum and energy. This indicates that free propagator cannot gain any
corrections and hence fermion velocities and tilting parameter are invariant under RG analysis.
\begin{widetext}
\begin{eqnarray}
\frac{d\lambda_0}{dl}
\!\!\!&=&\!\!\!-\lambda_0\!+\!\frac{\left[\zeta^{2}\left(\lambda_{0}^2+\lambda_{1}^2+\lambda_{2}^{2}
+\lambda_{3}^2-2\lambda_{0}\lambda_{2}\right)
+2\left(\zeta^*-1\right)\lambda_{0}(\lambda_{1}-\lambda_{2})\right]}
{2\pi v_{1}v_{2}\zeta^{2}\zeta^*},\label{Eq_RG-type-I-lambda-0}\\
\frac{d\lambda_1}{dl}
\!\!\!&=&\!\!\!-\lambda_1\!+\!\frac{2\zeta^{2}[\lambda_{1}(2\lambda_0+2\lambda_1
-\lambda_2-\lambda_3)-\lambda_0\lambda_3]\!+\!(\zeta^*\!-\!1)[\lambda^2_0+5\lambda^2_1
+\lambda^2_2+\lambda^2_3+2\lambda_1(\lambda_0-\lambda_2-\lambda_3)
-2\lambda_0\lambda_3]}{2\pi v_{1}v_{2}\zeta^{2}\zeta^*},\\
\frac{d\lambda_2}{dl}
\!\!\!&=&\!\!\!-\lambda_2
\!+\!\frac{\left\{\left(1-\zeta^*\right)
\left[\lambda_{0}^2+\lambda_{1}^{2}+5\lambda_{2}^{2}+\lambda_{3}^{2}
+2\lambda_{2}(\lambda_{0}-\lambda_{1}-\lambda_{3})
-2\lambda_{0}\lambda_{3}\right]
-\zeta^{2}\left(\lambda_{0}^2+\lambda_{1}^{2}+\lambda_{2}^{2}+\lambda_{3}^{2}-
2\lambda_{0}\lambda_{2}\right)\right\}}{2\pi v_{1}v_{2}\zeta^{2}\zeta^*},\\
\frac{d\lambda_3}{dl}
\!\!\!&=&\!\!\!-\lambda_3\!+\!
\frac{2\left\{\zeta^{2}\left[
(\lambda_{1}+\lambda_{2}-2\lambda_{3})\lambda_{3}-\lambda_{0}\lambda_{1}\right]
-\left(\zeta^*-1\right)\lambda_{0}\left(\lambda_{1}-\lambda_{2}\right)
\right\}}{2\pi v_{1}v_{2}\zeta^2\zeta^*},\label{Eq_RG-type-I-lambda-3}
\end{eqnarray}
\end{widetext}
for type-I tilted Dirac fermions, and
\begin{eqnarray}
\frac{d\lambda_0}{dl}
\!\!\!&=&\!\!\!-\lambda_0\!+\!\frac{2\zeta^\star\lambda_{0}\lambda_{1}}
{\pi^2\zeta|\zeta|v_{1}v_{2}},\label{Eq_RG-type-II-lambda-0}\\
\frac{d\lambda_1}{dl}
\!\!\!&=&\!\!\!-\lambda_1\!+\!\frac{\zeta^\star(\lambda^2_0+\lambda^2_1
+\lambda^2_2+\lambda^2_3)}
{2\pi^2\zeta|\zeta| v_{1}v_{2}},\\
\frac{d\lambda_2}{dl}
\!\!\!&=&\!\!\!
-\lambda_2\!+\!\frac{2\zeta^\star[\lambda_{0}\lambda_{3}\!-\!\lambda_{2}(\lambda_{0}-\lambda_{1}
+2\lambda_{2}-\lambda_{3})]}
{\pi^2\zeta|\zeta|v_{1}v_{2}},\\
\frac{d\lambda_3}{dl}
\!\!\!&=&\!\!\!-\lambda_3\!+\!\frac{2\zeta^\star\lambda_{0}\lambda_{2}}
{\pi^2\zeta|\zeta|v_{1}v_{2}}.\label{Eq_RG-type-II-lambda-3}
\end{eqnarray}
for type-II tilted Dirac fermions. Two coefficients $\zeta^*$ and $\zeta^\star$ are designated as
\begin{eqnarray}
\zeta^*\equiv\sqrt{1-\zeta^2}, \hspace{0.5cm} \zeta^\star\equiv2|\zeta|
-\sqrt{\zeta^2-1},\label{Eq_zeta-star}
\end{eqnarray}
to write above RG equations more compactly. We hereby would like to highlight that
several approximations are exploited during the derivations of the RG equations for
type-II tilted DSMs as provided in Appendix~\ref{Appendix_1L-corrections}.
Next, we can extract the low-energy behaviors that are influenced or even
dictated by the fermion-fermion interactions from above coupled RG evolutions of
all interaction parameters.

\begin{figure*}[htbp]
\centering
\includegraphics[width=2.12in]{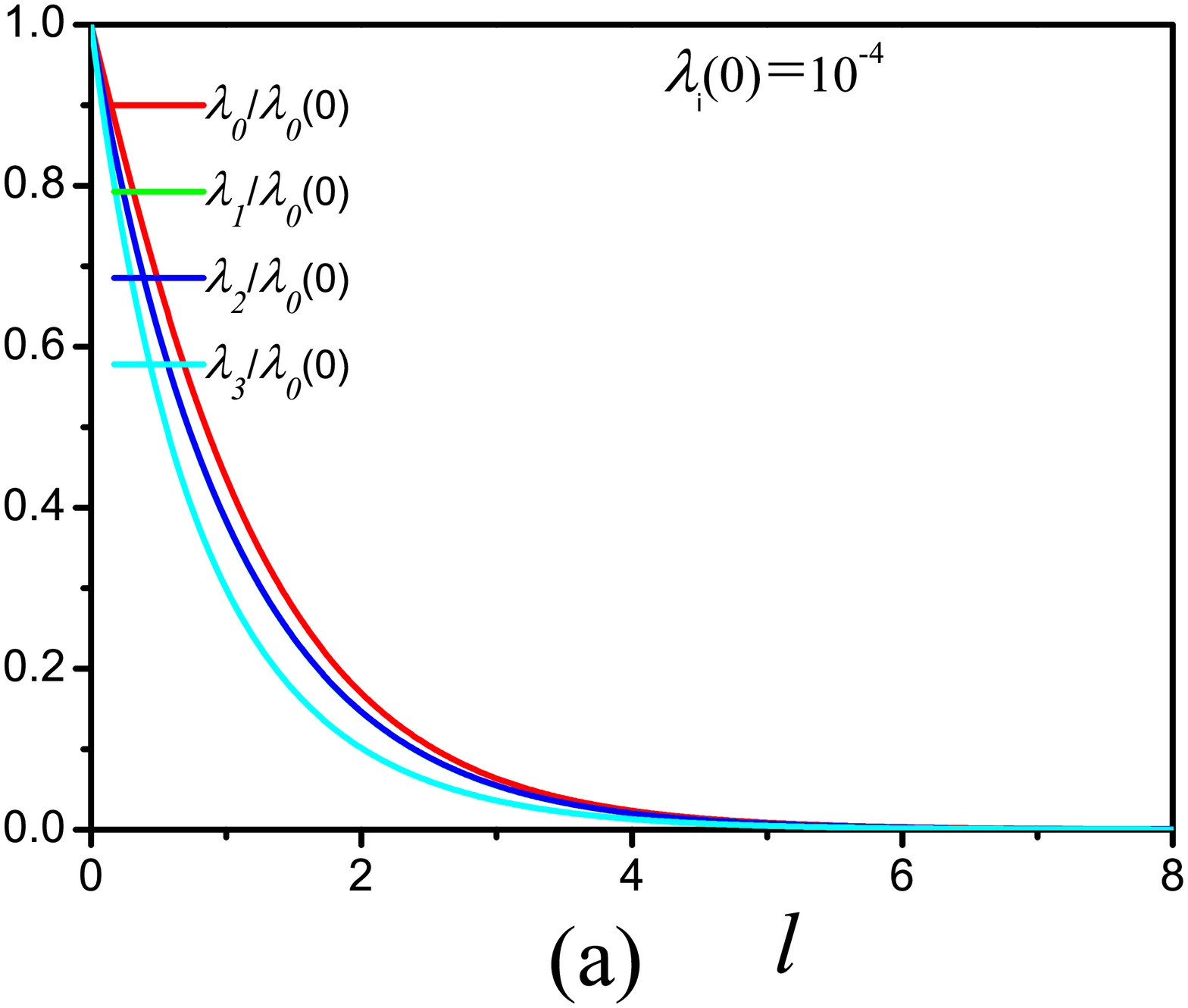}
\hspace{-1.43cm}
\includegraphics[width=2.12in]{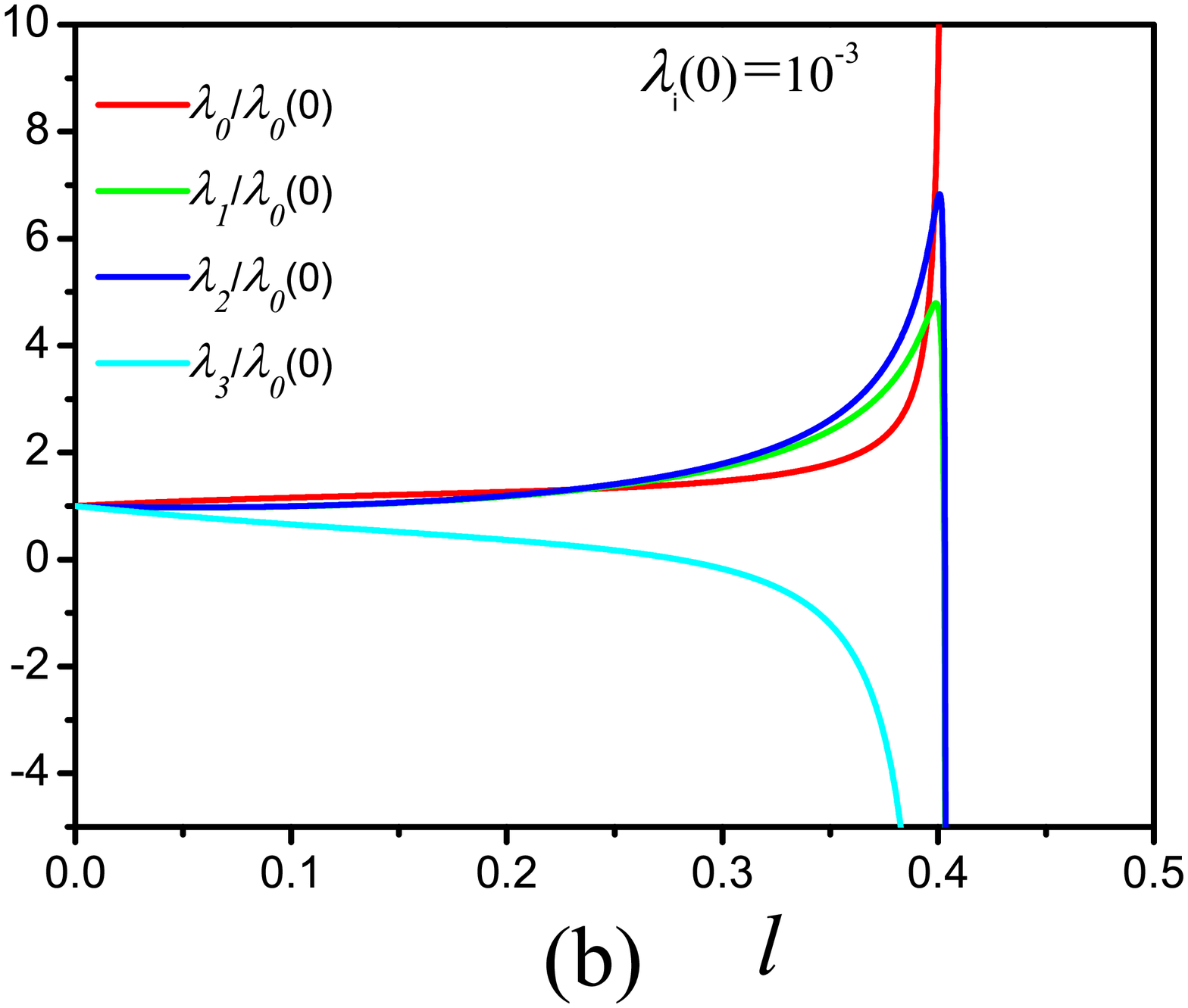}
\hspace{-1.43cm}
\includegraphics[width=2.12in]{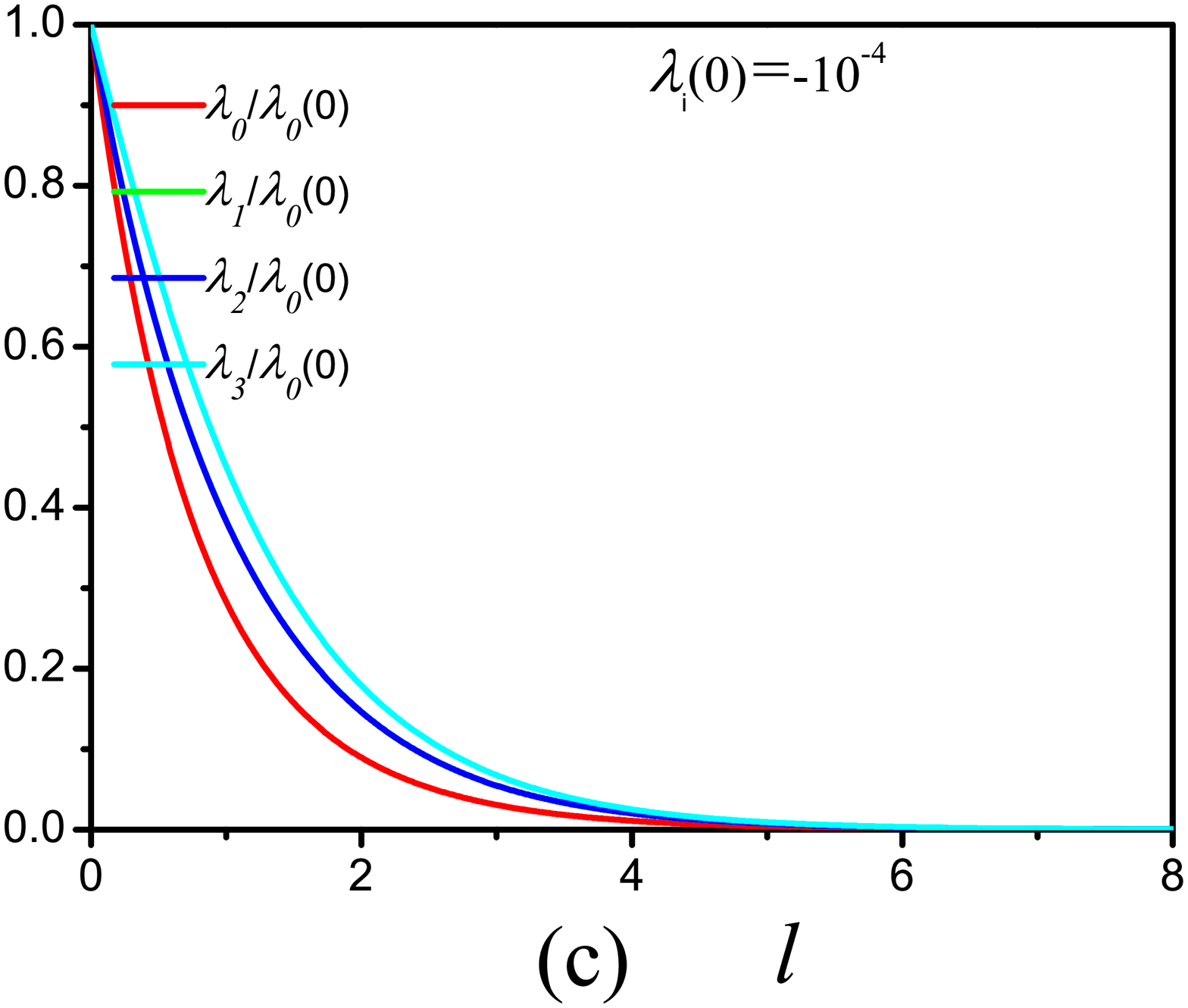}
\hspace{-1.43cm}
\includegraphics[width=2.12in]{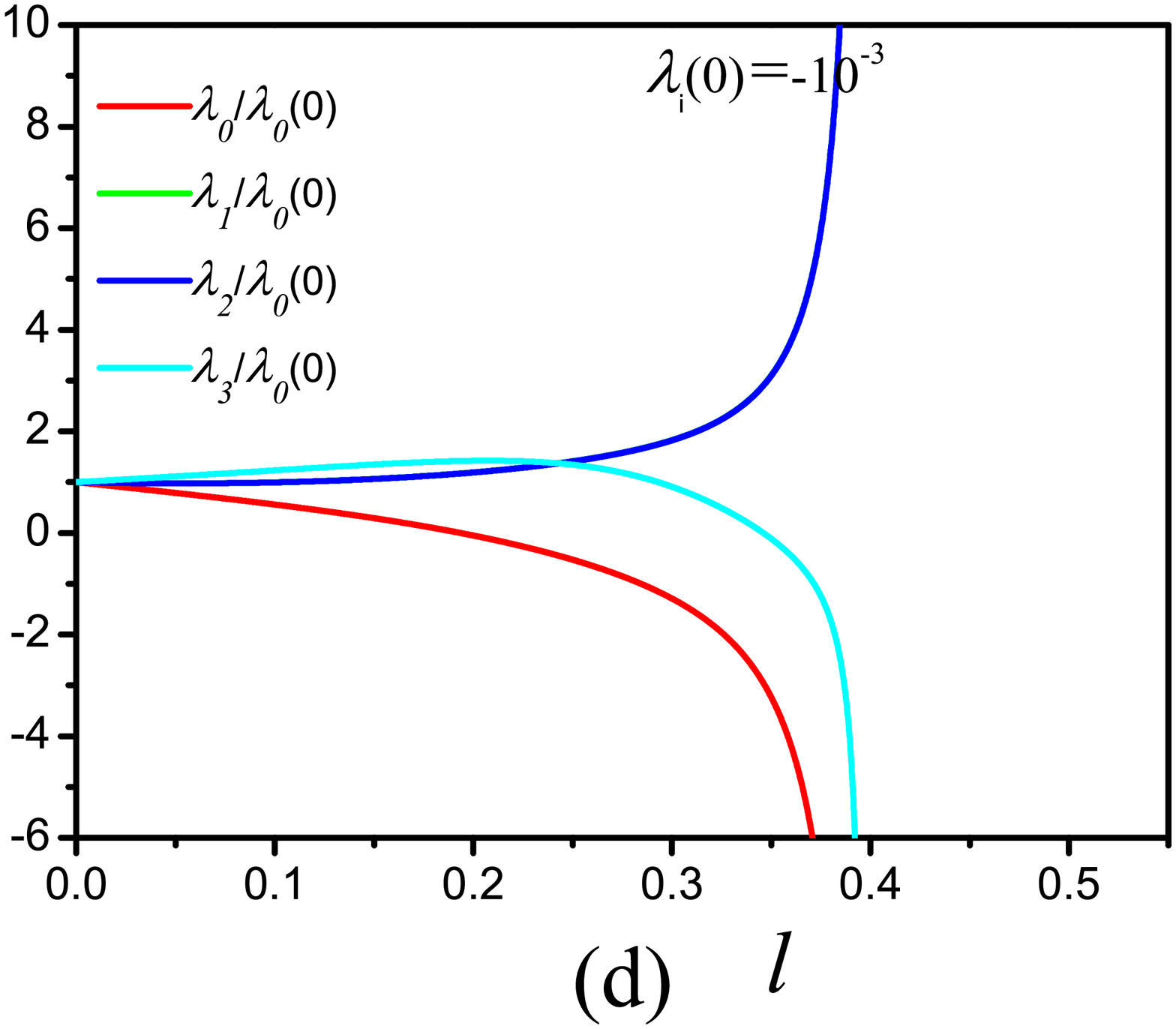}\\
\vspace{-0.05cm}
\caption{(Color online) Energy-dependent evolutions of fermion-fermion strengths
$\lambda_i(l)/\lambda_0(l=0)$ with distinct typical values of $\lambda_i(l=0)$
and tilting parameter $\zeta=0.10$ as well as $v_1=v_2=10^{-2}$.}\label{Fig-2}
\end{figure*}

\section{Low-energy fates of type-I tilted-Dirac semimetals}\label{Sec_type-I}

To proceed, we are going to investigate the low-energy
properties of tilted Dirac semimetals caused by the
fermion-fermion interactions in the low-energy regime by virtue of
their RG evolutions. Within this section, we endeavor to put our
focus on the type-I tilted Dirac fermions and defer the type-II
situation in the looming section.

%
%
%


In order to exactly capture the low-energy behaviors,
one however is required to perform numerical analysis of RG equations
~(\ref{Eq_RG-type-I-lambda-0})-(\ref{Eq_RG-type-I-lambda-3}) in that all
interaction couplings are not independent but intimately entangled.

\subsection{Roles of tilting parameter and starting values of
fermion-fermion interactions}

Learning from RG equations
~(\ref{Eq_RG-type-I-lambda-0})-(\ref{Eq_RG-type-I-lambda-3}) for type-I tilted
fermions (i.e., $|\zeta|<1$), it is interesting to point out
that the tilting parameter $\zeta$ always presents in the term of $\zeta^2$.
As a result, one only needs to take into account either $\zeta>0$ or $\zeta<0$
as they share with the same coupled RG evolutions.
In this respect, we hereafter consider the type-I tilted Dirac fermions with
$\zeta>0$. In addition, the tendencies of RG equations
would be closely associated with the beginning values of fermion-fermion interactions.
Without loss of generality, we assume all four types of fermion-fermion interactions to
host an equivalent initial value dubbed $\lambda_i(0)$ with $i=0,1,2,3$. Moreover, the
fermion velocities
do not flow and hence can be regarded as certain
constant (the basic results are insusceptible to its specific value).
To reiterate, both $\zeta$ and $\lambda_i(0)$ are crucial facets to
determine the low-energy physics.

Therefore, it is of considerable significance to explore how these
two parameters govern the low-energy physical behaviors of type-I tilted Dirac
fermions. To proceed, we obtain several interesting results after carrying out the
numerical analysis of entangled RG flows
~(\ref{Eq_RG-type-I-lambda-0})-(\ref{Eq_RG-type-I-lambda-3})
and adopting several representative values for fermion velocities, tilting parameter,
and $\lambda_i(0)$.

At first, we pick out an representative starting value of fermioinc interactions,
such as $|\lambda_{i}(0)|=10^{-4}$, which is insufficient large to induce
any instabilities for un-tilted Dirac fermions (i.e., $\zeta=0$). Learning from
Fig.~\ref{Fig-1} with this fixed $|\lambda_i(0)|$ and distinct tilting parameters,
we figure out that fermion-fermion couplings irrespective of repulsive or attractive
interactions $\lambda_i$ would gradually go towards Gaussian fixed point once the
system is slightly tilted with a small $\zeta$. However, while $\zeta$ is adequate large,
fermion-fermion interactions can be driven to divergence at certain critical energy scale.
These single out that some instability accompanied by potential phase transition can be
expected in the low-energy
regime~\cite{Murray2014PRB,Wang2017PRB_QBCP} as long as the Dirac system is
sufficiently tilted. As a consequence, one can draw a conclusion that the type-I
tilted Dirac system is more preferable to trigger instability compared to conventional
Dirac fermions since the possibility of instability is of proportional relevance to
the tilting parameter.

Subsequently, we take the tilting parameter as a fixed constant and
inspect the role of starting values of fermion-fermion couplings
in kindling possible instability. Specifically, choosing a typical value
$\zeta=0.1$ and performing analogous evaluations give rise to the key results
delineated in Fig.~\ref{Fig-2}. According to Fig.~\ref{Fig-2}, we realize
that the fermion-fermion couplings for type-I tilted Dirac fermions
progressively climb down and are apparently attracted by the Gaussian fixed
point (FP) once the beginning value $|\lambda_i(0)|$ is small. On the contrary,
it is manifest that the Gaussian FP can be completely sabotaged and accordingly
some instability would be activated by increasing the starting values of fermionic
couplings. This implies that the initial values of fermion-fermion couplings, besides
the tilting parameter $\zeta$, are also very prone to produce the instability in the
low-energy regime. It is worth stressing again that the basic results are insusceptible
to the signs of fermion-fermion
interactions.

To be brief, both the tilting parameter $\zeta$ and fermionic starting value
$\lambda_i(0)$ are helpful to the development of instability for the type-I
tilted Dirac fermions in the low-energy regime. In addition, it is of particular
interest to ask which of them takes a leading responsibility for pinpointing the
low-energy states of type-I tilted Dirac fermions. To this end, we are going to
concentrate on this question in the next subsection.

\subsection{Competition between tilting parameter and starting values of
fermion-fermion interactions}\label{Sec_type-I-B}

In last subsection, we show that both tilting parameter $\zeta$ and
initial values of fermion-fermion interactions $|\lambda_i(0)|$ are closely
linked to potential instability. We hereby endeavor to explore how they
compete and which of them is more favorable to spark certain instability in the
type-I tilted Dirac fermions at the lowest-energy limit.

\begin{figure}
\hspace{-0.8cm}
\includegraphics[width=2.1in]{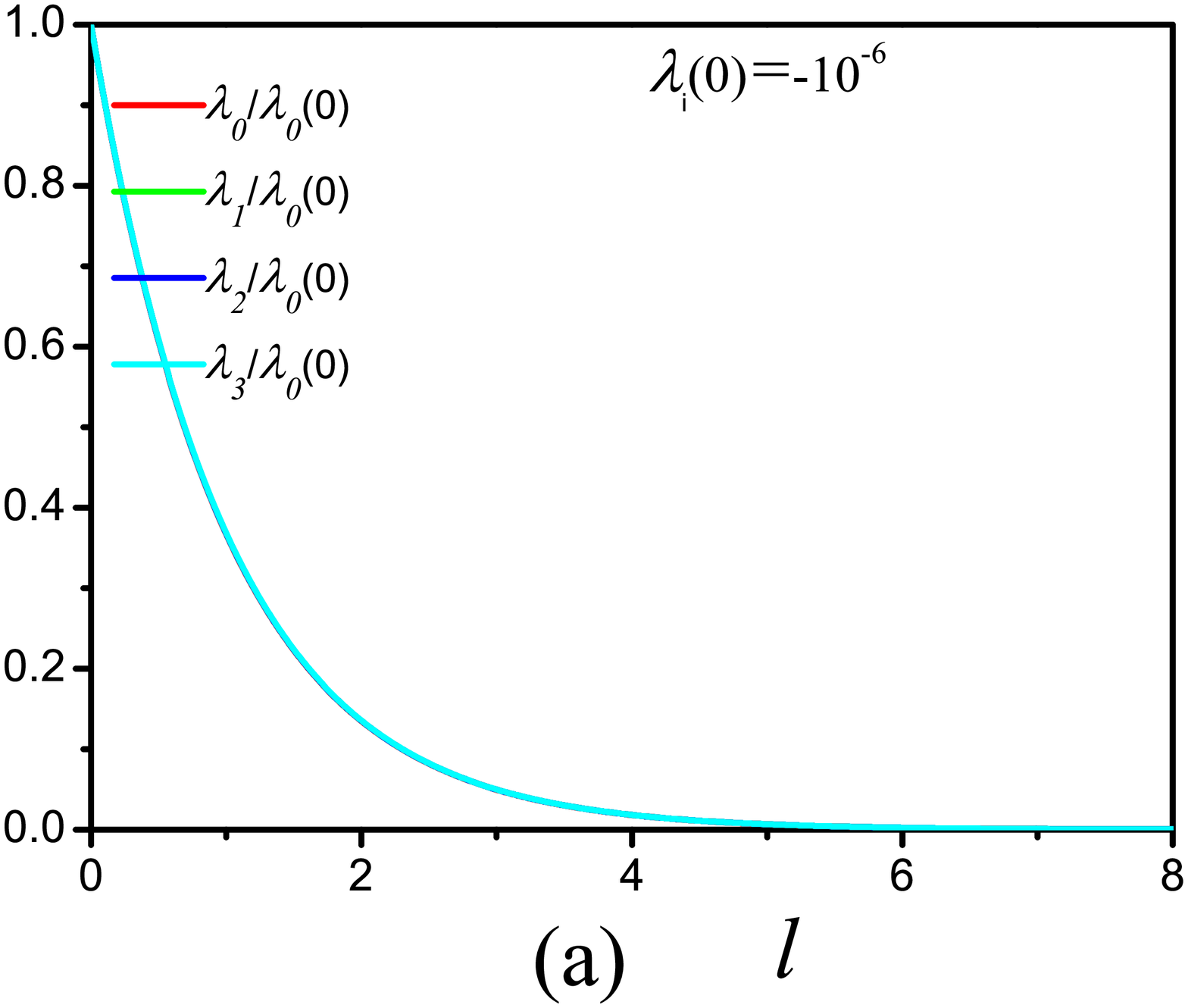}
\hspace{-1.6cm}
\includegraphics[width=2.1in]{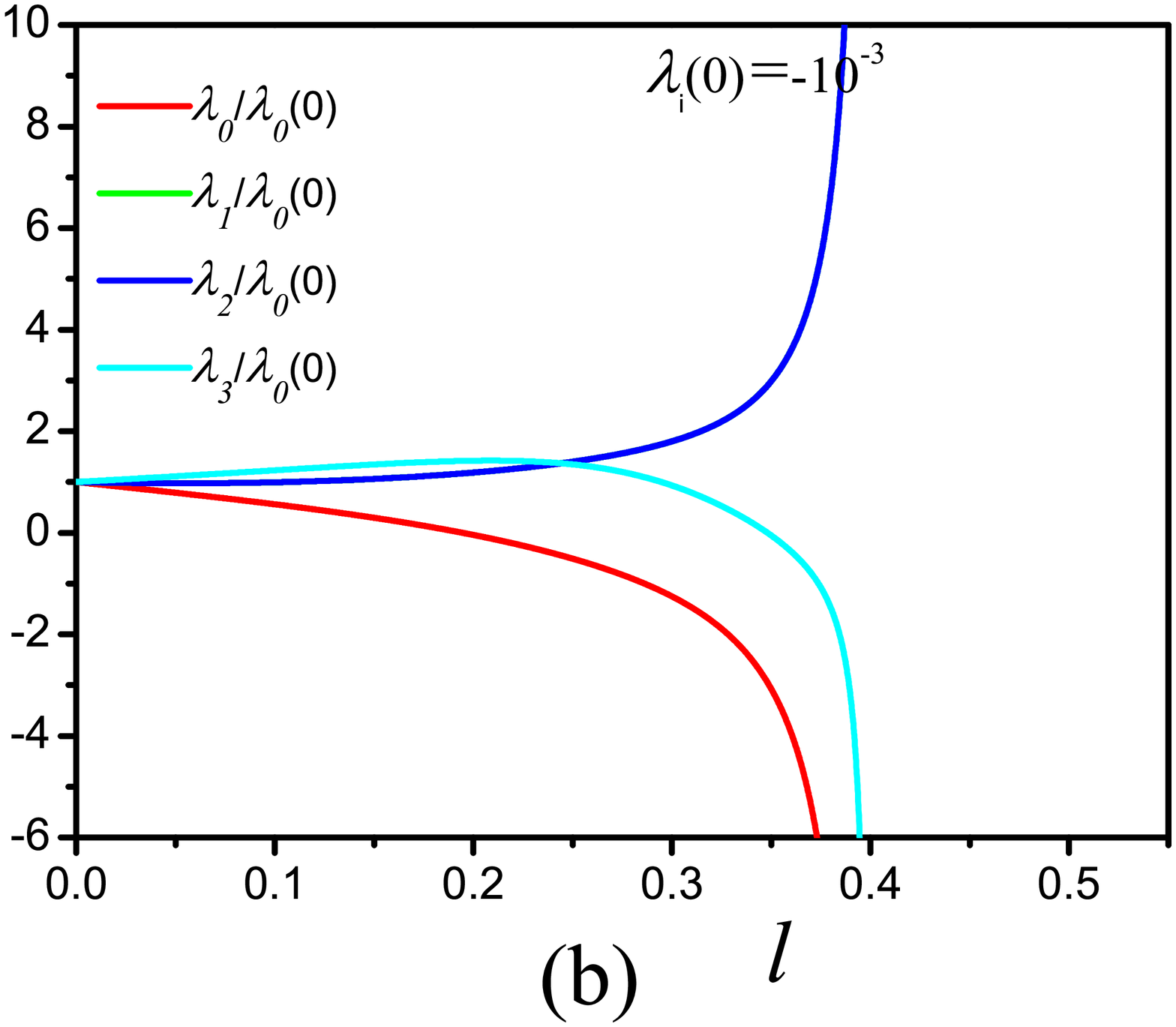} 
\\
\vspace{-0.05cm}
\caption{(Color online) Energy-dependent evolutions of fermion-fermion strengths
$\lambda_i(l)/\lambda_0(l=0)$ for (a) $\lambda_i(0)=-10^{-6}$ and (b) $\lambda_i(0)=-10^{-3}$
at tilting parameter $\zeta\rightarrow0$
and $v_1=v_2=10^{-2}$ (the basic results for $\lambda_i(0)>0$
are similar and hence not shown here).}\label{Fig-3}
\end{figure}

For type-I case, the tilting parameter $\zeta$ is restricted to $|\zeta|<1$,
whose concrete value principally dictates the structure of Dirac cones and low-energy excitations.
In order to simplify the analysis, we divide the tilting parameter into three subregimes, namely Zone-I, Zone-II, and Zone-III,  which correspond to $|\zeta|\rightarrow 0$, $|\zeta|\rightarrow 1$,
and $|\zeta|\in$ other values, respectively. At Zone-I, carrying out the similar procedures apparently indicates that the potential instabilities can only be produced once the initial fermion-fermion interactions are adequately large to exceed certain critical value regardless of repulsive or attractive interactions as displayed in Fig.~\ref{Fig-3}. Clearly, these results are in well agreement with untilted Dirac fermions in that tilted Dirac fermions naturally reduce to conventional Dirac fermions
at $\zeta\rightarrow0$ and hence the potential instability can only be
generated once the initial fermion-fermion interactions exceed certain critical value~\cite{Nandkishore2013PRB,Potirniche2014PRB,Wang2017PRB}.

\begin{figure}
\hspace{-0.8cm}
\includegraphics[width=2.1in]{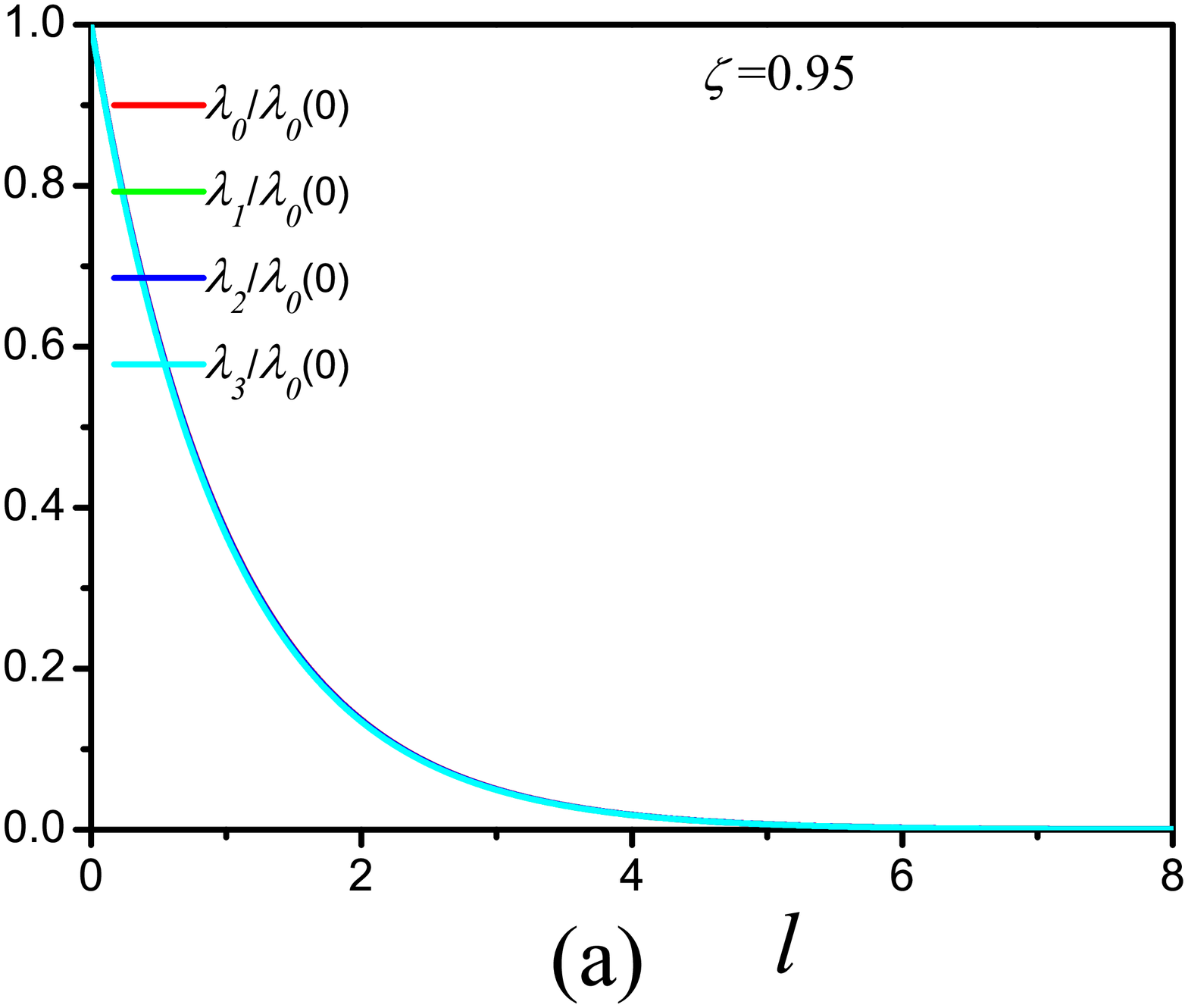}
\hspace{-1.6cm}
\includegraphics[width=2.1in]{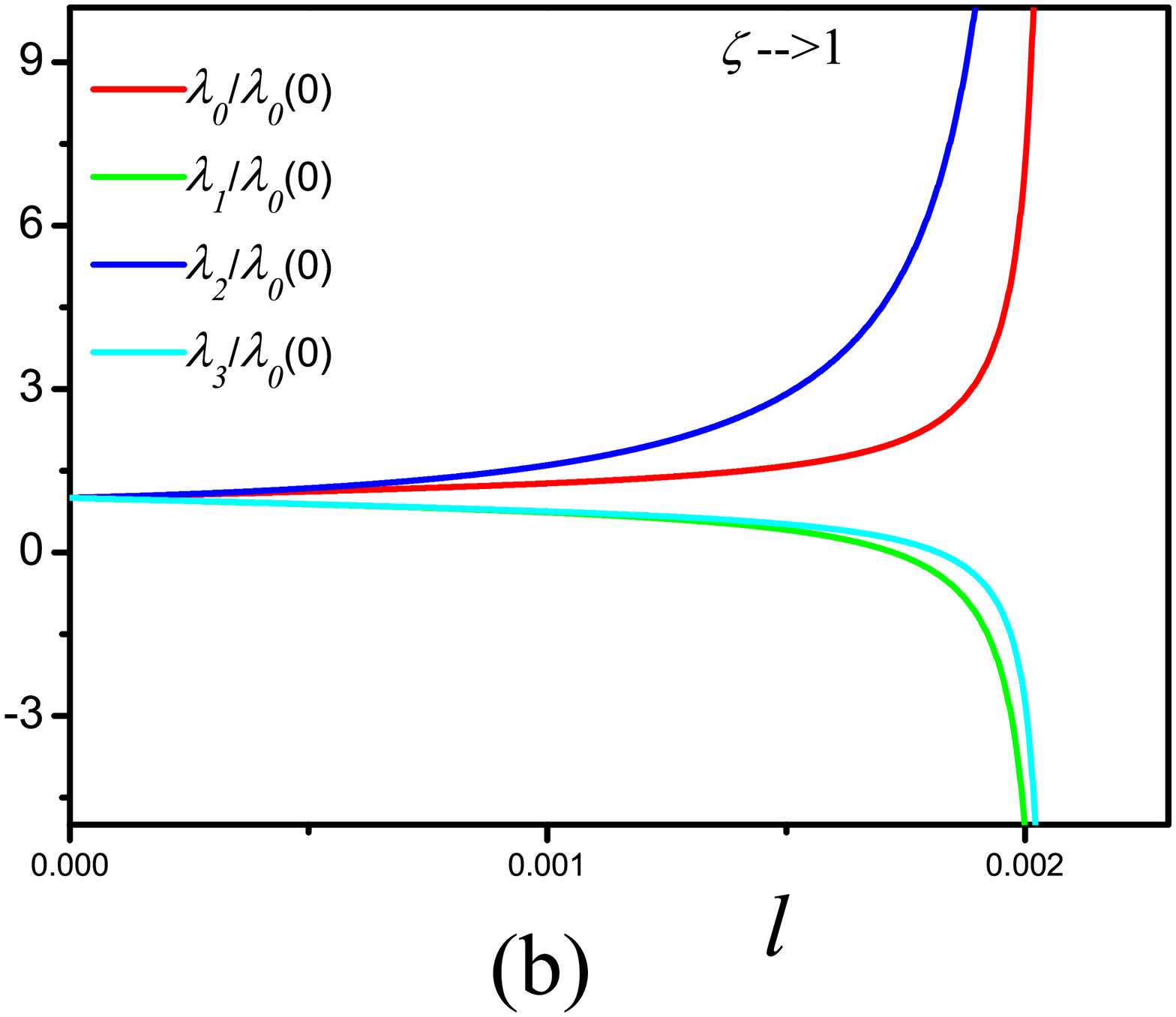} 
\\
\vspace{-0.05cm}
\caption{(Color online) Energy-dependent evolutions of fermion-fermion strengths $\lambda_i(l)/\lambda_0(l=0)$
for (a) $\zeta=0.95$ and (b) $\zeta\rightarrow1$ at $\lambda_i(0)=10^{-6}$ and $v_1=v_2=10^{-2}$
(the principal conclusions for $\lambda_i(0)<0$
are analogous and hence not shown here).}\label{Fig-4}
\end{figure}

Afterwards, we move to Zone-III. Following the related steps,
the tilted Dirac fermions at this regime unambiguously exhibit analogous low-energy behaviors
to their counterparts in Zone-I owing to similar Fermi surfaces and Dirac cones.
A slight distinction between these two situations is that the critical value
of fermion-fermion interaction, beyond which the instability is triggered,
would be progressively decreased while the tilting parameter $\zeta$ is tuned up.
With these respects, the instability can be inescapably induced by a
sufficiently large $|\lambda_i(0)|$ no matter what a specific value of $\zeta$
is assigned. Accordingly, the initial values of fermion-fermion couplings $|\lambda_i(0)|$
are more significant than the tilting parameter $\zeta$ at both Zone-I and
Zone-III. In sharp distinction to Zone-I and Zone-III, as demonstrated in
Fig~\ref{Fig-4}, the type-I tilted Dirac fermion can be driven into
certain instability by a small fermion-fermion interaction that is much smaller
than critical value $|\lambda_i(0)|$ while the Zone-II is approached.
In other words, an underlying instability generally sets in if the
Lifshitz phase point ($|\zeta|=1$)~\cite{Lifshitz1942JP, Dzyaloshinski1964ZETF, Goshen1974IJM, Hornreich1975PRL}, which separates type-I and type-II tilted Dirac
fermions~\cite{Lee2018PRB,Lee2019PRB}, is accessed even though the
starting values of fermion-fermion strengths are very small.
One therefore naturally expects the topological changes of Fermi surfaces
may be responsible for this unique phenomenon. In this regard, one may
ascribe this consequence to the singular effects caused by
the very Lifshtiz phase point~\cite{Lee2018PRB}.
At this stage, one can deduce that the tilting parameter $\zeta$ is more
important at Zone-II as $\zeta$ is directly associated with the Fermi
surface's structure of tilted Dirac fermions.

To reiterate, either the increase of starting values of fermion-fermion strengths
$|\lambda_i(0)|$ or tilting parameter $\zeta$ for type-I tilted Dirac fermions
is helpful to develop certain instability in the low-energy regime.
In particular, $|\lambda_i(0)|$ and $\zeta$ play a more crucial role in
igniting the potential instability at Zone-I (or Zone-III) and
Zone-II, respectively. Fig.~\ref{Fig-5}(a) and Table~\ref{Type-I}
briefly summarize our primary conclusions.

\section{Low-energy fates of type-II tilted-Dirac semimetals}\label{Sec_type-II}

Within this section, we dwell on the physical properties of
type-II tilted Dirac fermions caused by fermion-fermion interactions
in the low-energy. Hereby, we focus on the $\zeta>0$ case at first and then
present our discussions for the $\zeta<0$ situation at the end of
this section.

\subsection{Roles of tilting parameter and starting values of
fermion-fermion interactions}\label{Sec_type-II-role}

\begin{figure}
\hspace{-0.8cm}
\includegraphics[width=2.05in]{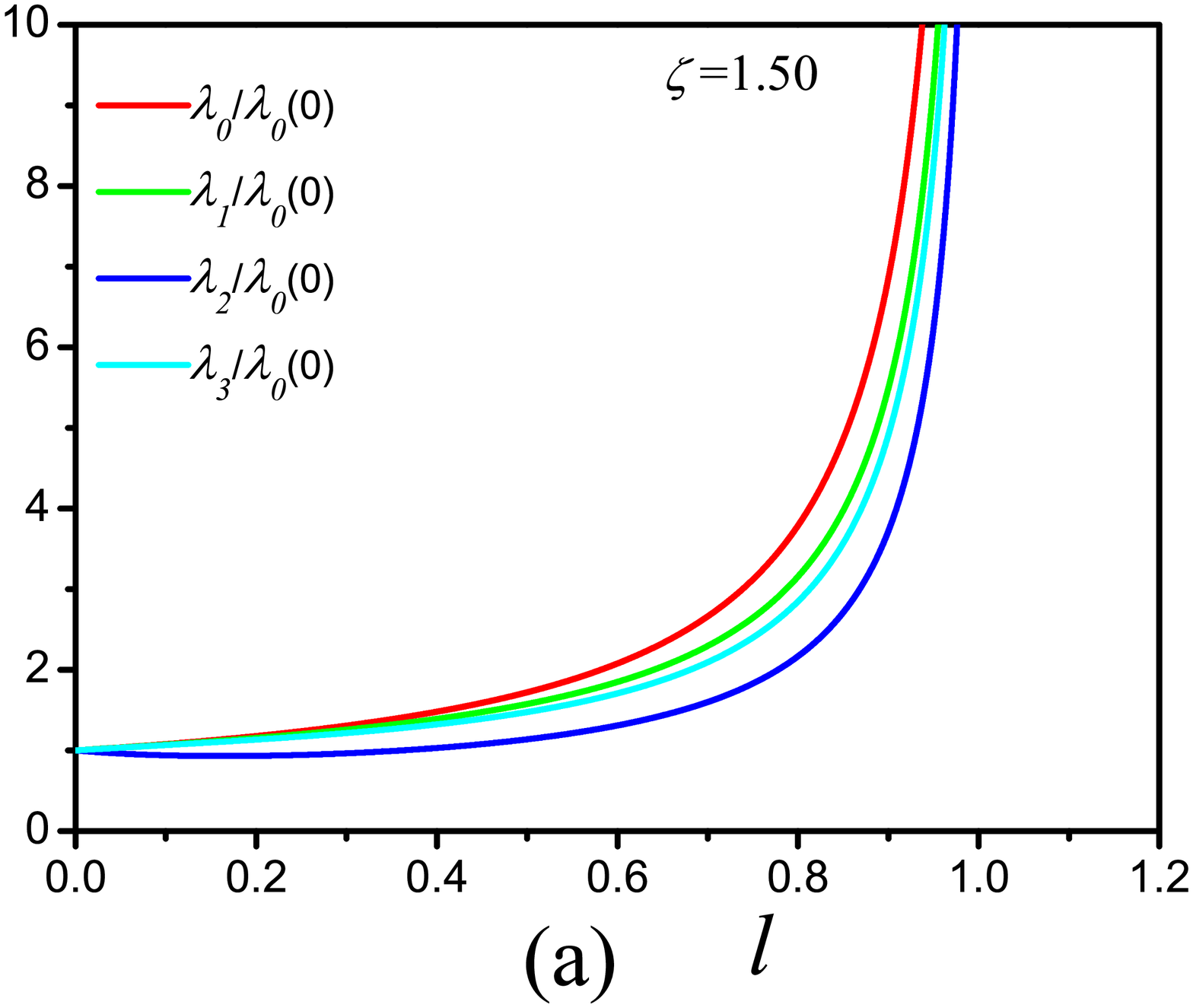}\hspace{-1.3cm}
\includegraphics[width=2.05in]{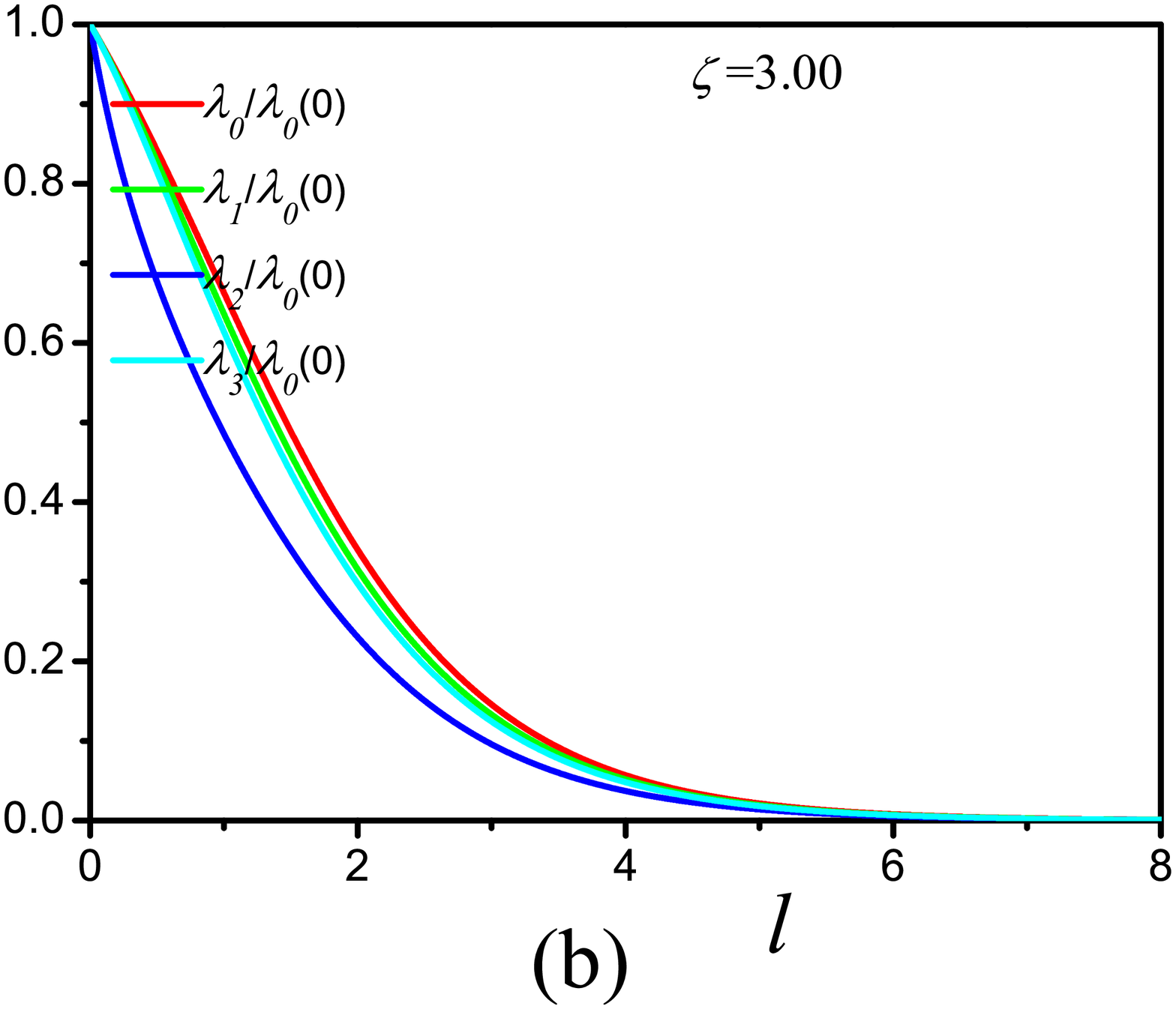}\\
\vspace{-0.05cm}
\caption{(Color online) Energy-dependent evolutions of fermion-fermion strengths
$\lambda_i(l)/\lambda_0(l=0)$ for (a) $\zeta=1.5$ and (b) $\zeta=3.0$ at $\lambda_i(0)=10^{-3}$
and $v_1=v_2=10^{-2}$ (the basic results for $\lambda_i(0)<0$
are similar and the tendency of divergences are analogous to
Fig.~\ref{Fig-7}. Hence they are not shown here).}\label{Fig-6}
\end{figure}

In order to capture the effects of $|\lambda_i(0)|$
and $\zeta$ on the type-II system, we follow the procedures in previous section, namely taking $|\lambda_i(0)|$ as a fixed value and adjusting the values of $\zeta$ to observe the impact induced by $\zeta$ and vive versa. Without loss of generality, we choose $|\lambda_i(0)|=10^{-3}$ and show the results in Fig.~\ref{Fig-6} with variation of $\zeta$. In the light of these, we figure out that the potential instability triggered at a small $\zeta$ would be destroyed and replaced by the Gaussian FP as the tilting parameter is increased. In apparent distinction to type-I case, this signals that the increase of $\zeta$ brings detriments to the emergence of instability for type-II tilted Dirac fermions. Despite the basic conclusions are robust irrespective of signs of $\lambda_i(0)$, there are different divergence trends for $\lambda_{i}(0)>0$ and $\lambda_{i}(0)<0$, which will be studied in Sec.~\ref{Sec_instability}.

\begin{figure}
\hspace{-0.8cm}
\includegraphics[width=2.05in]{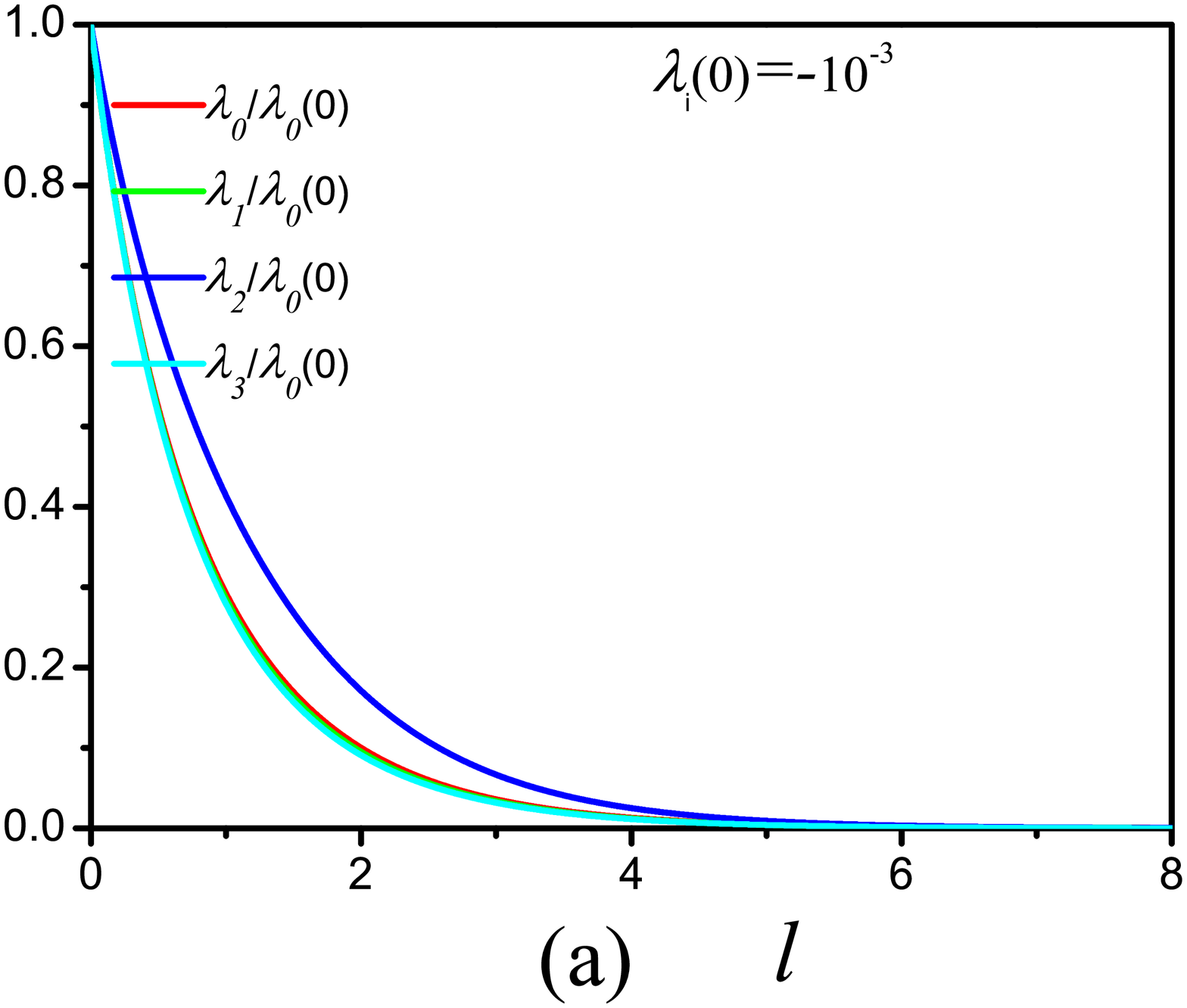}\hspace{-1.3cm}
\includegraphics[width=2.05in]{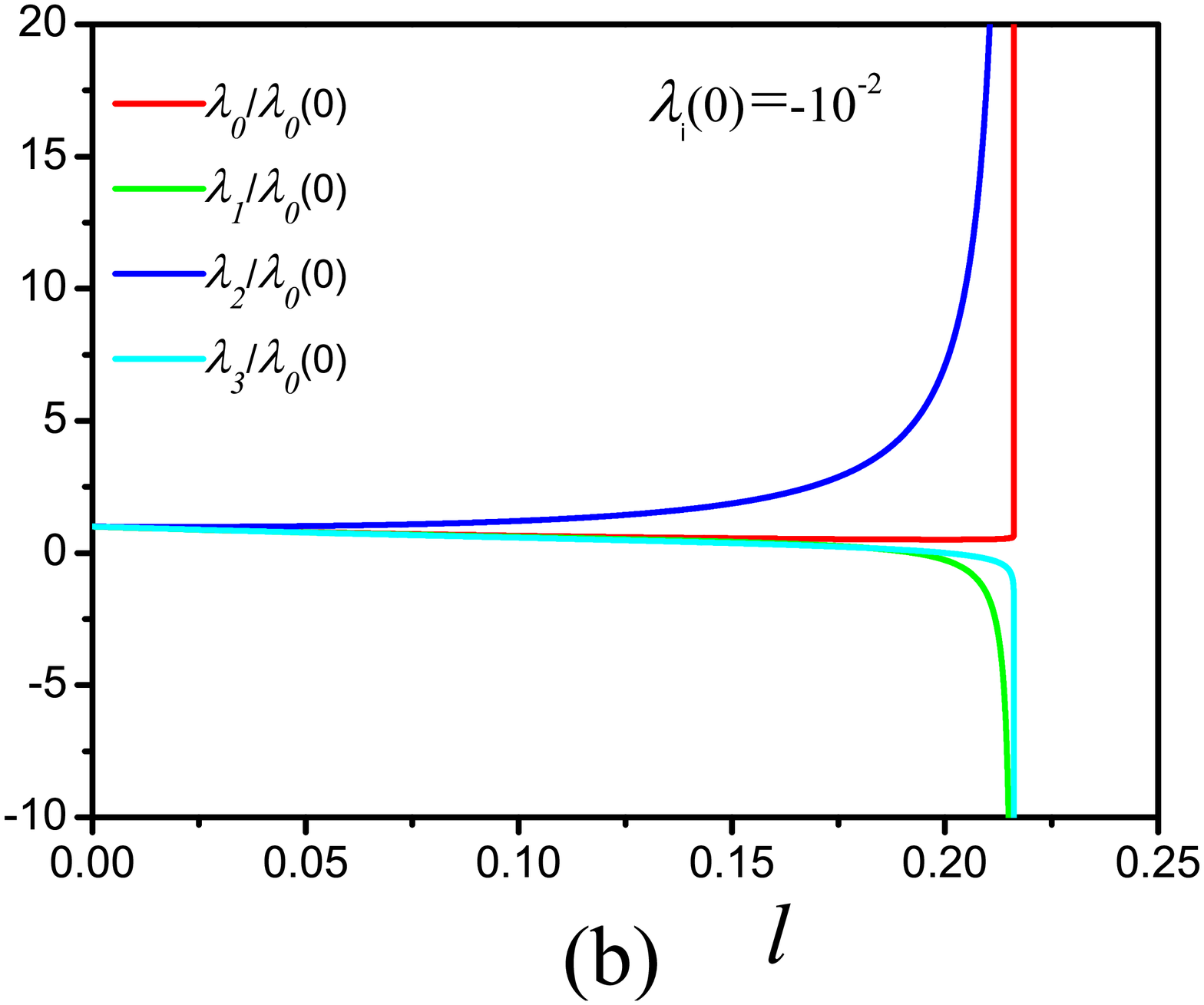}\\
\vspace{-0.05cm}
\caption{(Color online) Energy-dependent evolutions of fermion-fermion strengths
$\lambda_i(l)/\lambda_0(l=0)$ for (a) $\lambda_i(0)=-10^{-3}$ and (b) $\lambda_i(0)=-10^{-2}$
at tilting parameter $\zeta=5$ and $v_1=v_2=10^{-2}$ (the basic results for $\lambda_i(0)>0$
are similar and the tendency of divergences are analogous to
Fig.~\ref{Fig-6}. Hence they are not shown here).}\label{Fig-7}
\end{figure}

In addition, we consider the tilting parameter as a constant, for instance
$\zeta=5$, adjust the starting values of fermion-fermion couplings, and plot the
numerical results in Fig.~\ref{Fig-7} after performing numerical analysis of coupled RG equations ~(\ref{Eq_RG-type-II-lambda-0})-(\ref{Eq_RG-type-II-lambda-3}).
Fig.~\ref{Fig-7} unambiguously delivers that trajectories of $\lambda_i$ can be converted
from Gaussian FP to instability via tuning up the value of $|\lambda_{0}(0)|$. This implies
that there always exists a critical value $|\lambda_{i}^{c}(0)|$, beyond which certain
instability is conventionally expected. It indicates that the increase of initial values
of fermion-fermion interactions is in favor of the generation of instability in the
type-II tilted fermionic system. This conclusion is well in line with the type-I case.

\begin{figure}
\hspace{-0.7cm}
\includegraphics[width=1.6in]{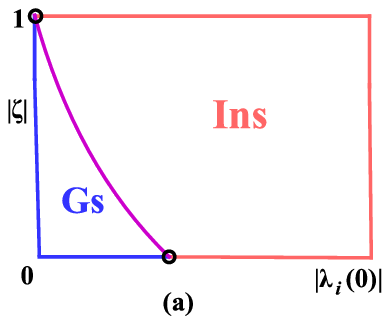}\hspace{0.3cm}
\includegraphics[width=1.6in]{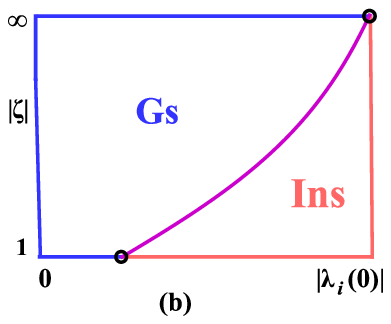}\\
\vspace{0.8cm}
\includegraphics[width=3.2in]{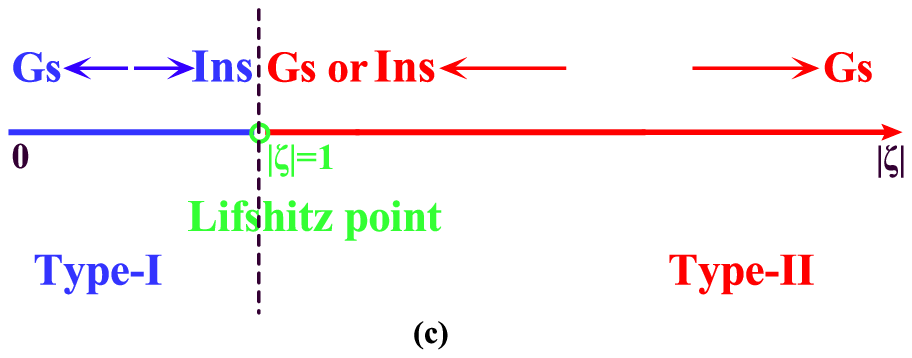}\\
\vspace{0.15cm}
\caption{(Color online) Schematic phase diagrams in the $\zeta-\lambda_i(0)$
plane for (a) type-I and (b) type-II tilted Dirac semimetals. Subfigure (c) summarizes
both (a) and (b) by schematically displaying the key points separated by the Lifshtiz
phase point. For convenience, ``Gs" and ``Ins"  are exploited to serve
as Gaussian FP and certain instability, respectively
(The dominant phases induced by these instabilities
are carefully examined in Sec.~\ref{Subsection_phase} and
schematically exhibited in Fig.~\ref{Fig-14}).} \label{Fig-5}
\end{figure}

To reiterate, we find that the effects
of two parameters $|\lambda_i(0)|$ and $\zeta$ on the type-II Dirac
fermions are opposite. Specifically, the increase of
$|\lambda_i(0)|$ prefers to switch on the instability in the low-energy.
Rather, tuning up $\zeta$ is harmful to the development of instability.


\subsection{Competition between tilting parameter and starting values of
fermion-fermion interactions}\label{Sec_type-II-B}

In last subsection, we deliver that the initial values of fermion-fermion
couplings $|\lambda_i(0)|$ and tilting parameter $\zeta$ in principle promote
and suppress the instability of the type-II tilted Dirac fermions, respectively.
At this stage, it is now tempting to inquire which of these two parameters plays
a leading role in dictating low-energy states within certain parameter-space region.
Motivated by this, we hereby are going to response this question. For convenience,
the tilting parameter $\zeta$ is at first assumed to be positive.

\begin{figure}
\centering
\includegraphics[width=2.3in]{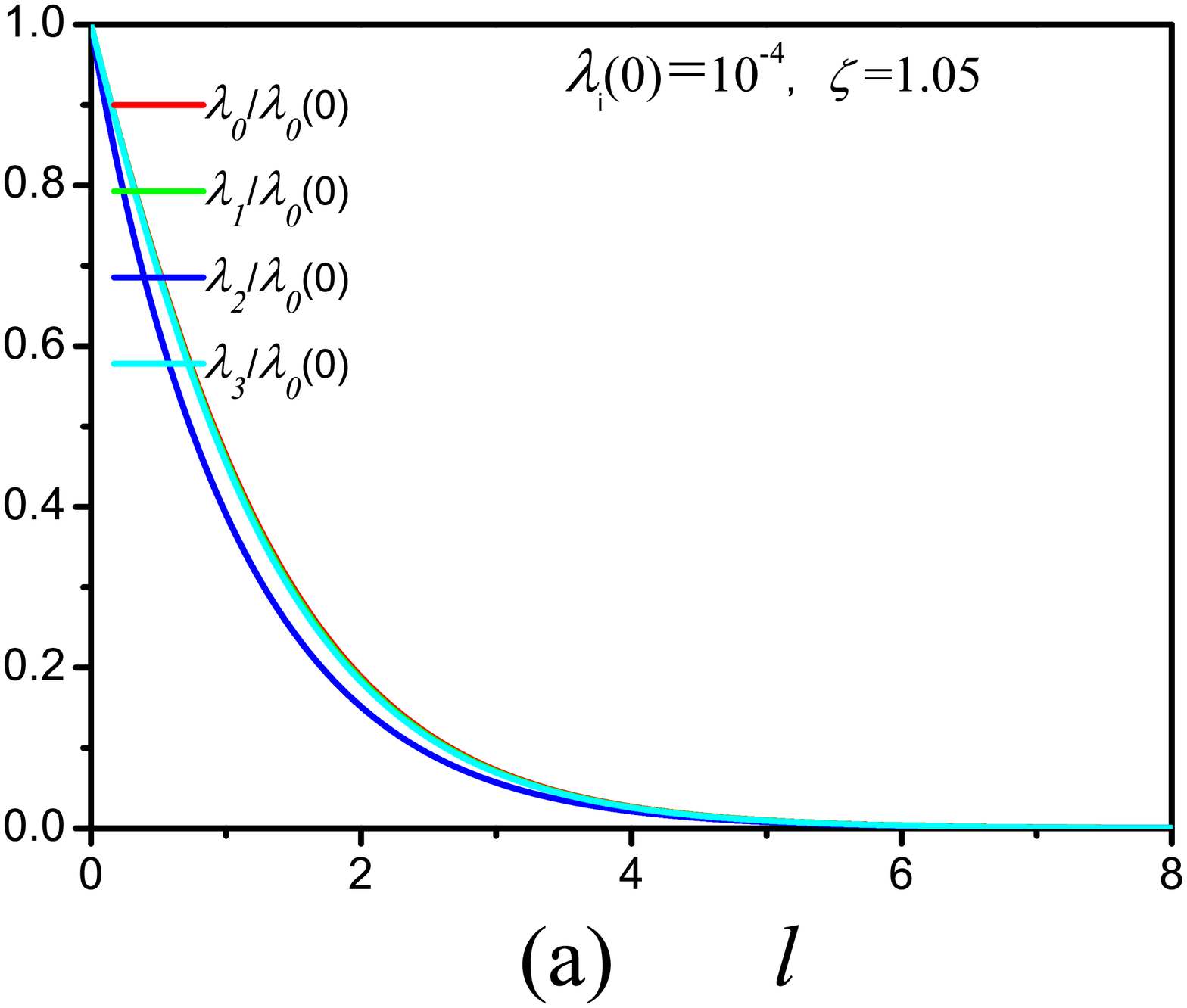}\\
\vspace{-0.2cm}
\includegraphics[width=2.3in]{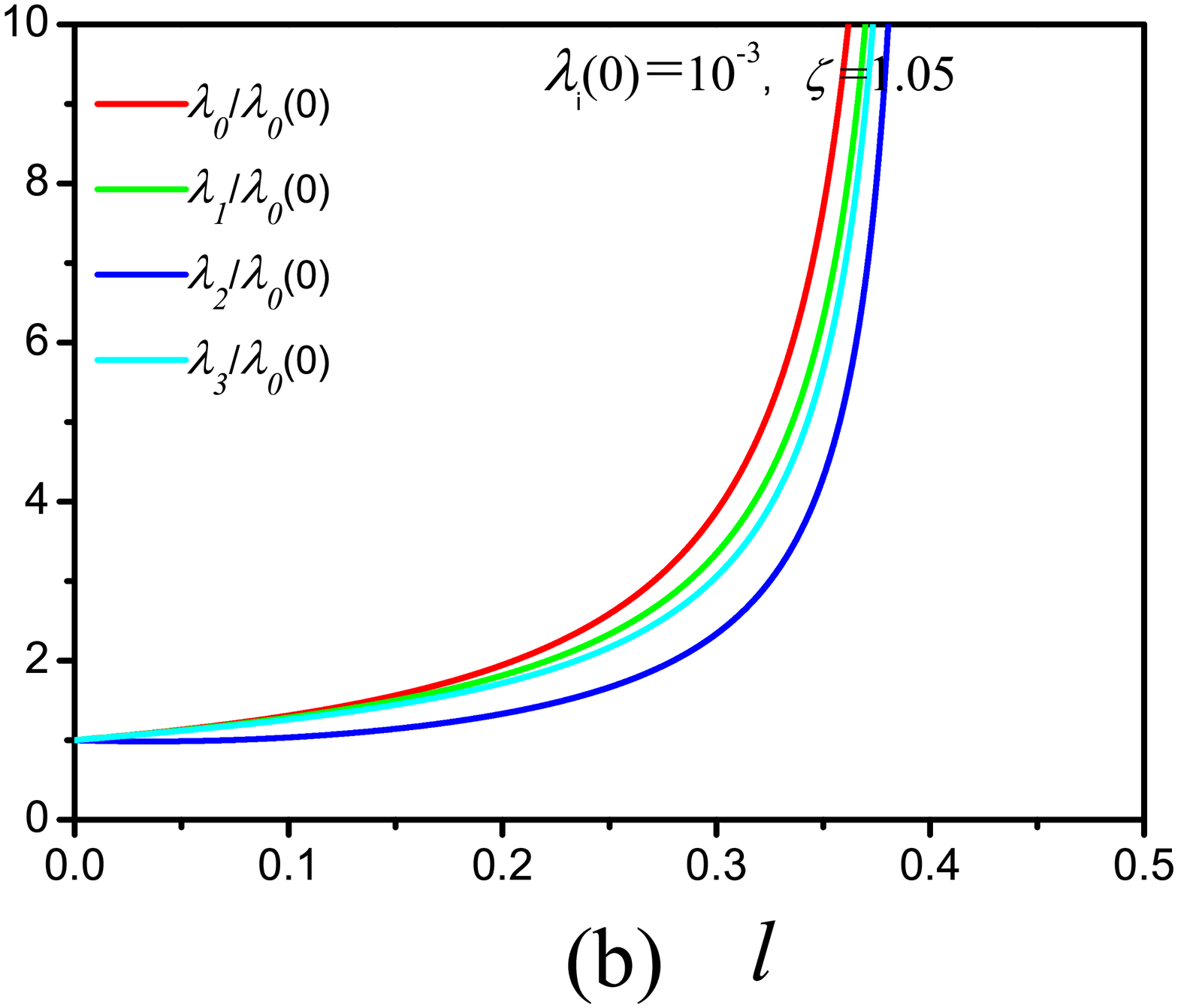}\\
\vspace{-0.2cm}
\includegraphics[width=2.3in]{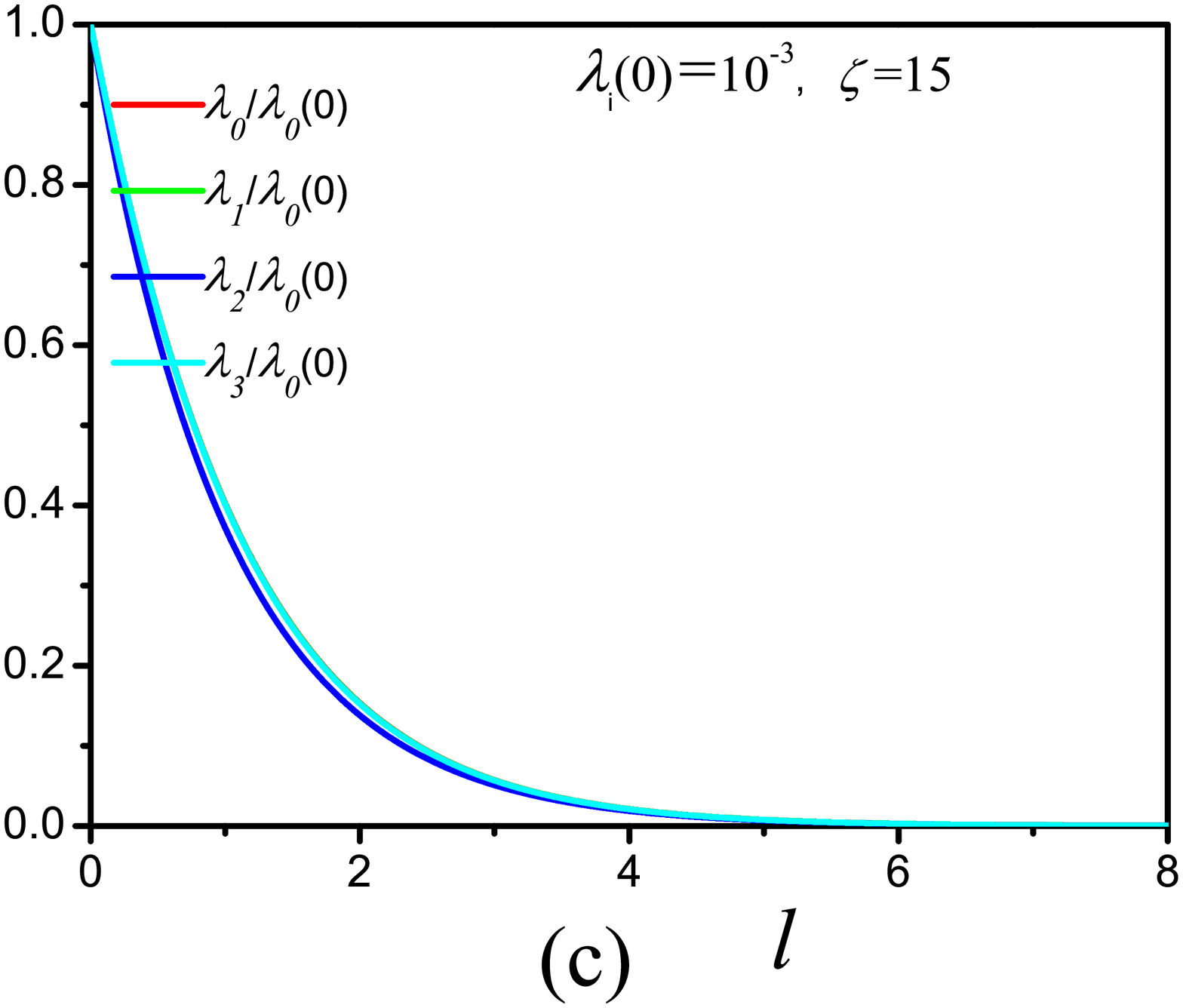}
\vspace{-0.05cm}
\caption{(Color online) Energy-dependent evolutions of fermion-fermion strengths
$\lambda_i(l)/\lambda_0(l=0)$ for (a) $\lambda_i(0)=10^{-4}$, $\zeta=1.05$;
(b) $\lambda_i(0)=10^{-3}$, $\zeta=1.05$; and (c)  $\lambda_i(0)=10^{-3}$, $\zeta=15$ at
$v_1=v_2=10^{-2}$ (the basic results for $\lambda_i(0)<0$
are similar and the tendency of divergences are analogous to
Fig.~\ref{Fig-7}. Hence they are not shown here).}\label{Fig-8}
\end{figure}

To proceed, the tilting parameter $\zeta$, in analogous to Type-I case, can also
cluster into three sets, namely $\mathrm{Zone-I}$ with $|\zeta|\rightarrow1$, $\mathrm{Zone-II}$ with $|\zeta|\rightarrow\infty$, and $\mathrm{Zone-III}$ with $|\zeta|\in$ other
values, respectively. At $\zeta\rightarrow\mathrm{Zone-I}$, each of the coupled RG
equations~(\ref{Eq_RG-type-II-lambda-0})-(\ref{Eq_RG-type-II-lambda-3})
can be formally rewritten as $d\lambda=\lambda(1-c\lambda)$ with $c$ being a
finite constant and $1/c$ the critical value $\lambda^{c}(0)$ that is the minimum
value to generate an instability. In other words, $\lambda$ is only increased once
$\lambda(0)$ exceeds $1/c$ with $\lambda(0)>0$. Accordingly, in a sharp contrast to
type-I case where an instability is always produced at $\zeta\rightarrow1$, we realize
that $\zeta$ is no longer the key factor to trigger an instability at $\mathrm{Zone-I}$.
Rather, $|\lambda_i(0)|$ solely pins down whether certain instability can be induced.
To be specific, an instability sets in once $|\lambda_i(0)|$ is large enough to go
beyond some critical value. Otherwise, the system directly evolves to Gaussian FP.
Fig~\ref{Fig-8}(a) and (b) clearly illustrate the behaviors approaching the
$\mathrm{Zone-I}$. Consequently, we would like to stress that the qualitative
results are insusceptible to the concrete values of $|\lambda_i(0)|$. As is
analogous to type-I situation, one can infer that the basic conclusions are
well consistent with $\mathrm{Zone-I}$'s when the tilting parameter $\zeta$
belongs to $\mathrm{Zone-III}$. However, a bigger $|\lambda_i(0)|$ is required
to activate a potential instability in that $\zeta$ prefers to hinder the development
of instability as pointed out in Sec.~\ref{Sec_type-II-role}.
Further, we move to $\mathrm{Zone-II}$ of type-II case. Implementing and
practicing aforementioned strategies when $\zeta$ approaches $\mathrm{Zone-II}$,
one can examine and realize that the threshold of fermion-fermion couplings goes
towards infinity, i.e., $|\lambda_i^{c}(0)|\rightarrow\infty$. As a consequence,
the type-II tilted fermion inescapably flows towards the Gaussian FP at the
lowest-energy limit. The numerical evaluations of coupled RG equations
corroborate these analysis as depicted in Fig.~\ref{Fig-8}(c). Learning from
Fig.~\ref{Fig-8}, we can also find that the underlying instability can be
switched off once the tilting parameter $\zeta$ is increased. Given that
fermion-fermion interactions cannot be taken too strong, the Gaussian FP would
be always expected as long as $\zeta$ is sufficient large. In a word, the low-energy
fates of the type-II tilted Dirac fermions are firmly rooted in the considerable
competition between $|\lambda_i(0)|$ and $\zeta$. The tilting parameter $\zeta$
becomes a major ingredient that is able to be responsible for the low-energy
states as long as it is sufficiently large or restricted to $\mathrm{Zone-II}$. This
signals that any instabilities are not allowed in the low-energy regime.
However, $|\lambda_i(0)|$ dominates over the tilting parameter in case
$\zeta$ is small or accessing $\mathrm{Zone-I}$.

Before closing this section, we finally address several comments on the $\zeta<0$
situation. In apparent variance with type-I case at which the coupled RG evolutions
keep invariant under the transformation $\zeta\rightarrow-\zeta$, it is henceforth
worth highlighting that the tilting parameter $\zeta$ in the coupled RG equations~(\ref{Eq_RG-type-II-lambda-0})-(\ref{Eq_RG-type-II-lambda-3})
can appear in terms of either $\zeta$, $|\zeta|$, or $\zeta^2$.
At the first sight, this signals that we need to study $\zeta>0$ and $\zeta<0$
separately. However, after revisiting the coupled RG evolutions of type-II tilted
Dirac fermions in more details, it is interesting to point out that the RG equations
for $\zeta>0$ with $\lambda_i(0)>0$ and $\lambda_i(0)<0$ exactly correspond to their
$\zeta<0$ counterparts with $\lambda_i(0)<0$ and $\lambda_i(0)>0$, respectively.
In this sense, it is sufficient to investigate $\zeta>0$ case as the results
for $\zeta<0$ can be easily obtained via replacing $\lambda>0$ with $\lambda<0$.
In short, our primary conclusions for type-II tilted Dirac fermions are schematically
displayed and vigilantly collected by Fig.~\ref{Fig-5}(b) and Table~\ref{Type-II}.

\section{Instabilities and dominant phases}\label{Sec_instability}

On the basis of previous analysis, several potential instabilities can be
generated for both type-I and type-II tilted Dirac semimetals attesting to
the competitions among distinct sorts of fermion-fermion interactions in the
low-energy regime. Gathering all principal results presented in Sec.~\ref{Sec_type-I}
and Sec.~\ref{Sec_type-II}, we figure out that the potential instabilities
are broken into several distinguished sorts with variations of the tilting parameter
and starting values of fermion-fermion interactions. Subsequently, we endeavor to
classify these underlying instabilities and then determine the dominant phases
around them one by one.

\subsection{Classifications of instabilities}\label{Subsection_instability}


\begin{table}
\centering
\caption{Low-energy fates of type-I tilted Dirac fermions with variations of
tilting parameter $\zeta$ and beginning values of fermion-fermion
couplings $\lambda_i(0)$. To be convenient, ``Gs" and ``Ins" stand for
Gaussian FP and certain instability, respectively. In addition,
we have already introduced $\mathrm{Zone-I}$, $\mathrm{Zone-II}$,
and $\mathrm{Zone-III}$ in Sec.~\ref{Sec_type-I-B} to classify
values of $\zeta$ for type-I tilted DSMs. Moreover, Ins-I, Ins-II, Ins-III,
and Ins-IV designating distinct types of instabilities and the leading phases
around these instabilities are carefully investigated in Sec.~\ref{Sec_instability}
and schematically displayed in Fig.~\ref{Fig-14}.}\label{Type-I}
\vspace{0.5cm}
\renewcommand{\arraystretch}{1.2}
\begin{tabular}{p{1.6cm}<{\centering}|p{1.6cm}<{\centering}|p{1.5cm}<{\centering}|
p{1.5cm}<{\centering}|p{1.5cm}<{\centering}}
\hline 
\hline
&\multirow{2}{1.5cm}{\centering$|\lambda_{i}(0)|$}&\multicolumn{3}{c}{$\zeta$}\\
\cline{3-5}
& &$\mathrm{Zone-I}$& $\mathrm{Zone-III}$ &$\mathrm{Zone-II}$\\
\hline
\multirow{3}{1.6cm}{$\lambda_{i}(0)>0$}&Small&Gs&Gs&Ins-I\\
\cline{2-5}
&Medium&Gs&Ins-I&Ins-I\\
\cline{2-5}
&Large&Ins-IV&Ins-III&Ins-I\\
\hline
\multirow{3}{1.6cm}{$\lambda_{i}(0)<0$}&Small&Gs&Gs&Ins-II\\
\cline{2-5}
&Medium&Gs&Gs&Ins-II\\
\cline{2-5}
&Large&Ins-II&Ins-II&Ins-II\\
\hline
\hline
\end{tabular}
\end{table}

\begin{figure*}[htbp]
\centering
\includegraphics[width=1.75in]{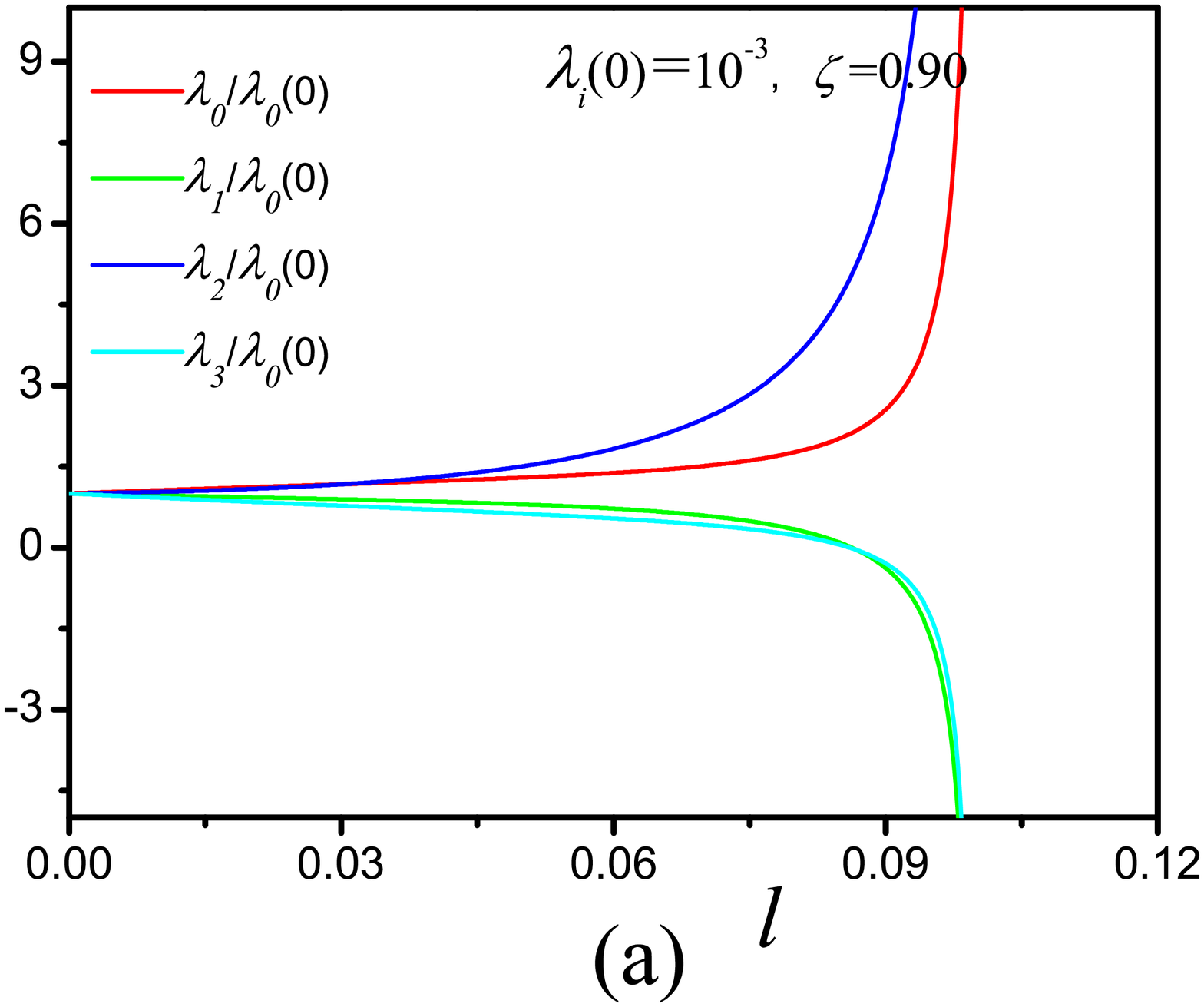}
\hspace{-1.30cm}
\includegraphics[width=1.75in]{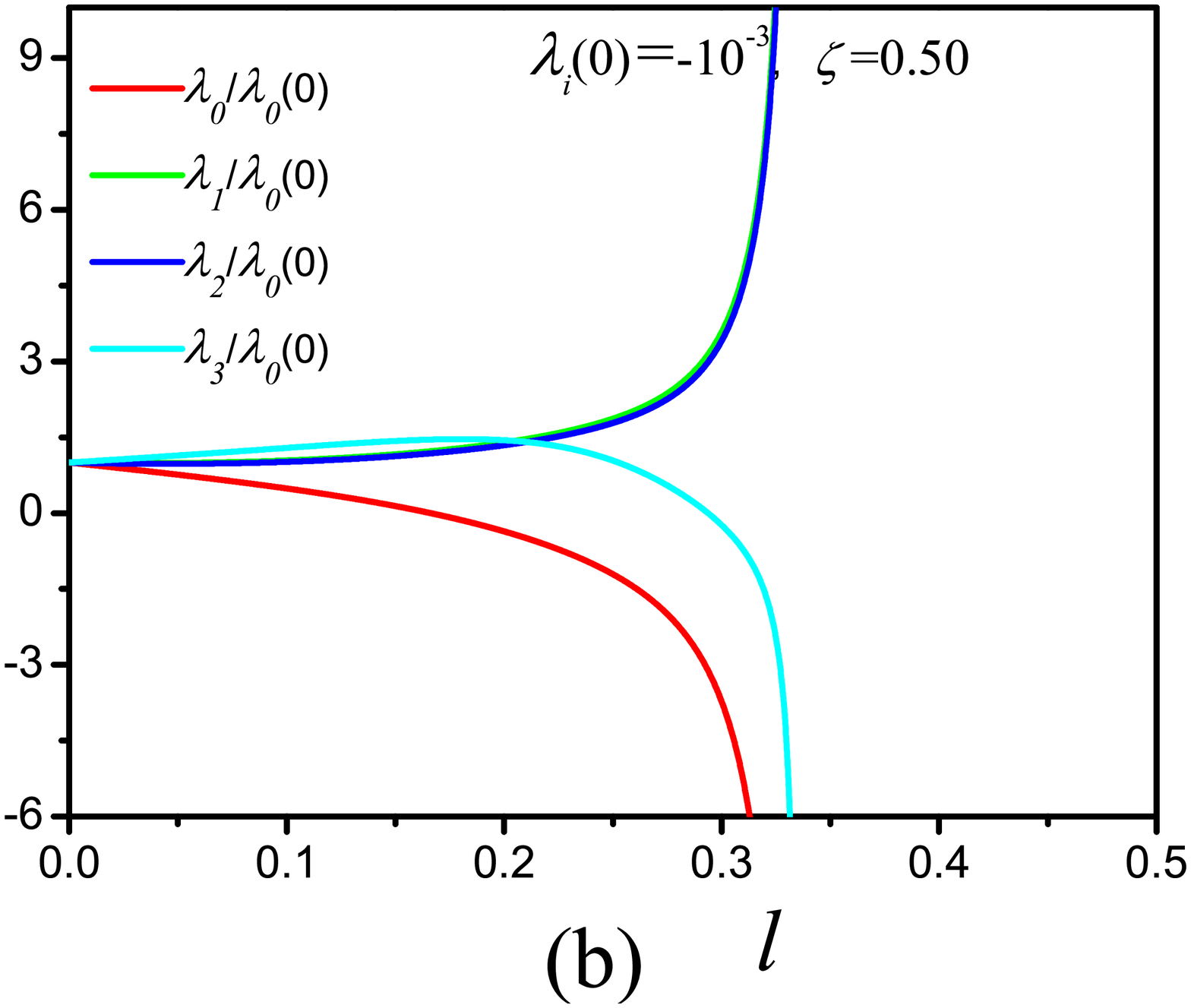}
\hspace{-1.30cm}
\includegraphics[width=1.75in]{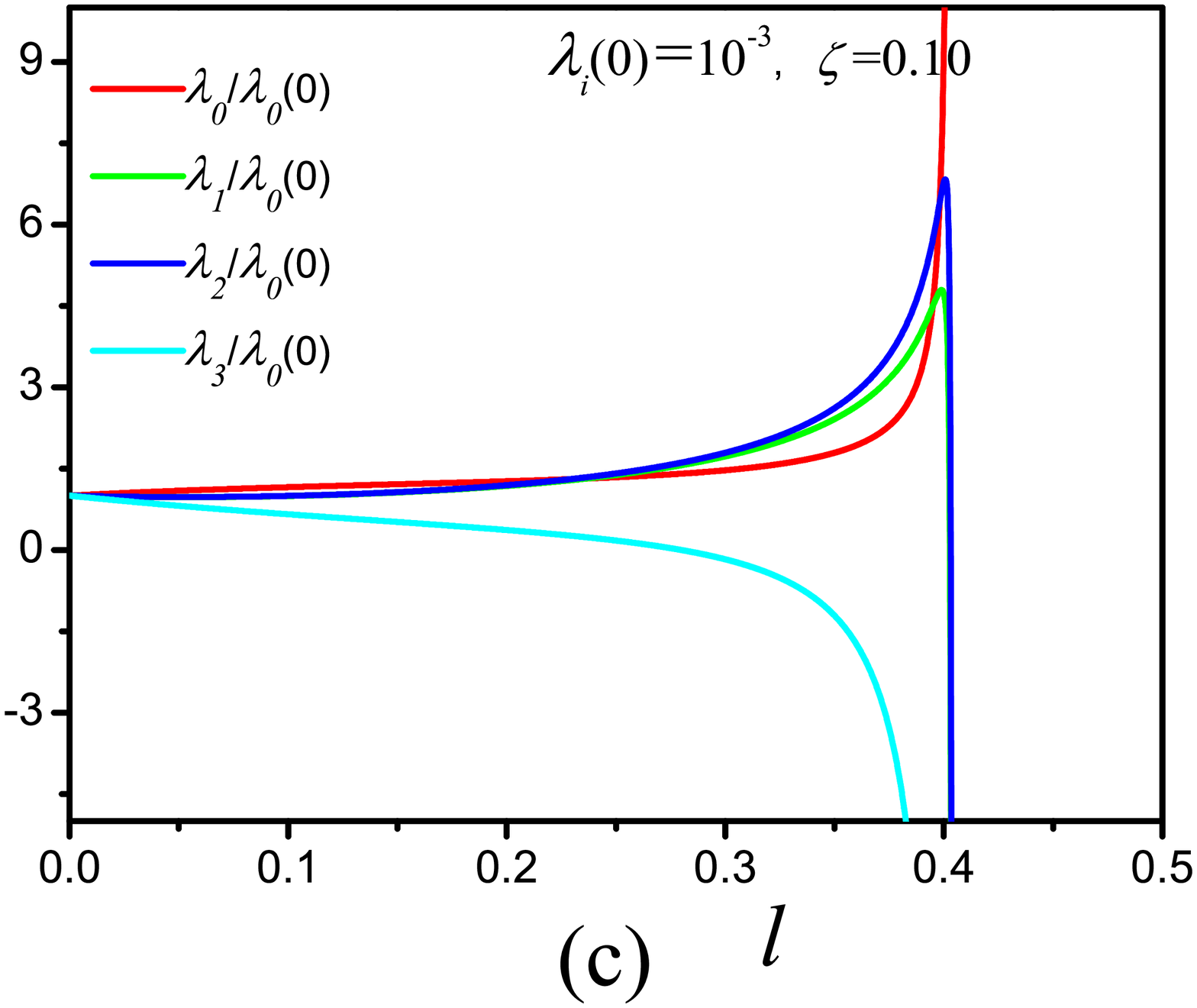}
\hspace{-1.30cm}
\includegraphics[width=1.75in]{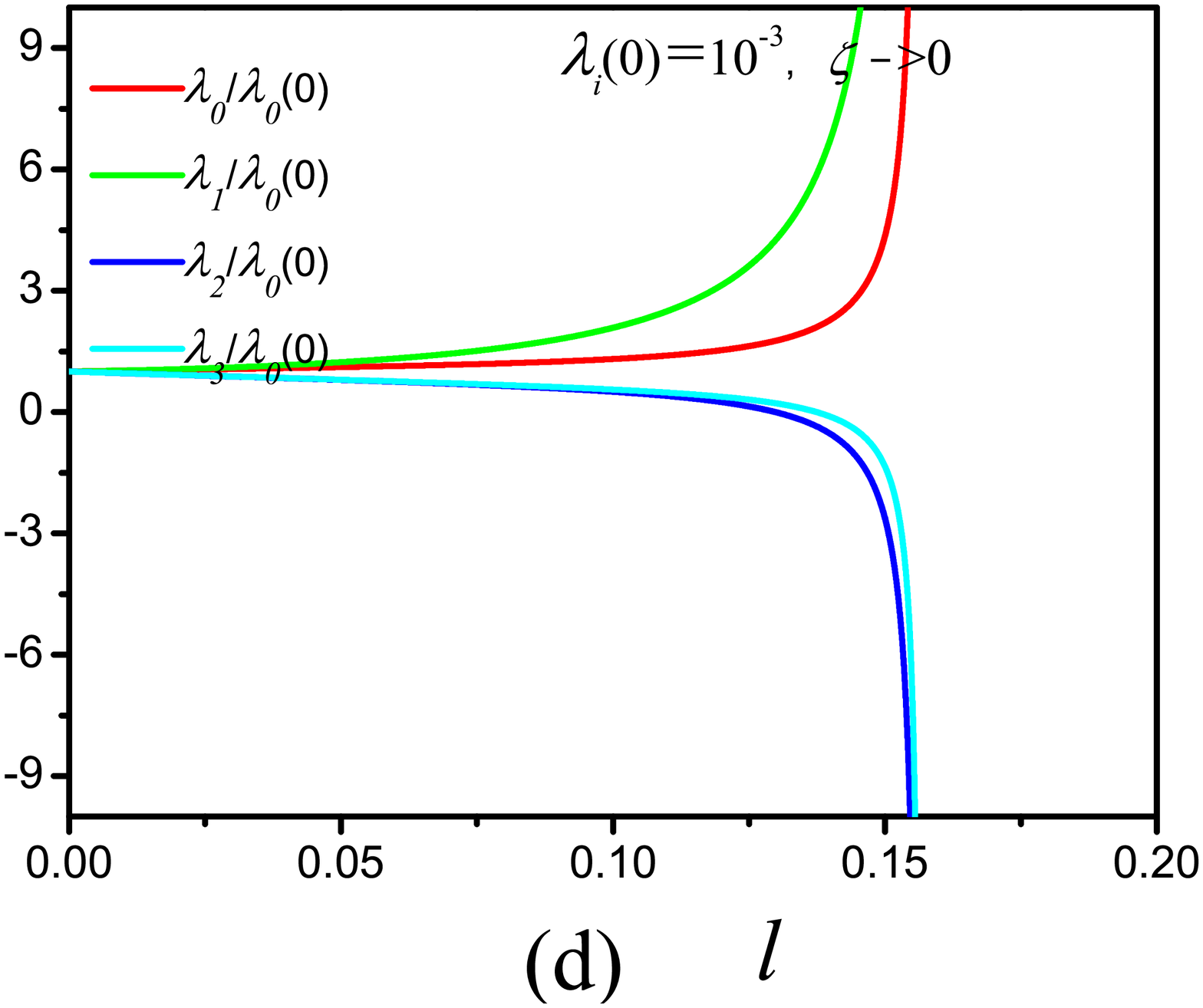}
\hspace{-1.30cm}
\includegraphics[width=1.75in]{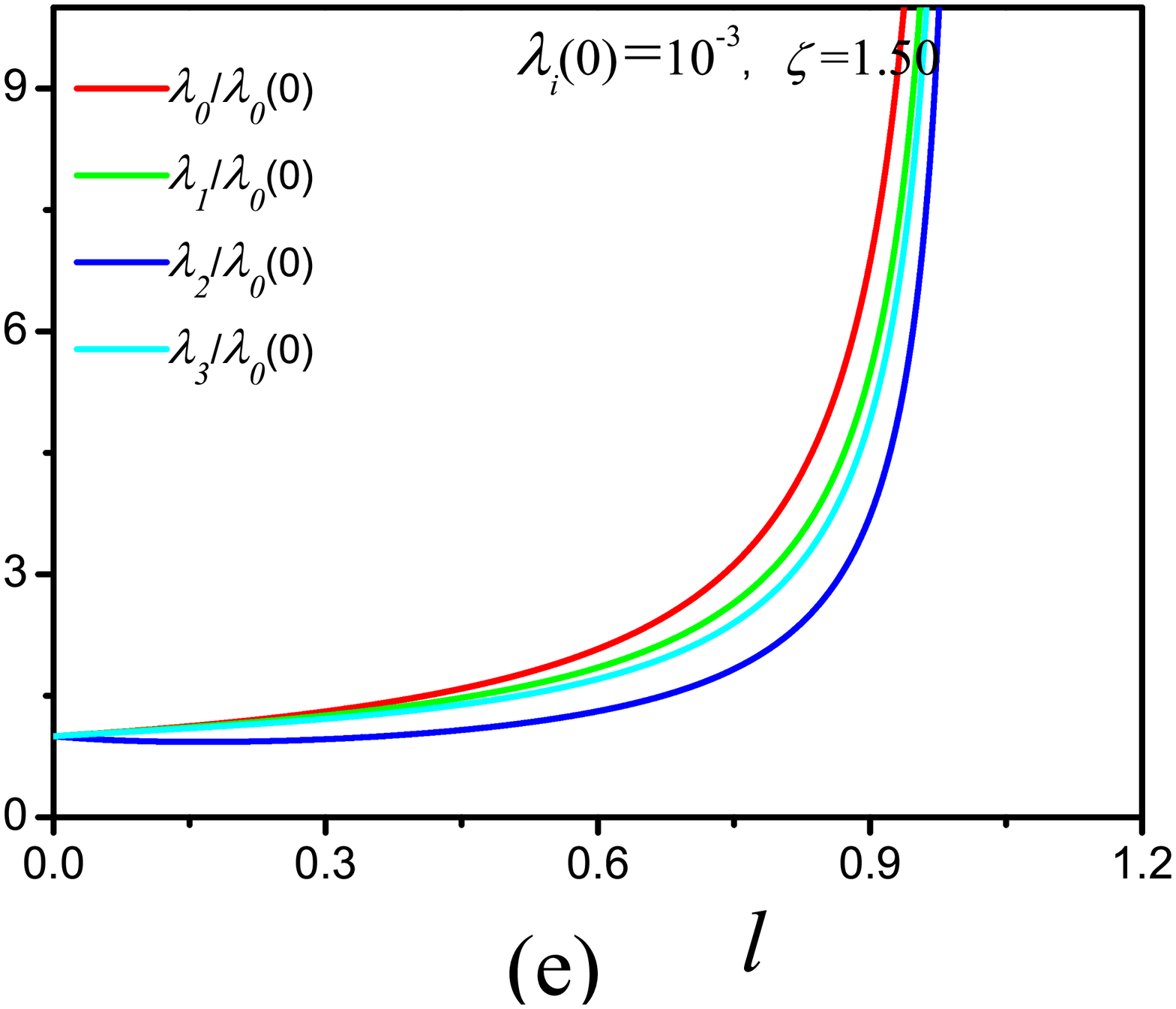}\\
\vspace{0.1cm}
\includegraphics[width=1.75in]{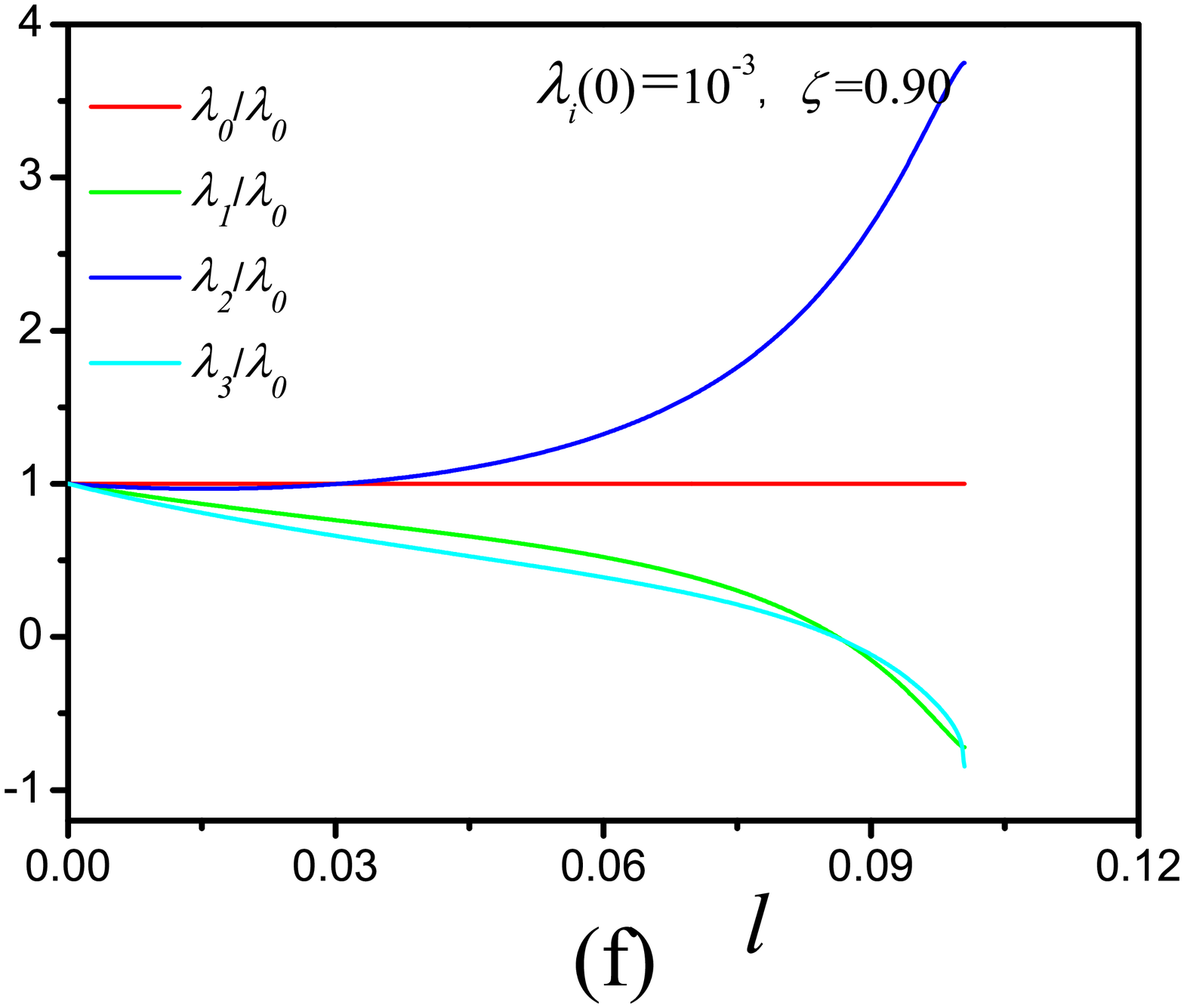}
\hspace{-1.30cm}
\includegraphics[width=1.75in]{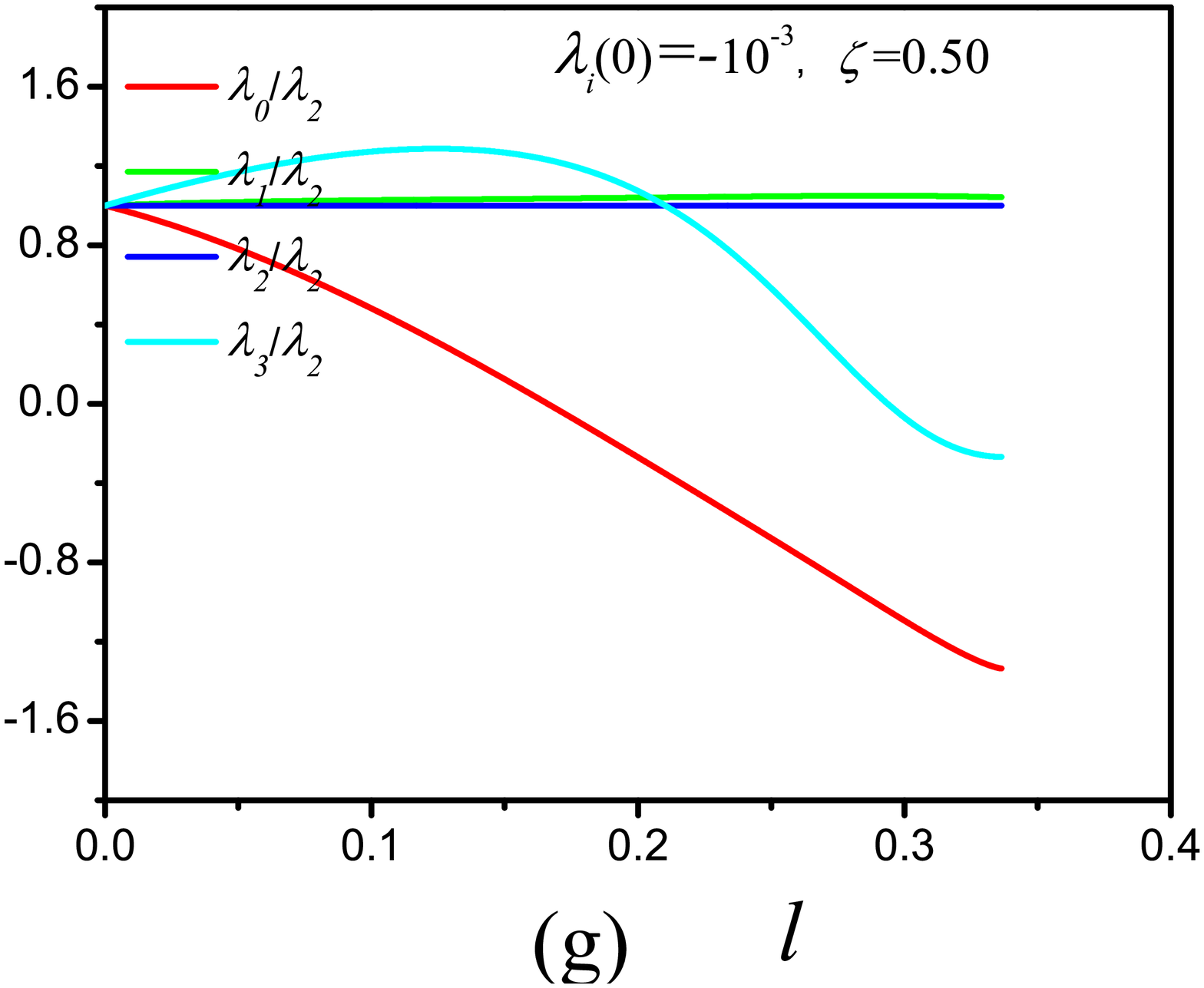}
\hspace{-1.30cm}
\includegraphics[width=1.75in]{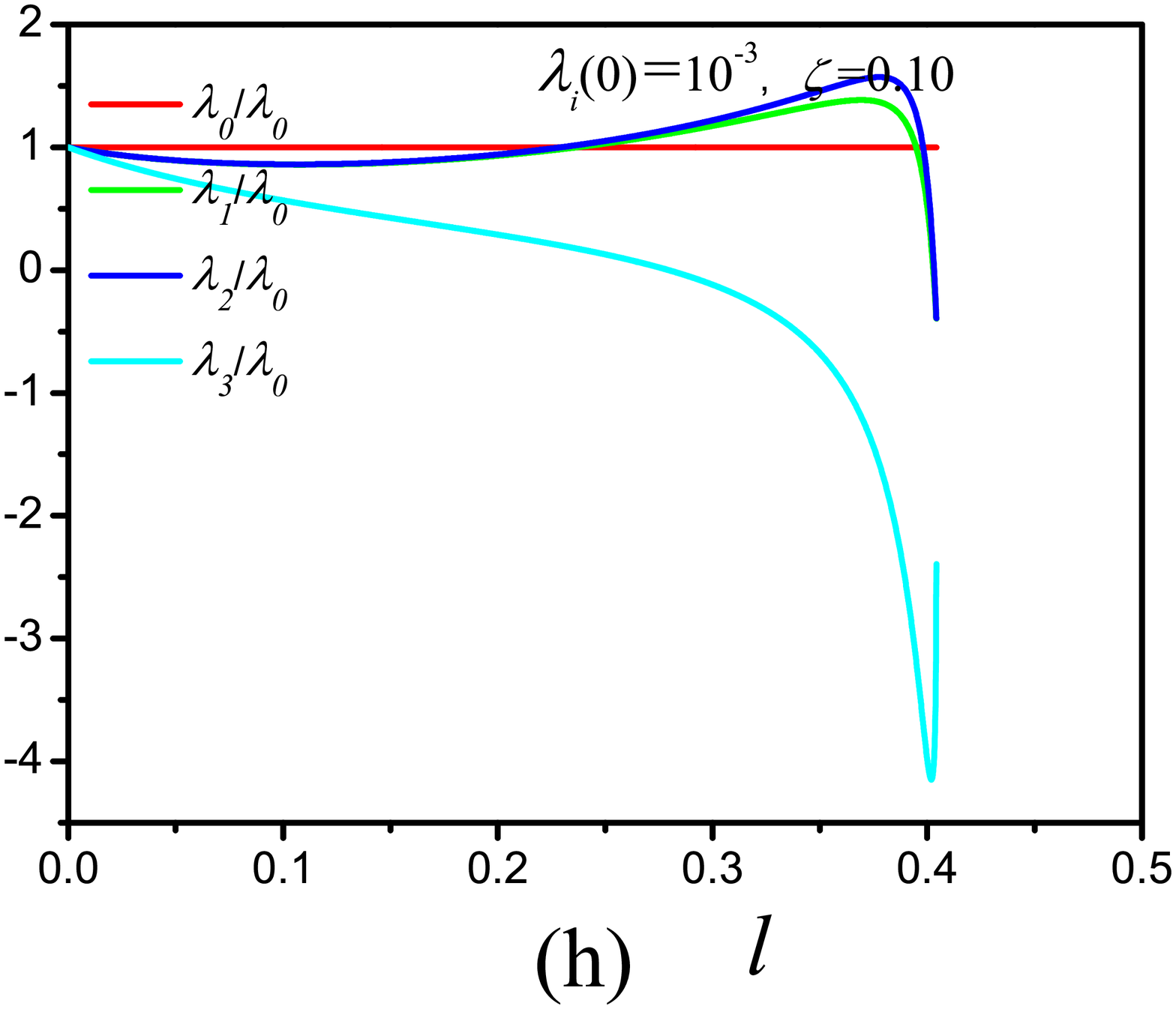}
\hspace{-1.30cm}
\includegraphics[width=1.75in]{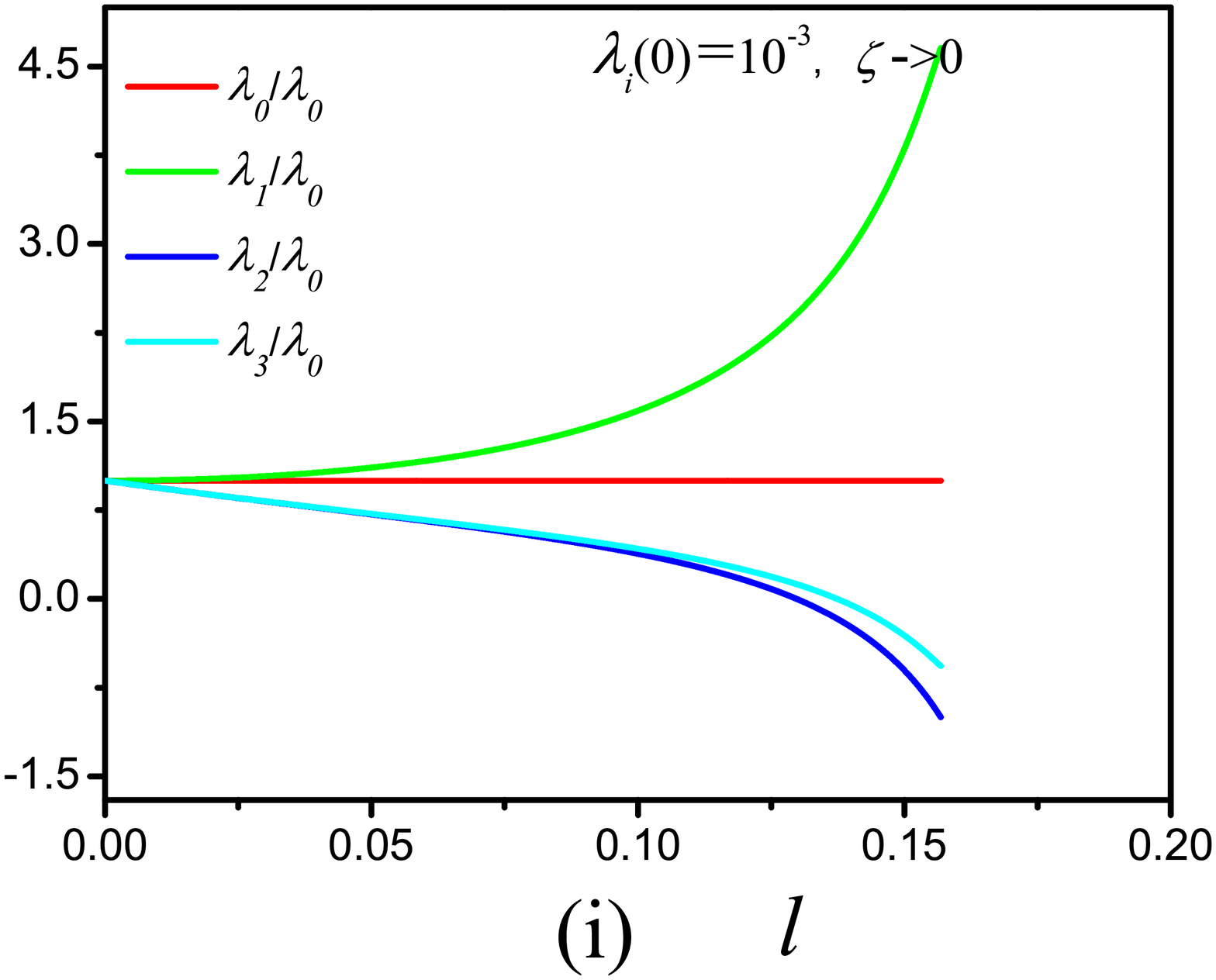}
\hspace{-1.30cm}
\includegraphics[width=1.75in]{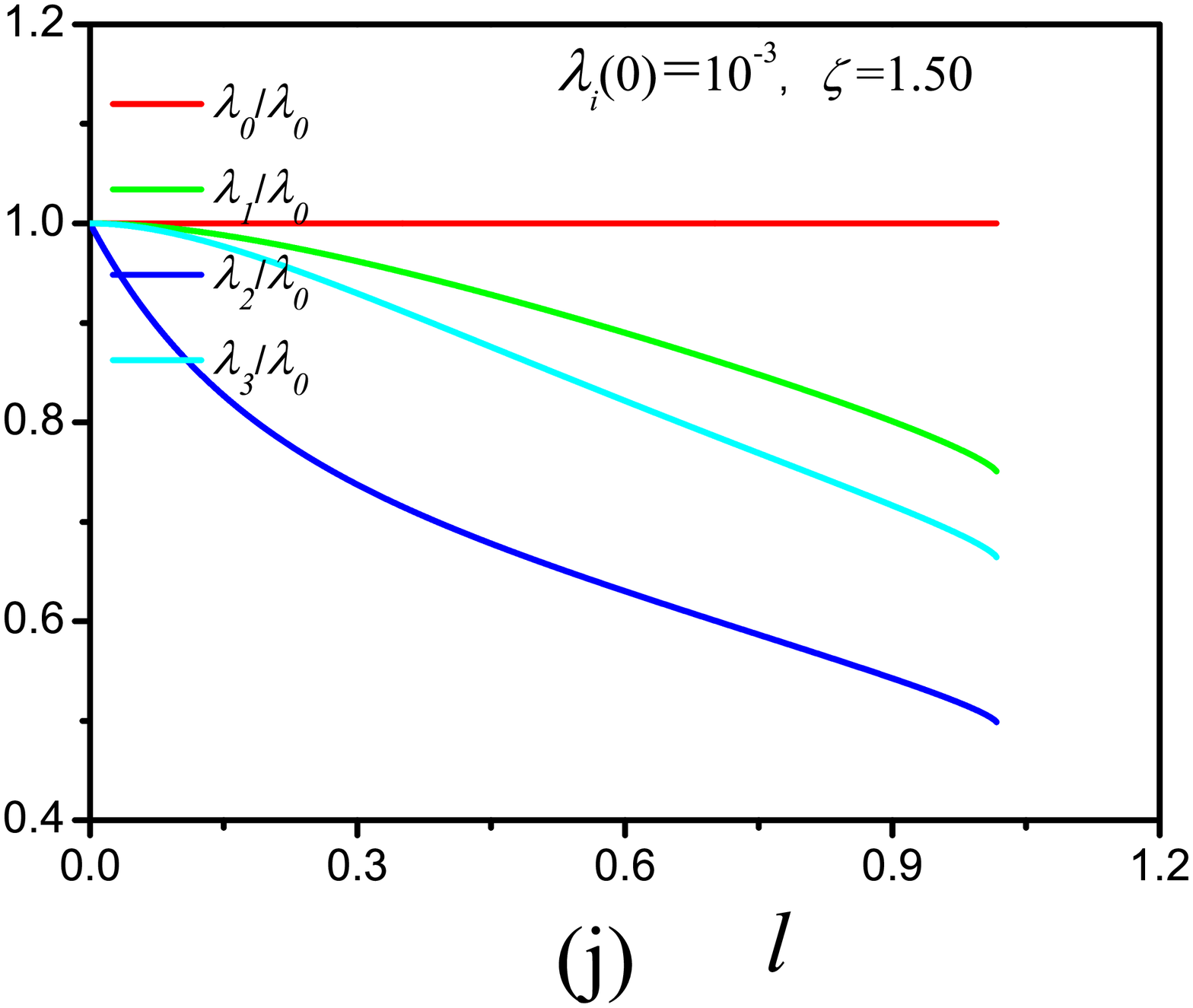}\\
\vspace{-0.0cm}
\caption{(Color online) Potential five distinct types of instabilities
for (a) Ins-I (with $\lambda_i(0)=10^{-3}$ and $\zeta=0.9$); (b) Ins-II (with $\lambda_i(0)=-10^{-3}$
and $\zeta=0.5$); (c) Ins-III (with $\lambda_i(0)=10^{-3}$
and $\zeta=0.1$); (d) Ins-IV (with $\lambda_i(0)=10^{-3}$
and $\zeta\rightarrow0$); and (e) Ins-V (with $\lambda_i(0)=10^{-3}$
and $\zeta=1.5$) at $v_1=v_2=10^{-2}$, whose RFPs correspond to subfigures
(f), (g), (h), (i) and (j), respectively.}\label{Fig-9}
\end{figure*}

At first, we consider type-I tilted Dirac fermions. On one side, Fig.~\ref{Fig-9}(a) clearly
exhibits an instability with $\lambda_i(0)>0$. In this case, both fermion-fermion couplings
$\lambda_0$ and $\lambda_2$ go up monotonously and diverge at critical energy scale without
sign change. In comparison, $\lambda_1$ and $\lambda_3$ gradually increase but quickly decrease,
change signs and eventually flow infinity in close proximity to the critical energy scale.
To be convenient, we designate this kind of instability as the first case of instability (Ins-I).
On the other side, one can obviously realize that the instability
illustrated in Fig.~\ref{Fig-9}(b) features qualitative differences of evolutions
compared to its Ins-I counterpart. Specifically, both $\lambda_1$ and
$\lambda_2$ flow towards the strong couplings without sign change at
the critical energy scale. However, the divergences of $\lambda_0$ and $\lambda_3$
at the critical energy scale are accompanied with sign changes.
Analogously, this kind of instability is denominated as the
second class of instability (Ins-II). What is more, there exists a unique kind of instability as
demonstrated apparently in Fig~\ref{Fig-9}(c), which is of remarkable distinction from both
Ins-I and Ins-II. To be concrete, the fermion-fermion coupling $\lambda_0$ quickly increases
with lowering the energy scale and directly tends to diverge at some critical point. In contrast,
both $\lambda_1$ and $\lambda_2$ progressively climb up and then diverge abruptly in the
opposite direction near the critical point. However, the interaction parameter $\lambda_3$
is monotonously decreased until diverges negatively in the proximity of the critical point.
In other words, there is only one fermion-fermion interaction that diverges positively and
all the others negatively. In order to distinguish this instability from Ins-I and Ins-II,
we thereby nominate it as the third class of instability (Ins-III) to specify these particular
low-energy behaviors of fermion-fermion couplings. What is more, there exists another class
of instability as illustrated in Fig.~\ref{Fig-9}(d).
In this circumstance, both fermion-fermion strengths $\lambda_0$
and $\lambda_1$ are monotonically and rapidly increased but instead
$\lambda_2$ and $\lambda_3$ are drastically pulled down with
the energy scales lowering. At last, all of them flow towards the
strong couplings at a critical point. Analogously, we consider this kind of
instability as the fourth class of instability (Ins-IV).

Next, we move to the type-II tilted Dirac fermions. At $\lambda_i(0)<0$, we
find the possible instability in type-II tilted system corresponds to Ins-I.
In comparison, another type of instability, namely Ins-II, is developed at $\lambda_i(0)>0$.
Other than Ins-I, Ins-II, Ins-III, and Ins-IV, Fig.~\ref{Fig-9}(e) is unambiguously indicative of an outlandish instability. In this circumstance, all the energy-dependent trajectories of $\lambda_{i}/\lambda_0(0)$ prefer to gradually increase and diverge in the same direction
without any sign change. Given that these behaviors of fermion-fermion couplings are distinct
from all the other four sorts of instabilities, the fifth class of instability (Ins-V) is
employed to characterize this kind of unique phenomenon. Conventionally, instabilities of
fermion-fermion interactions in the low-energy correspond to the relatively fixed pints (RFPs)~\cite{Murray2014PRB,Wang2017PRB_QBCP,Wang2018}, at which
phase transitions are usually accompanied.
In order to capture more information of these instabilities,
it is therefore interesting to judge whether tilted Dirac fermions
possess any RFPs of fermion-fermion couplings at the lowest-energy limit.
To this end, we can measure all four-fermion interaction parameters with
one of them whose sign is unchanged during the RG process~\cite{Murray2014PRB,Wang2017PRB_QBCP}.
For instance, we pick out $\lambda_0$ to examine whether the tilted Dirac fermion
hosts any RFPs in the parameter space, which are described by the evolutions of $\lambda_i/\lambda_0$.
To proceed, we derive and plot the trajectories of $\lambda_i/\lambda_0$ (or $\lambda_i/\lambda_2$) approaching the corresponding RFPs (possible instabilities) upon lowering the energy scale as clearly characterized in Fig.~\ref{Fig-9}(f)-(j). These results manifestly show the distinctions among
different classes of instabilities. With the help of these RFPs, it would be very helpful to investigate
the physical implications on the tendency of strong couplings for the fermion-fermion interactions~\cite{Murray2014PRB,Wang2017PRB_QBCP,Wang2018}.

\begin{table}
\centering
\caption{Low-energy fates of type-II tilted Dirac fermions with variations of
tilting parameter $\zeta$ and beginning values of fermion-fermion
couplings $\lambda_i(0)$. Hereby, ``Gs" and ``Ins"  are again adopted
to characterize Gaussian FP and certain instability, respectively.
In addition, Ins-I, Ins-V specify distinct types of instabilities, which are
addressed and denominated in Sec.~\ref{Sec_instability}. Further,
$\mathrm{Zone-I}$, $\mathrm{Zone-II}$, and $\mathrm{Zone-III}$ are
designated in Sec.~\ref{Sec_type-II-B} to discriminate values of
$\zeta$ for type-II tilted DSMs (In Sec.~\ref{Subsection_phase},
the leading phases generated by Ins-I and Ins-V are carefully investigated
and schematically displayed in Fig.~\ref{Fig-14}).}\label{Type-II}
\vspace{0.5cm}
\renewcommand{\arraystretch}{1.2}
\begin{tabular}{p{1.6cm}<{\centering}|p{1.6cm}<{\centering}|p{1.5cm}<{\centering}|
p{1.5cm}<{\centering}|p{1.5cm}<{\centering}}
\hline 
\hline
&\multirow{2}{1.5cm}{\centering$|\lambda_{i}(0)|$}&\multicolumn{3}{c}{$\zeta$}
\\
\cline{3-5}
& &$\mathrm{Zone-I}$& $\mathrm{Zone-III}$ &$\mathrm{Zone-II}$\\
\hline
\multirow{3}{1.6cm}{\centering$\lambda_{i}(0)>0$}  &Small&Gs &Gs
&Gs\\
\cline{2-5}
&Medium&Ins-V&Gs&Gs \\
\cline{2-5}
&Large&Ins-V&Ins-V&Gs \\
\hline
\multirow{3}{1.6cm}{\centering$\lambda_{i}(0)<0$}
&Small&Gs &Gs&Gs\\
\cline{2-5}
&Medium&Ins-I&Gs&Gs\\
\cline{2-5}
&Large&Ins-I&Ins-I&Gs\\
\hline
\hline
\end{tabular}
\end{table}

\begin{figure}
\centering
\includegraphics[width=2.3in]{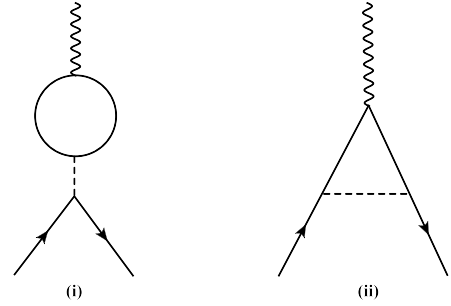}
\vspace{-0.05cm}
\caption{One-loop corrections to the bilinear fermion-source terms. The solid, dash,
and wave lines serve as
the fermion, fermion-fermion interaction and source term field, respectively.}\label{Fig-10}
\end{figure}

To be brief, the five distinct classes of instabilities, namely Ins-I,
Ins-II, Ins-III, Ins-IV, and Ins-V, can be activated attesting to the
intimate interplays among four kinds of fermion-fermion interactions in tandem with
the tilting parameter in the low-energy regime of tilted Dirac semimetals.
Specifically, Ins-I, Ins-II, Ins-III, and Ins-IV are expected in the type-I tilted
system. Rather, the type-II tilted Dirac fermions host Ins-I and Ins-V.
Table~\ref{Type-I} together with Table~\ref{Type-II} as well as
Fig.~\ref{Fig-5} present the full information of all five sorts of instabilities and
schematically exhibit physical pictures of both type-I and type-II
tilted Dirac fermions.

\begin{figure}
\centering
\includegraphics[width=3.3in]{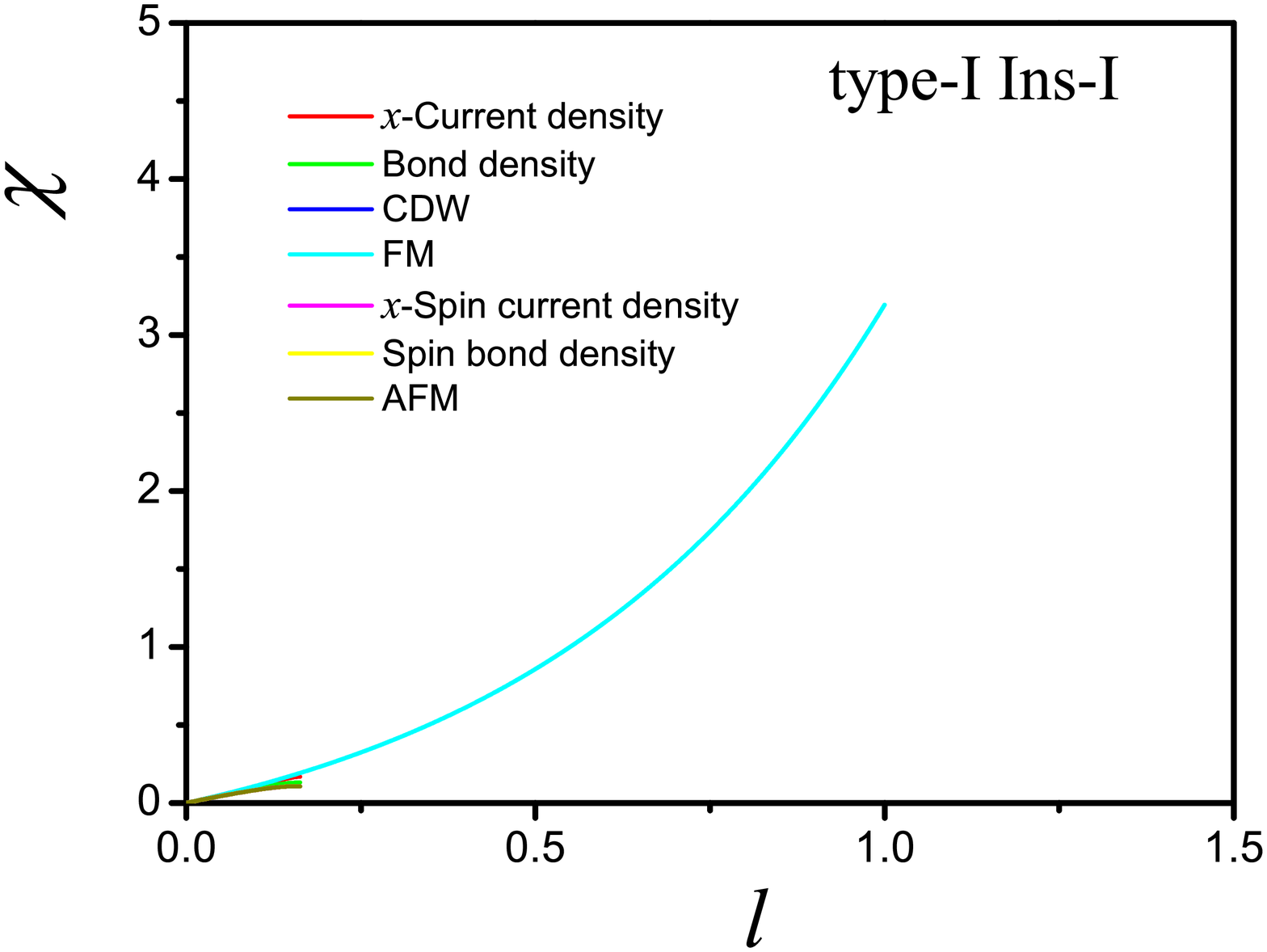}\\
\vspace{-3.9cm}
\hspace{3.1cm}\includegraphics[width=1.5in]{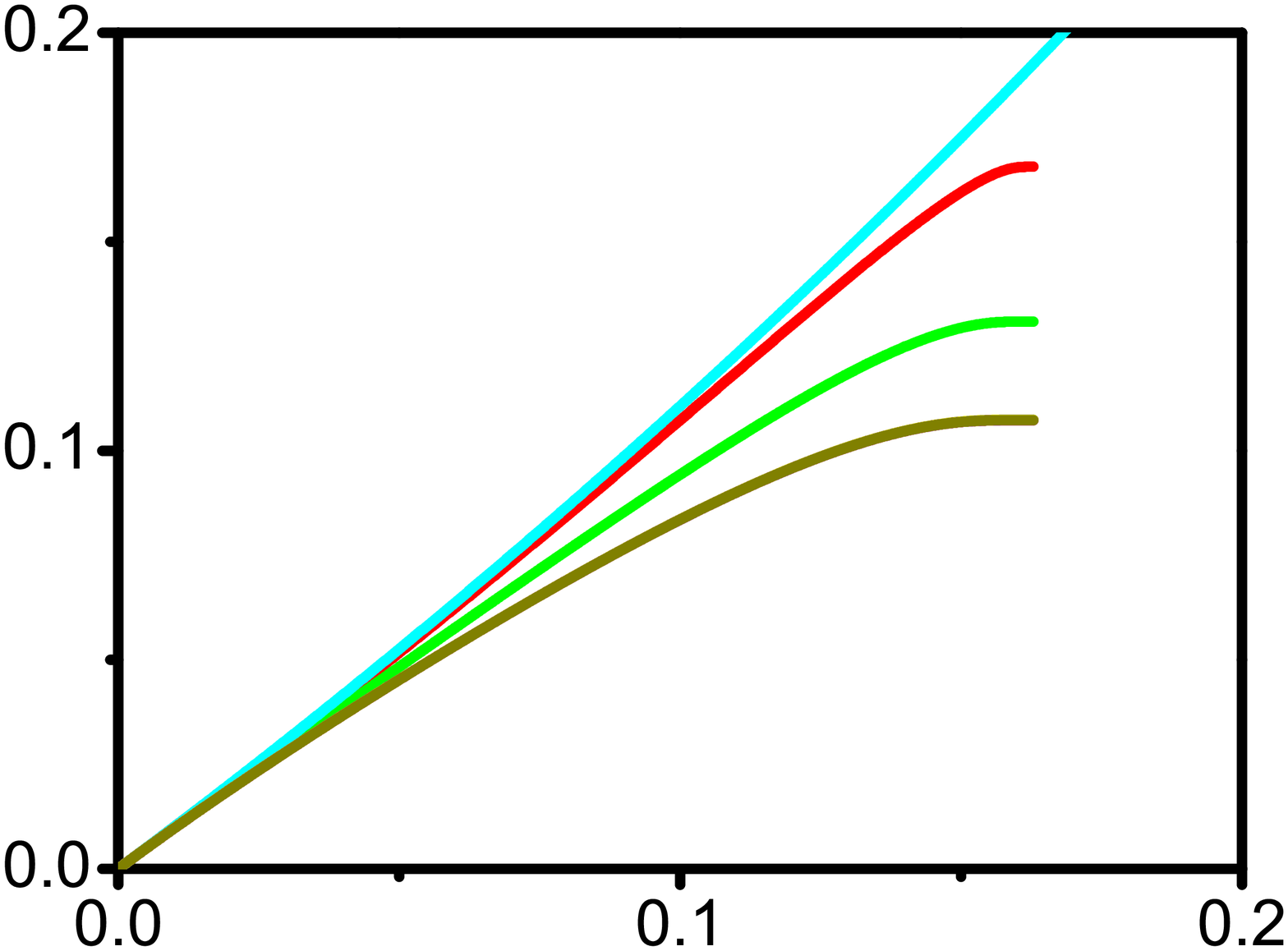}
\vspace{0.70cm}
\caption{(Color online) Susceptibilities approaching the Ins-I for type-I tilted DSMs.
The inset shows
the flows of beginning regime (the basic results for the Ins-IV are similar to
Ins-I and hence are not shown).}\label{Fig-11}
\end{figure}

\subsection{Dominant phases}\label{Subsection_phase}

To proceed, we endeavor to determine the potential leading phases in the vicinity
of these distinct types of instabilities. For this purpose, we add the following
source terms into the effective action
~(\ref{Eq_S-eff})~\cite{Murray2014PRB,Roy2017PRB-96,Roy2018PRX2}
\begin{eqnarray}
S_{\mathrm{source}}
&=&\int d\tau\int d^2\mathbf{x}\sum_{i}\Delta_{i}\psi^{\dag}\mathcal{G}_{i}\psi,
\end{eqnarray}
where the matrix $\mathcal{G}_i$ defines the various fermion bilinears and
$\Delta_i$ refers to the strength of corresponding fermion-source term. The
susceptibility of potential phase is closely related to $\Delta_i$ according to~\cite{Murray2014PRB,Wang2017PRB_QBCP}
\begin{eqnarray}
\delta \chi_i=-\frac{\partial^2\delta f}{\partial\Delta_i(0)\partial\Delta^*(0)},\label{Eq_chi}
\end{eqnarray}
with $f$ being the free energy density. To be specific,  the matrices $\mathcal{G}_1=\sigma_1$, $\mathcal{G}_2=\sigma_2$, $\mathcal{G}_3=\sigma_3$, $\mathcal{G}_4=\tau_{k}\otimes\sigma_{0}$
($\tau_k$ with $k=1,2,3$ acts on the spin space), $\mathcal{G}_5=\tau_{k}\otimes\sigma_{1}$, $\mathcal{G}_6=\tau_{k}\otimes\sigma_{2}$, and $\mathcal{G}_7=\tau_{k}\otimes\sigma_{3}$ correspond
to $x$-current density, bond density, charge density wave, ferromagnet, $x$-spin current density,
spin bond density, and antiferromagnet, respectively~\cite{Roy2017PRB-96,Roy2018PRX}.
Here, it is necessary to point out that the $x$-current and $x$-spin current density
would break certain symmetry. However, these conversed symmetry currents cannot be
regarded as some true order parameters. For completeness, we
also list them as potential stable states. In addition, the bond density and spin bond density are
associated with spin-independence anisotropic and spin-dependence anisotropic modulations of the nearest-neighbor hopping amplitude~\cite{Roy2018PRX}.

\begin{figure}
\centering
\includegraphics[width=3.3in]{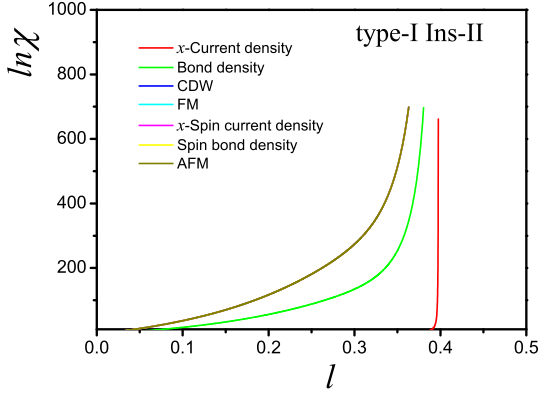}
\vspace{-0.20cm}
\caption{(Color online) Susceptibilities approaching the Ins-II for type-I tilted DSMs.
The inset shows
the flows of beginning regime (The basic results for the Ins-III are similar
to Ins-II and hence are not shown).}\label{Fig-12}
\end{figure}

\begin{figure}
\centering
\includegraphics[width=3.3in]{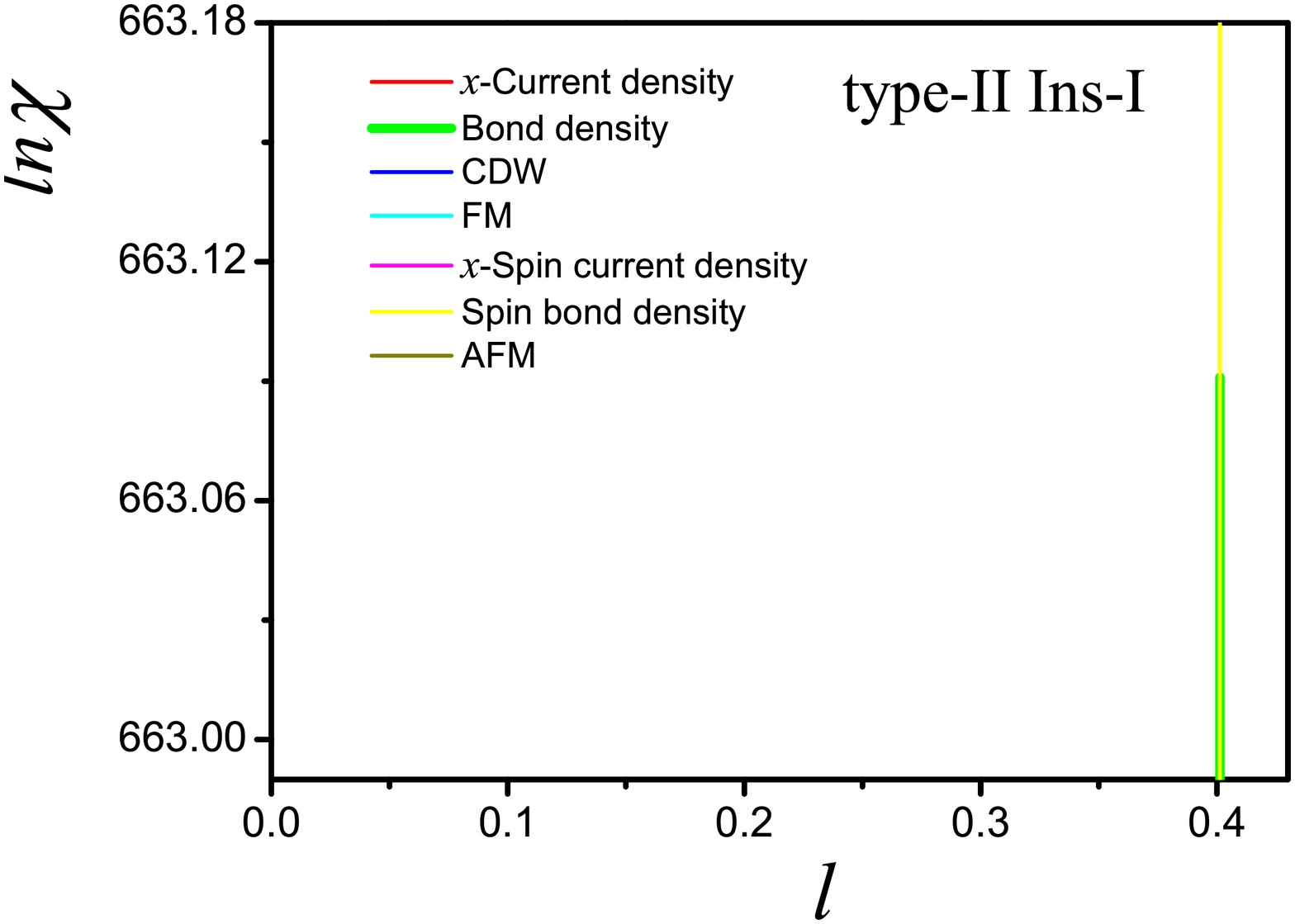}\\
\vspace{-3.6cm}
\hspace{2.1cm}\includegraphics[width=1.5in]{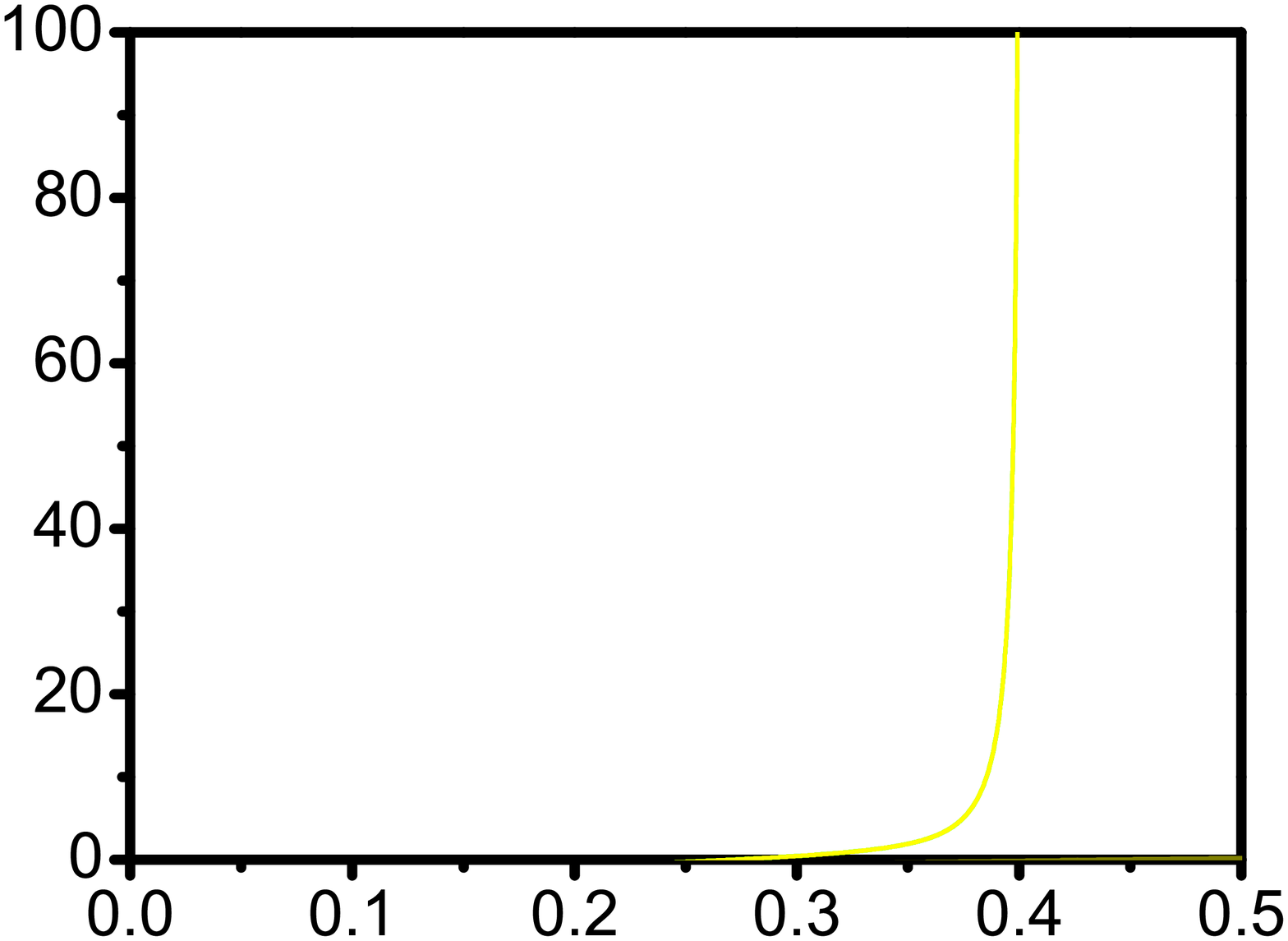}
\vspace{0.50cm}
\caption{(Color online) Susceptibilities approaching the Ins-I for type-II tilted DSMs.
The inset shows
the flows of full regime.}\label{Fig-13}
\end{figure}

Subsequently, we move our focus to one-loop corrections to the
source terms shown in Fig.~\ref{Fig-10}. After performing long but
straightforward calculations~\cite{Murray2014PRB,Wang2017PRB_QBCP,Roy2017PRB-96,
Roy2018PRX} and employing the RG scalings~(\ref{Eq_scaling-1})-(\ref{Eq_scaling-2}),
we obtain the energy-dependent source terms,
\begin{eqnarray}
\frac{d\Delta_1}{dl}
\!\!&=&\!\!\left[1+\frac{(\lambda_{0}-\lambda_{1}-\lambda_{2}-\lambda_{3})(1-\zeta^*)}
{2\pi v_{1}v_{2}\zeta^{2}}\right]\!\!\Delta_1,\label{Eq_Delta_c1}\\
\frac{d\Delta_2}{dl}
\!\!&=&\!\!\left[1+\frac{(\lambda_{0}-\lambda_{1}-\lambda_{2}-\lambda_{3})(1-\zeta^*)}
{2\pi\zeta^{2}\zeta^* v_{1}v_{2}}\right]\!\!\Delta_2,\\
\frac{d\Delta_3}{dl}
\!\!&=&\!\!\left[1+\frac{(\lambda_{0}-\lambda_{1}-\lambda_{2}-\lambda_{3})
}{2\pi v_{1}v_{2}\zeta^*}\right]\!\!\Delta_3,\\
\frac{d\Delta_4}{dl}
\!\!&=&\!\!\Delta_4,\label{Eq_Delta_c1}\\
\frac{d\Delta_5}{dl}
\!\!&=&\!\!\left[1+\frac{(3\lambda_{0}+\lambda_{1}-3\lambda_{2}-3\lambda_{3})
(1-\zeta^*)}{3\pi v_{1}v_{2}\zeta^2}\right]\!\!\Delta_5,\\
\frac{d\Delta_6}{dl}
\!\!&=&\!\!\left[1+\frac{(3\lambda_{0}-3\lambda_{1}+\lambda_{2}-3\lambda_{3})
(1-\zeta^*)}{3\pi v_{1}v_{2}\zeta^{2}\zeta^*}\right]\!\!\Delta_6,\\
\frac{d\Delta_7}{dl}
\!\!&=&\!\!\left[1+\frac{(3\lambda_{0}-3\lambda_{1}-3\lambda_{2}+\lambda_{3})
}{3\pi v_{1}v_{2}\zeta^*}\right]\!\!\Delta_7.\label{Eq_Delta_c2}
\end{eqnarray}
for type-I tilted DSMs, and
\begin{eqnarray}
\frac{d\Delta_j}{dl}
\!\!&=&\!\!\Delta_j\,\,\, \mathrm{with}\,\,j=1,3-5,7,\label{Eq_Delta_s1}\\
\frac{d\Delta_2}{dl}
\!\!&=&\!\!\left[1+\frac{\zeta^\star(\lambda_{2}-\lambda_{0}+\lambda_{1}+\lambda_{3})}{\pi^2\zeta|\zeta| v_{1}v_{2}}\right]\!\!\Delta_3,\\
\frac{d\Delta_6}{dl}
\!\!&=&\!\!\left[1-\frac{2\zeta^\star(3\lambda_{0}-3\lambda_{1}+\lambda_{2}-3\lambda_{3})}
{3\pi^2\zeta|\zeta| v_{1}v_{2}}\right]\!\!\Delta_7,\label{Eq_Delta_s2}
\end{eqnarray}
for type-II case. Here $\zeta^*$ and $\zeta^\star$ are denominated in Eq.~(\ref{Eq_zeta-star}).

Generally, the dominant phase nearby an instability is accompanied by the leading
susceptibility~\cite{Metzner2000PRL,Chubukov2016PRX,Murray2014PRB,Roy2017PRB-96,
Roy2018PRX}. Accordingly, we are suggested to evaluate the susceptibilities of all
these eight potential phases approaching the five different instabilities and sort
out the dominant one. To this end, we need to combine the RG flow equations of
interaction parameters in Eqs.~(\ref{Eq_RG-type-I-lambda-0})-(\ref{Eq_RG-type-II-lambda-3})
and energy-dependent evolutions of source terms~(\ref{Eq_Delta_c1})-(\ref{Eq_Delta_s2})
as well as the relationships between susceptibilities and source terms~(\ref{Eq_chi}).
The primary results are presented in Figs.~\ref{Fig-11}, ~\ref{Fig-12} and ~\ref{Fig-13}
after carrying out the numerical analysis.

At first, we consider type-I tilted DSMs at which Ins-I, Ins-II, Ins-III, and Ins-IV
can be induced.  All eight types of susceptibilities are increased while the Ins-I is
accessed. In particular, the ferromagnet(FM) gets the largest susceptibility as clearly
depicted in Fig.~\ref{Fig-11}. Although the susceptibility is not divergent, we would
like to expect that these two phases are more preferable than others around the Ins-I.
Once the system approaches the Ins-II, Fig.~\ref{Fig-12} manifestly exhibits that
the antiferromagnet (AFM) susceptibility climbs up very quickly and firstly goes
towards infinity. This apparently signals that the AFM is the dominant phase among
all eight potential candidates in proximity to the Ins-II. It is therefore interesting
to point out that the leading phase around Ins-II is roughly consistent with the one
caused by Coulomb interaction in the untilted DSMs~\cite{Herbut-2}. Performing the
similar steps, we find that the basic results for approaching Ins-III and Ins-IV
are qualitatively consistent with their Ins-II and Ins-I counterparts. Next, we
move to the type-II case at which only Ins-I and Ins-V are allowed to be triggered.
Learning from Fig.~\ref{Fig-13}, we can reach that the spin bond density is
the most competitive one among all possible phases in the vicinity of Ins-I for
type-II tilted DSMs. In addition, we address that the Ins-V shares the similar conclusion with Ins-I.
To be brief, we conclude that FM, AFM, and spin bond density are more preferable
as the instabilities are approached. Concretely, which of them is the
leading one would heavily rely upon the tilting parameter and concrete types of instabilities.


\begin{figure}
\centering
\includegraphics[width=3.4in]{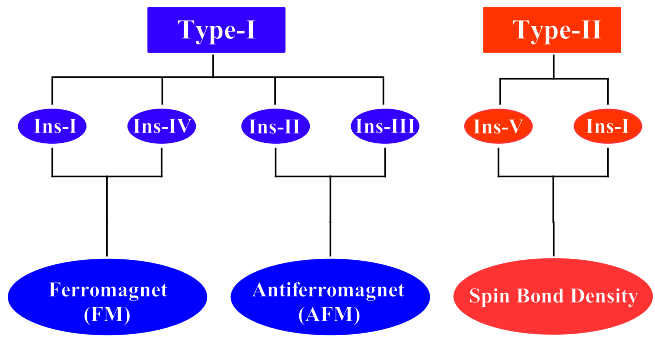}
\vspace{0.1cm}
\caption{(Color online) Schematic diagrams sketching the leading phases
accompanied by distinct sorts of instabilities for both type-I and type-II
tilted DSMs. The details of instabilities including Ins-I to Ins-V are
presented in Table~\ref{Type-I}, Table~\ref{Type-II} and Fig.~\ref{Fig-5}.}\label{Fig-14}
\end{figure}

\section{Summary}\label{Sec_summary}

In summary, we investigate the low-energy states of both type-I and type-II tilted DSMs in the presence of four distinct types of fermion-fermion interactions. By virtue of the powerful RG approach~\cite{Shankar1994RMP}, we derive the energy-dependent coupled flow equations of all interaction parameters. Carrying out detailed analysis of these evolutions signals that the effects of fermionic interaction on low-energy behaviors of tilted DSMs heavily hinge
upon the tilting parameter $\zeta$ and initial value of fermion-fermion interaction $\lambda_{i}(0)$. Several interesting low-energy properties of tilted DSMs are manifestly delivered.


Concretely, we find the tilted DSMs with lowering the energy scale can either tend to the Gaussian FP or exhibit an instability that is linked to certain phase transition. At the outset, we realize that the parameters $\zeta$ and $\lambda_{i}(0)$ play different roles in pining down the low-energy fates of type-I and type-II tilted DSMs. For type-I tilted DSMs, either increase of $\zeta$ or $\lambda_{i}(0)$ would be profitable to trigger an instability in the low-energy regime. In a contrast, tuning up the initial value of the four-fermion interaction is still favorable to the generation of instability for type-II tilted DSMs. However, the tilting parameter $\zeta$ is not proportional to the instability. This implies that the system unavoidably flows towards the Gaussian FP once the $\zeta$ is sufficiently large. In addition, we figure out the parameters $\zeta$ and $\lambda_{i}(0)$ strongly compete within distinct zones of $\zeta$. In the type-I tilted DSMs, $\lambda_{i}(0)$ wins the
competition at both Zone-I ($\zeta\rightarrow0$) and Zone-II ($\zeta\rightarrow1$). Whereas the tilting parameter $\zeta$ plays a more crucial role in igniting the potential instability at Zone-III characterized by $\zeta\in$ other values. In comparison, the parameter $|\lambda_{i}(0)|$ dominates over the tilting parameter if $\zeta$ is small for type-II tilted DSMs. Once $\zeta$ is large, it becomes a leading facet to determine the low-energy state. In particular, any instability is not allowed at $\zeta\rightarrow\infty$. Moreover, all of underlying instabilities induced by the fermion-fermion interactions via appropriately adjusting
the two parameters $\zeta$ and $\lambda_{i}(0)$ can be clustered into five distinct classes, which own qualitatively different energy-dependent trajectories and RFPs at the critical energy scale and are nominated as Ins-I, Ins-II, Ins-III, Ins-IV, and Ins-V, respectively. The detailed information is provided in Table~\ref{Type-I}, Table~\ref{Type-II} and Fig.~\ref{Fig-9}.
Based on our studies, Ins-I, Ins-II, Ins-III, and Ins-IV can be expected in the type-I tilted DSMs. Rather, the type-II tilted Dirac fermions only host Ins-I and Ins-V. After computing and comparing susceptibilities of eight potential phases around these instabilities, we find that the dominant phases are closely related to the tilting parameter and concrete instabilities. To be brief, the type-I
and the type-II tilted Dirac fermions exhibit distinct low-energy physical
behaviors under the effects of fermion-fermion interactions.

In principle, all of these interesting physical behaviors for type-I and type-II tilted DSMs caused by the four-fermion interactions in the low-energy regime would be remarkably instructive to further study the physical properties affected by instabilities, detect the positions of potential phase transitions and so on in the tilted DSMs or other tilted systems. In a word, we expect our efforts can play an important positive role in both future theoretical and experimental explores in tilted fermionic systems.


\section*{ACKNOWLEDGEMENTS}

J.W. is partially supported by the National Natural Science Foundation of China under Grant 11504360.

\section*{AUTHOR CONTRIBUTION STATEMENTS}

J. W. initiated and supervised the project as well as performed the numerical analysis and wrote the manuscript. J.Q.L. carried out the analytical calculations. Z.D.X. participated in some discussions and provided several useful suggestions.

\vspace{0.5cm}

\appendix

\section{One-loop corrections}\label{Appendix_1L-corrections}

After carrying out long but straightforward one-loop calculations~\cite{Murray2014PRB,Wang2017PRB_QBCP,Wang2018}
and collecting all corrections, we are left with the following results,
\begin{widetext}
\begin{eqnarray}
S_{\xi}^{\lambda_{0}}
&=&\int\frac{dp_{0}^{\prime}dp_{0}^{\prime\prime}dp_{0}^{\prime\prime\prime}}{(2\pi)^{3}}
\frac{d^{2}\mathbf{p}^{\prime}
d^{2}\mathbf{p}^{\prime\prime}d^{2}\mathbf{p}^{\prime\prime\prime}}{(2\pi)^{6}}
[\psi_{\xi\alpha}^{\dagger}(p_{0}^{\prime},\mathbf{p}^{\prime})\sigma_{0}\psi_{\xi\alpha}
(p_{0}^{\prime\prime},\mathbf{p}^{\prime\prime})\psi_{\xi^{\prime}
\alpha^{\prime}}^{\dag}(p_{0}^{\prime\prime\prime},\mathbf{p}^{\prime\prime\prime})\sigma_{0}
\psi_{\xi^{\prime}\alpha^{\prime}}(p_{0}^{\prime}+p_{0}^{\prime\prime}-p_{0}^{\prime\prime\prime},
\mathbf{p}^{\prime}+\mathbf{p}^{\prime\prime}-\mathbf{p}^{\prime\prime\prime})]\nonumber\\
&&\times\frac{\left[\zeta^{2}\left(\lambda_{0}^2+\lambda_{1}^2+\lambda_{2}^{2}
+\lambda_{3}^2-2\lambda_{0}\lambda_{2}\right)
+2\left(\zeta^{\ast}-1\right)\lambda_{0}(\lambda_{1}-\lambda_{2})
\right] l}{2\pi v_{1}v_{2}\zeta^{2}\zeta^{\ast}},\\
S_{\xi}^{\lambda_{1}}
&=&\int\frac{dp_{0}^{\prime}dp_{0}^{\prime\prime}dp_{0}^{\prime\prime\prime}}{(2\pi)^{3}}
\frac{d^{2}\mathbf{p}^{\prime}
d^{2}\mathbf{p}^{\prime\prime}d^{2}\mathbf{p}^{\prime\prime\prime}}{(2\pi)^{6}}
[\psi_{\xi\alpha}^{\dagger}(p_{0}^{\prime},\mathbf{p}^{\prime})\sigma_{1}\psi_{\xi\alpha}
(p_{0}^{\prime\prime},\mathbf{p}^{\prime\prime})\psi_{\xi^{\prime}
\alpha^{\prime}}^{\dag}(p_{0}^{\prime\prime\prime},\mathbf{p}^{\prime\prime\prime})\sigma_{1}
\psi_{\xi^{\prime}\alpha^{\prime}}(p_{0}^{\prime}+p_{0}^{\prime\prime}-p_{0}^{\prime\prime\prime},
\mathbf{p}^{\prime}+\mathbf{p}^{\prime\prime}-\mathbf{p}^{\prime\prime\prime})]\nonumber\\
&&\times\frac{\left(\zeta^{\ast}-1\right)\Bigr[\lambda^2_0+5\lambda^2_1+\lambda^2_2
+\lambda^2_3+2\lambda_1(\lambda_0-\lambda_2-\lambda_3)-2\lambda_0\lambda_3\Bigr]
+2\zeta^{2}\left[\lambda_{1}(2\lambda_0+2\lambda_1-\lambda_2-\lambda_3)
-\lambda_0\lambda_3\right]}{2\pi v_{1}v_{2}\zeta^{2}\zeta^{\ast}}l,\\
S_{\xi}^{\lambda_{2}}
&=&-\int\frac{dp_{0}^{\prime}dp_{0}^{\prime\prime}dp_{0}^{\prime\prime\prime}}{(2\pi)^{3}}
\frac{d^{2}\mathbf{p}^{\prime}
d^{2}\mathbf{p}^{\prime\prime}d^{2}\mathbf{p}^{\prime\prime\prime}}{(2\pi)^{6}}
[\psi_{\xi\alpha}^{\dagger}(p_{0}^{\prime},\mathbf{p}^{\prime})\sigma_{2}\psi_{\xi\alpha}
(p_{0}^{\prime\prime},\mathbf{p}^{\prime\prime})\psi_{\xi^{\prime}
\alpha^{\prime}}^{\dag}(p_{0}^{\prime\prime\prime},\mathbf{p}^{\prime\prime\prime})\sigma_{2}
\psi_{\xi^{\prime}\alpha^{\prime}}(p_{0}^{\prime}+p_{0}^{\prime\prime}-p_{0}^{\prime\prime\prime},
\mathbf{p}^{\prime}+\mathbf{p}^{\prime\prime}-\mathbf{p}^{\prime\prime\prime})]\nonumber\\
&&\times\frac{\left\{\left(\zeta^{\ast}-1\right)
\left[\lambda_{0}^2+\lambda_{1}^{2}+5\lambda_{2}^{2}+\lambda_{3}^{2}
+2(\lambda_{0}-\lambda_{1}-\lambda_{3})\lambda_{2}
-2\lambda_{0}\lambda_{3}\right]
+\zeta^{2}\left(\lambda_{0}^2+\lambda_{1}^{2}+\lambda_{2}^{2}+\lambda_{3}^{2}-
2\lambda_{0}\lambda_{2}\right)\right\}l}{2\pi v_{1}v_{2}\zeta^{2}\zeta^{\ast}},\\
S_{\xi}^{\lambda_{3}}
&=&-\int\frac{dp_{0}^{\prime}dp_{0}^{\prime\prime}dp_{0}^{\prime\prime\prime}}{(2\pi)^{3}}
\frac{d^{2}\mathbf{p}^{\prime}
d^{2}\mathbf{p}^{\prime\prime}d^{2}\mathbf{p}^{\prime\prime\prime}}{(2\pi)^{6}}
[\psi_{\xi\alpha}^{\dagger}(p_{0}^{\prime},\mathbf{p}^{\prime})\sigma_{3}
\psi_{\xi\alpha}(p_{0}^{\prime\prime},\mathbf{p}^{\prime\prime})\psi_{\xi^{\prime}
\alpha^{\prime}}^{\dag}(p_{0}^{\prime\prime\prime},\mathbf{p}^{\prime\prime\prime})
\sigma_{3}\psi_{\xi^{\prime}\alpha^{\prime}}(p_{0}^{\prime}+p_{0}^{\prime\prime}
-p_{0}^{\prime\prime\prime},\mathbf{p}^{\prime}+\mathbf{p}^{\prime\prime}
-\mathbf{p}^{\prime\prime\prime})]\nonumber\\
&&\times\frac{\left\{-\zeta^{2}\left[
(\lambda_{1}+\lambda_{2}-2\lambda_{3})\lambda_{3}-\lambda_{0}\lambda_{1}\right]
+\left(\zeta^{\ast}-1\right)\lambda_{0}\left(\lambda_{1}-\lambda_{2}\right)
\right\}l}{\pi v_{1}v_{2}\zeta^2\zeta^{\ast}},
\end{eqnarray}
for type-I tilted-Dirac semimetals and
\begin{eqnarray}
S_{\xi}^{\lambda_{0}}
&=&\!\!-\!\int_{-\infty}^{+\infty}\frac{dp_{0}
^{\prime}dp_{0}^{\prime\prime}dp_{0}^{\prime\prime\prime}}{(2\pi)^{3}}\int\frac{d^{2}\mathbf{p}^{\prime}
d^{2}\mathbf{p}^{\prime\prime}d^{2}\mathbf{p}^{\prime\prime\prime}}{(2\pi)^{6}}
[\psi_{\xi\alpha}^{\dagger}(p_{0}^{\prime},\mathbf{p}^{\prime})\sigma_{0}\psi_{\xi\alpha}
(p_{0}^{\prime\prime},\mathbf{p}^{\prime\prime})\psi_{\xi^{\prime}
\alpha^{\prime}}^{\dag}(p_{0}^{\prime\prime\prime},\mathbf{p}^{\prime\prime\prime})\sigma_{0}
\psi_{\xi^{\prime}\alpha^{\prime}}(p_{0}^{\prime}+p_{0}^{\prime\prime}-p_{0}^{\prime\prime\prime},
\mathbf{p}^{\prime}+\mathbf{p}^{\prime\prime}-\mathbf{p}^{\prime\prime\prime})]\nonumber\\
&&\times\left[-(\lambda_{0}^2+\lambda_{1}^{2}+\lambda_{2}^{2}
+\lambda_{3}^{2})L_{3}+2\lambda_{0}\lambda_{1}L_{2}
+2\lambda_{0}\lambda_{2}L_{1}\right],\\
S_{\xi}^{\lambda_{1}}
&=&\int_{-\infty}^{+\infty}\frac{dp_{0}^{\prime}dp_{0}^{\prime\prime}
dp_{0}^{\prime\prime\prime}}{(2\pi)^{3}}\int\frac{d^{2}\mathbf{p}^{\prime}
d^{2}\mathbf{p}^{\prime\prime}d^{2}\mathbf{p}^{\prime\prime\prime}}
{(2\pi)^{6}}[\psi_{\xi\alpha}^{\dagger}(p_{0}^{\prime},\mathbf{p}^{\prime})\sigma_{1}
\psi_{\xi\alpha}(p_{0}^{\prime\prime},\mathbf{p}^{\prime\prime})\psi_{\xi^{\prime}
\alpha^{\prime}}^{\dag}(p_{0}^{\prime\prime\prime},\mathbf{p}^{\prime\prime\prime})
\sigma_{1}\psi_{\xi^{\prime}\alpha^{\prime}}(p_{0}^{\prime}+p_{0}^{\prime\prime}-p_{0}
^{\prime\prime\prime},\mathbf{p}^{\prime}+\mathbf{p}^{\prime\prime}-\mathbf{p}
^{\prime\prime\prime})]\nonumber\\
&&\times\left\{2\left[\lambda_{1}\left(\lambda_{0}+2\lambda_{1}-2\lambda_{2}-\lambda_{3}\right)
-\lambda_{0}\lambda_{3}\right]L_{1}-\left(\lambda_{0}^{2}+\lambda_{1}^{2}
+\lambda_{2}^{2}+\lambda_{3}^{2}\right)L_{2}
+2\lambda_{0}\lambda_{1}L_{3}\right\},\\
S_{\xi}^{\lambda_{2}}
&=&\int_{-\infty}^{+\infty}\frac{dp_{0}^{\prime}dp_{0}^{\prime\prime}
dp_{0}^{\prime\prime\prime}}{(2\pi)^{3}}\int\frac{d^{2}\mathbf{p}^{\prime}
d^{2}\mathbf{p}^{\prime\prime}d^{2}\mathbf{p}^{\prime\prime\prime}}
{(2\pi)^{6}}
[\psi_{\xi\alpha}^{\dagger}(p_{0}^{\prime},\mathbf{p}^{\prime})\sigma_{2}
\psi_{\xi\alpha}(p_{0}^{\prime\prime},\mathbf{p}^{\prime\prime})\psi_{\xi^{\prime}
\alpha^{\prime}}^{\dag}(p_{0}^{\prime\prime\prime},\mathbf{p}^{\prime\prime\prime})
\sigma_{2}\psi_{\xi^{\prime}\alpha^{\prime}}(p_{0}^{\prime}+p_{0}^{\prime\prime}-p_{0}
^{\prime\prime\prime},\mathbf{p}^{\prime}+\mathbf{p}^{\prime\prime}-\mathbf{p}
^{\prime\prime\prime})]\nonumber\\
&&\times\left\{-\left(\lambda_{0}^2+\lambda_{1}^2+\lambda_{2}^2+\lambda_{3}^2\right)
L_{1}+2\left[\lambda_{2}\left(\lambda_{0}-\lambda_{1}+2\lambda_{2}
-\lambda_{3}\right)-\lambda_{0}\lambda_{3}\right]L_{2}+2\lambda_{0}\lambda_{2}L_{3}
\right\},\\
S_{\xi}^{\lambda_{3}}
&=&\int_{-\infty}^{+\infty}\frac{dp_{0}^{\prime}dp_{0}^{\prime\prime}
dp_{0}^{\prime\prime\prime}}{(2\pi)^{3}}\int\frac{d^{2}\mathbf{p}^{\prime}
d^{2}\mathbf{p}^{\prime\prime}d^{2}\mathbf{p}^{\prime\prime\prime}}
{(2\pi)^{6}}[\psi_{\xi\alpha}^{\dagger}(p_{0}^{\prime},\mathbf{p}^{\prime})\sigma_{3}
\psi_{\xi\alpha}(p_{0}^{\prime\prime},\mathbf{p}^{\prime\prime})\psi_{\xi^{\prime}
\alpha^{\prime}}^{\dag}(p_{0}^{\prime\prime\prime},\mathbf{p}^{\prime\prime\prime})
\sigma_{3}\psi_{\xi^{\prime}\alpha^{\prime}}(p_{0}^{\prime}+p_{0}^{\prime\prime}-p_{0}
^{\prime\prime\prime},\mathbf{p}^{\prime}+\mathbf{p}^{\prime\prime}-\mathbf{p}
^{\prime\prime\prime})]\nonumber\\
&&\times\left\{-2\lambda_{0}\lambda_{1}L_{1}-2\lambda_{0}\lambda_{2}L_{2}
+2\left(\lambda_{1}+\lambda_{2}-2\lambda_{3}\right)
\lambda_{3}L_{3}\right\},
\end{eqnarray}
for type-II tilted-Dirac semimetals, respectively.
Here, the related coefficients are nominated as
\begin{eqnarray}
L_{1}\!\!\!&=&\!\!\!\frac{1}{16\pi^2 v_{1}v_{2}}
\int_{\Lambda/s}^{\Lambda}\frac{dE}{(\zeta^{2}-1)^{1/2}}
\left[\mathcal{D}_{1}(\xi,\zeta)+\mathcal{D}_{1}(-\xi,\zeta)\right]
+\frac{1}{16\pi^2 v_{1}v_{2}}
\int_{-\Lambda}^{-\Lambda/s}\!\!\!\frac{dE}{(\zeta^{2}-1)^{1/2}}
\left[\mathcal{D}_{1}(-\xi,\zeta)+\mathcal{D}_{1}(\xi,\zeta)\right],~\label{L_{1}}\\
L_{2}\!\!\!&=&\!\!\!\frac{1}{16\pi^2 v_{1}v_{2}}
\int_{\Lambda/s}^{\Lambda}\frac{dE}{(\zeta^{2}-1)^{1/2}}
\left[\mathcal{D}_{2}(\xi,\zeta)+\mathcal{D}_{2}(-\xi,\zeta)\right]+\frac{1}{16\pi^2 v_{1}v_{2}}
\int_{-\Lambda}^{-\Lambda/s}\!\!\!\frac{dE}{(\zeta^{2}-1)^{1/2}}
\left[\mathcal{D}_{2}(-\xi,\zeta)+\mathcal{D}_{2}(\xi,\zeta)\right],~\label{L_{2}}\\
L_{3}\!\!\!&=&\!\!\!\frac{1}{16\pi^2 v_{1}v_{2}}
\int_{\Lambda/s}^{\Lambda}\frac{dE}{(\zeta^{2}-1)^{1/2}}
\left[\mathcal{D}_{3}(\xi,\zeta)+\mathcal{D}_{3}(-\xi,\zeta)\right]+\frac{1}{16\pi^2 v_{1}v_{2}}
\int_{-\Lambda}^{-\Lambda/s}\!\!\!\frac{dE}{(\zeta^{2}-1)^{1/2}}
\left[\mathcal{D}_{3}(-\xi,\zeta)+\mathcal{D}_{3}(\xi,\zeta)\right],~\label{L_{3}}
\end{eqnarray}
with
\begin{eqnarray}
\mathcal{D}_{1}(\xi,\zeta)
\!\!\!&=&\!\!\!\int_{-\infty}^{\infty}\!\!\!d\theta
\frac{(|\zeta|\cosh\theta+\eta_{\zeta}\xi)\left(\zeta^{2}-1\right)\sinh^{2}\theta}
{\left(\xi\zeta\cosh\theta+1\right)^{3}},
\mathcal{D}_{1}(-\xi,\zeta)
=\!\!\!\int_{-\infty}^{\infty}\!\!\!d\theta
\frac{(|\zeta|\cosh\theta-\eta_{\zeta}\xi)\left(\zeta^{2}-1\right)\sinh^{2}\theta}
{\left(-\xi\zeta\cosh\theta+1\right)^{3}},\\
\mathcal{D}_{2}(\xi,\zeta)
\!\!\!&=&\!\!\!\int_{-\infty}^{\infty}\!\!\!d\theta
\frac{(|\zeta|\cosh\theta+\eta_{\zeta}\xi)\left(\xi\zeta+\cosh\theta\right)^{2}}
{\left(\xi\zeta\cosh\theta+1\right)^{3}},\,\,\,
\mathcal{D}_{2}(-\xi,\zeta)
=\!\!\!\int_{-\infty}^{\infty}\!\!\!d\theta
\frac{(|\zeta|\cosh\theta-\eta_{\zeta}\xi)\left(-\xi\zeta+\cosh\theta\right)^{2}}
{\left(-\xi\zeta\cosh\theta+1\right)^{3}},\\
\mathcal{D}_{3}(\xi,\zeta)
\!\!\!&=&\!\!\!\int_{-\infty}^{\infty}\!\!\!d\theta
\frac{(|\zeta|\cosh\theta+\eta_{\zeta}\xi)}
{\left(\xi\zeta\cosh\theta+1\right)},\hspace{2.3cm}
\mathcal{D}_{3}(-\xi,\zeta)
=\!\!\!\int_{-\infty}^{\infty}\!\!\!d\theta
\frac{(|\zeta|\cosh\theta-\eta_{\zeta}\xi)}
{\left(-\xi\zeta\cosh\theta+1\right)}.
\end{eqnarray}

In order to calculate the coefficients $\mathcal{D}_i$ with $i=1,2,3$
for type-II tilted Dirac fermions, one needs to introduce an UV cutoff
in $\theta$, which is designated as $\theta^\Lambda$ and determined by
the size of first Brillouin Zone~\cite{Lee2018PRB,Lee2019PRB}.
Following the strategy advocated in Ref.~\cite{Lee2019PRB}, we assume the maximum value
of $|p_2|$ is related to the lattice spacing $a_0$ at the scale $l$, namely
$|p_2|_{\mathrm{max}}=\pi/a_0$, and then arrive at for a specific
value of $E$ by resorting to the Eq.~(\ref{Eq_intergal-II}),
\begin{eqnarray}
\widetilde{p}_{2}=v_{2}p_{2}=\frac{|E|\sinh\theta_\Lambda}{\sqrt{\zeta^2-1}}
\approx\frac{|E|e^{\theta_\Lambda}}
{2\sqrt{\zeta^2-1}}\approx \frac{v_2\pi}{a_0}\equiv D,
\end{eqnarray}
with $\Lambda=O(D)$ (the basic results are insensitive to the ratio $D/\Lambda$~\cite{Lee2018PRB,Lee2019PRB}).
This yields to
\begin{eqnarray}
e^{\theta_{\Lambda}}=\frac{2\sqrt{\zeta^2-1}D}{|E|}.
\end{eqnarray}
Exploiting this approach and performing several calculations eventually
give rise to the compact one-loop corrections for type-II tilted Dirac fermions,
\begin{eqnarray}
S_{\xi}^{\lambda_{0}}
&=&\int_{-\infty}^{+\infty}\frac{dp_{0}
^{\prime}dp_{0}^{\prime\prime}dp_{0}^{\prime\prime\prime}}{(2\pi)^{3}}\int\frac{d^{2}\mathbf{p}^{\prime}
d^{2}\mathbf{p}^{\prime\prime}d^{2}\mathbf{p}^{\prime\prime\prime}}{(2\pi)^{6}}
[\psi_{\xi\alpha}^{\dagger}(p_{0}^{\prime},\mathbf{p}^{\prime})\sigma_{0}\psi_{\xi\alpha}
(p_{0}^{\prime\prime},\mathbf{p}^{\prime\prime})\psi_{\xi^{\prime}
\alpha^{\prime}}^{\dag}(p_{0}^{\prime\prime\prime},\mathbf{p}^{\prime\prime\prime})\sigma_{0}\nonumber\\
&&\times\psi_{\xi^{\prime}\alpha^{\prime}}(p_{0}^{\prime}+p_{0}^{\prime\prime}-p_{0}^{\prime\prime\prime},
\mathbf{p}^{\prime}+\mathbf{p}^{\prime\prime}-\mathbf{p}^{\prime\prime\prime})]
\frac{2\lambda_{0}\lambda_{1}\zeta^{\star}l}
{\pi^2\zeta|\zeta|v_{1}v_{2}},\\
S_{\xi}^{\lambda_{1}}
&=&\int_{-\infty}^{+\infty}\frac{dp_{0}^{\prime}dp_{0}^{\prime\prime}
dp_{0}^{\prime\prime\prime}}{(2\pi)^{3}}\int\frac{d^{2}\mathbf{p}^{\prime}
d^{2}\mathbf{p}^{\prime\prime}d^{2}\mathbf{p}^{\prime\prime\prime}}
{(2\pi)^{6}}[\psi_{\xi\alpha}^{\dagger}(p_{0}^{\prime},\mathbf{p}^{\prime})\sigma_{1}
\psi_{\xi\alpha}(p_{0}^{\prime\prime},\mathbf{p}^{\prime\prime})\psi_{\xi^{\prime}
\alpha^{\prime}}^{\dag}(p_{0}^{\prime\prime\prime},\mathbf{p}^{\prime\prime\prime})
\sigma_{1}\nonumber\\
&&\times\psi_{\xi^{\prime}\alpha^{\prime}}(p_{0}^{\prime}+p_{0}^{\prime\prime}-p_{0}
^{\prime\prime\prime},\mathbf{p}^{\prime}+\mathbf{p}^{\prime\prime}-\mathbf{p}
^{\prime\prime\prime})]\frac{(\lambda^2_0+\lambda^2_1
+\lambda^2_2+\lambda^2_3)\zeta^{\star}l}
{2\pi^2\zeta|\zeta| v_{1}v_{2}},\\
S_{\xi}^{\lambda_{2}}
&=&\int_{-\infty}^{+\infty}\frac{dp_{0}^{\prime}dp_{0}^{\prime\prime}
dp_{0}^{\prime\prime\prime}}{(2\pi)^{3}}\int\frac{d^{2}\mathbf{p}^{\prime}
d^{2}\mathbf{p}^{\prime\prime}d^{2}\mathbf{p}^{\prime\prime\prime}}
{(2\pi)^{6}}[\psi_{\xi\alpha}^{\dagger}(p_{0}^{\prime},\mathbf{p}^{\prime})\sigma_{2}
\psi_{\xi\alpha}(p_{0}^{\prime\prime},\mathbf{p}^{\prime\prime})\psi_{\xi^{\prime}
\alpha^{\prime}}^{\dag}(p_{0}^{\prime\prime\prime},\mathbf{p}^{\prime\prime\prime})
\sigma_{2}\nonumber\\
&&\times\psi_{\xi^{\prime}\alpha^{\prime}}(p_{0}^{\prime}+p_{0}^{\prime\prime}-p_{0}
^{\prime\prime\prime},\mathbf{p}^{\prime}+\mathbf{p}^{\prime\prime}-\mathbf{p}
^{\prime\prime\prime})]\frac{2[\lambda_{0}\lambda_{3}-\lambda_{2}\left(\lambda_{0}-\lambda_{1}
+2\lambda_{2}-\lambda_{3}\right)]\zeta^{\star}l}
{\pi^2\zeta|\zeta|v_{1}v_{2}},\\
S_{\xi}^{\lambda_{3}}
&=&\int_{-\infty}^{+\infty}\frac{dp_{0}^{\prime}dp_{0}^{\prime\prime}
dp_{0}^{\prime\prime\prime}}{(2\pi)^{3}}\int\frac{d^{2}\mathbf{p}^{\prime}
d^{2}\mathbf{p}^{\prime\prime}d^{2}\mathbf{p}^{\prime\prime\prime}}
{(2\pi)^{6}}[\psi_{\xi\alpha}^{\dagger}(p_{0}^{\prime},\mathbf{p}^{\prime})\sigma_{3}
\psi_{\xi\alpha}(p_{0}^{\prime\prime},\mathbf{p}^{\prime\prime})\psi_{\xi^{\prime}
\alpha^{\prime}}^{\dag}(p_{0}^{\prime\prime\prime},\mathbf{p}^{\prime\prime\prime})
\sigma_{3}\nonumber\\
&&\times\psi_{\xi^{\prime}\alpha^{\prime}}(p_{0}^{\prime}+p_{0}^{\prime\prime}-p_{0}
^{\prime\prime\prime},\mathbf{p}^{\prime}+\mathbf{p}^{\prime\prime}-\mathbf{p}
^{\prime\prime\prime})]\frac{2\lambda_{0}\lambda_{2}\zeta^{\star}l}
{\pi^2\zeta|\zeta|v_{1}v_{2}},
\end{eqnarray}
where the coefficients $\zeta^*$ and $\zeta^\star$ are introduced in Eq.~(\ref{Eq_zeta-star}).

\end{widetext}



\end{document}